\documentclass[a4paper,twoside]{report}
\usepackage[english]{babel}
\usepackage{footmisc}
\usepackage{epsfig}
\usepackage{float}
\usepackage{amssymb}
\usepackage{amsmath}
\usepackage{subfigure}
\usepackage{graphicx}
\begin{document}
\title{PWO Crystal Measurements and Simulation Studies of $\bar{\Lambda}$ Hyperon Polarisation for PANDA} 
\author{Licenciate Thesis \\Sophie Grape \\\\ 
\\\\\\\\\\\\ Department for Physics and Astronomy, Uppsala University} 
\date{January 2008}
\maketitle

\cleardoublepage
\chapter*{Abstract}
The Gesellschaft f\"ur Schwerionenforschung (GSI) facility in Darmstadt, Germany, will be upgraded to accommodate a new generation of physics experiments. The future accelerator facility will be called FAIR and one of the experiments at the site will be PANDA, which aims at performing hadron physics investigations by colliding anti-protons with protons. The licentiate thesis consists of three sections related to PANDA. The first contains energy resolution studies of $\mathrm{PbWO_4}$ crystals, the second light yield uniformity studies of $\mathrm{PbWO_4}$ crystals and the third reconstruction of the $\mathrm{\bar{\Lambda}}$-polarisation in the PANDA experiment.\\
\indent
Two measurements of the energy resolution were performed at MAX-Lab in Lund, Sweden, with an array of 3$\times$3 $\mathrm{PbWO_4}$ crystals using a tagged photon beam with energies between 19 and 56~MeV. For the April measurement, the crystals were cooled down to -15~$\mathrm{^{\circ}}$C and for the September measurement down to -25~$\mathrm{^{\circ}}$C. The measured relative energy resolution, $\sigma /E$, is decreasing from approximately 12\% at 20~MeV to 7\% at 55~MeV. In the standard energy resolution expression $\sigma/E= a/\sqrt{E} \oplus b/E \oplus c$, the three parameters $a$, $b$, $c$ seem to be strongly correlated and thus difficult to determine independently over this relative small energy range. The value of $a$ was therefore fixed to that one would expect from Poisson statistics of the light collection yield (50 phe/MeV) and the results from fits were $\sigma/E= 0.45\%/\sqrt{E_{GeV}} \oplus 0.18\%/E_{GeV}$ $\oplus 8.63\%$ and $\sigma/E= 0.45\%/\sqrt{E_{GeV}} \oplus 0.21\%/E_{GeV} \oplus 6.12\%$ for the April and September measurements, respectively. The data from the September measurement was also combined with previous data from MAMI for higher energies, ranging from approximately 64 to 715~MeV. The global fit over the whole range of energies gave an energy resolution expression of $\sigma/E= 1.6\%/\sqrt{E_{GeV}} \oplus 0.095\%/E_{GeV} \oplus 2.1\%$.\\
\indent
Light yield uniformity studies of five $\mathrm{PbWO_4}$ crystals, three tapered and two non-tapered ones, have also been performed. The tapered crystals delivered a light output which increased with increasing distance from the Photo Multiplier Tube (PM tube). Black tape was put on different sides of one tapered crystals, far from the PM tube to try to get a more constant uniformity profile. It was seen that the light output profile depends on the position of the tape. Generally, the steep increase in light output at large distances from the PM tube could be damped.\\ 
\indent
The third part of the thesis concerns the reconstruction of the $\mathrm{\bar{\Lambda}}$ polarisation in the reaction $\mathrm{\bar{p}p\rightarrow\bar{\Lambda}\Lambda}$. Events were generated using a modified generator from the PS185 experiment at LEAR. With a 100\% polarisation perpendicular to the scattering plane, a polarisation of (99$\pm$1.8)\% was reconstructed. Slight non-zero polarisations along the axis determined by the outgoing hyperon as well as the axis in the scattering plane, were also reconstructed. These were (4.1$\pm$2.1)\% and (2.6$\pm$2.0)\% respectively. From this investigation it was shown that the detector efficiency was not homogeneous and that slow pions are difficult to reconstruct.

\cleardoublepage
\tableofcontents
\cleardoublepage

\chapter{Introduction}
\label{intro}
The PANDA acronym stands for antiProton ANnihilation DArmstadt and it represents an international  physics collaboration consisting of more than 420 collaborators from 55 institutions in 17 countries. The detector was planned in the 1990's and is foreseen to start operating around year 2015. The main purpose of the PANDA detector is to do research with anti-protons and hadronic matter to gain better knowledge of the strong interaction.\\
\indent
The PANDA experiment will be carried out at FAIR, the Facility for Antiproton and Ion Research which will be built at the site of GSI, the Gesellschaft f\"{u}r Schwerionenforschung. The facility is located outside of Darmstadt in Germany. GSI was upgraded 15 years ago, allowing for a new heavy-ion accelerator. However, the future FAIR facility is more than an upgrade of GSI, it will allow for a whole new generation of medium energy physics experiments with anti-protons.\\
\indent
This licenciate thesis treats two topics of interest to the PANDA collaboration. The first concerns studies of interest to the electromagnetic calorimeter. It involves energy resolution measurements and light yield uniformity test for photons in PWO crystals. The second part involves simulations of the ability to correctly reconstruct $\mathrm{\bar{\Lambda}}$ hyperons and their polarisation. The two topics will be joined together via the electromagnetic calorimeter in future studies. The light $\mathrm{\Lambda}$ state is, in many cases, the decay product of heavier hyperons that either decay radiatively (emitting photons) or into particles which decay into photons. 
 
\chapter{Theoretical Background}
The Standard Model contains the theory of the electroweak interaction and the strong interaction (Quantum Chromo Dynamics, QCD). It incorporates the 12 fundamental particles we know of, three of their interactions as well as the carriers of these.

\section{Fundamental Particles}
There are two groups of fundamental particles carrying half-integer spin (fermions): quarks and leptons. The quarks are of six different flavours and are called the up-, down-, strange-, charm-, bottom- and top quarks. They are organised into three generations, depending on their mass and electric charge. Each generation consists of one positively and one negatively charged quark and includes particles which are lighter than the ones in the following generation. 

\begin{table}[htb]
\centering
\begin{tabular}{|l|c|c|c|c|c|}
\hline
Generation &Name &Charge (e) &Mass [GeV/$c^2$] &Spin\\
\hline
\hline
1 &Up (u) &+2/3 &0.0015-0.003 &1/2\\
 &Down (d)&-1/3 &0.003-0.007 &1/2\\
\hline
2 &Charm (c) &+2/3 &1.25$\pm$0.09 &1/2\\
 &Strange (s) &-1/3 &0.95$\pm$0.25 &1/2\\
\hline
3 &Top (t) &+2/3 &172-174 &1/2\\
 &Bottom (b) &-1/3 &4.2-4.7 &1/2\\
\hline
\end{tabular}
\caption{The six quarks and some of their properties \cite{pdg}.}
\label{tab:quarks}
\end{table}
\noindent
All quarks also carry the charge of the strong interaction which is called the colour charge. This charge comes in the varieties of red, green or blue. These charges solve the problem on how to separate identical fermions from each other according to the Pauli principle (which says that a fermion cannot be in the same quantum state as another fermion). If the colour charge did not exist, it would not be possible to separate the three s-quarks in the $\mathrm{\Omega^-}$-baryon or the u-quarks in $\mathrm{\Delta^{++}}$ from each other.\\
\indent
Individual quarks have never been found freely, they are always found in colour neutral configurations with two other quarks or one anti-quark. This feature is called confinement. The quarks are building blocks for so-called hadrons, strongly interacting particles, and they are divided into mesons and baryons. Mesons represent the quark-anti-quark ($\mathrm{q\bar{q}}$) configurations and have integer spin, while baryons are made up of three quarks and carry half-integer spin. There might be other configurations as well, but these two possibilities represent what is experimentally established today.\\ 
\indent
The leptons form the second group of these fundamental particles. They are also grouped into three generations.  

\begin{table}[htb]
\centering
\begin{tabular}{|l|c|c|c|c|c|}
\hline
Generation &Name &Charge (e) &Mass [MeV/$c^2$] &Spin\\
\hline
\hline
1 &Electron ($e^-$) &-1 &0.511 &1/2\\
 &Electron neutrino ($\nu_e$) &0 &$<2\cdot 10^{-6}$ &1/2\\
\hline
2 &Muon ($\mu^-$) &-1 &106.5 &1/2\\
 &Muon neutrino ($\nu_{\mu}$) &0 &$<0.19$ &1/2\\
\hline
3 &Tau ($\tau^-$) &-1 &1777 &1/2\\
 &Tau neutrino ($\nu_{\tau}$) &0 &$<0.018$ &1/2\\
\hline
\end{tabular}
\caption{The six leptons and some of their properties \cite{pdg}.}
\label{tab:leptons}
\end{table}

\section{Interactions}
There are four fundamental forces which govern the interactions in nature; the electromagnetic, the weak, the strong and the gravitational force. All but the last are incorporated into the Standard Model. The interactions are described by quantum field theory and their interactions are mediated by the quanta of the respective fields, the so-called gauge bosons. The gravitational force is much weaker than the other three forces and will not be considered here.\\
\indent
The electromagnetic force is mediated by the massless photon, making the range of the force infinite. This force keeps the electrons bound to the atomic nucleus and the atoms bound to other atoms in materials. Hadrons which decay with this type of interaction usually have life times of $\mathrm{10^{-16}-10^{-21}}$~s \cite{shaw}.\\
\indent
The weak force is mediated by the neutral Z boson and the flavour changing charged $\mathrm{W^{\pm}}$ bosons. Probably the most easily noticeable effect of this force is the radioactive decays where protons are transformed into neutrons or vice versa. Due to the heavy mass of these gauge bosons, the force only acts on small distances and the life times of decaying particles are typically $\mathrm{10^{-7}-10^{-13}}$~s \cite{shaw}.\\
\indent
The strong force is mediated by the massless gluons which carry both colour and anti-colour charge. At low energies it is useful to consider the hadronic degrees of freedom for the interaction instead of quarks and gluons. In this case the mediating particles are mesons, pions for short range interactions and omega for long range. The range of the strong force is about $\mathrm{10^{-15}}$~m and the decay times are typically $\mathrm{10^{-22}-10^{-24}}$~s \cite{shaw}. 
 
\section{Configurations and Symmetries}
\label{config}
There are certain rules that systems of quarks must obey. These rules are set by conservation laws of the so-called quantum numbers which characterise the system.\\
\indent
All interactions of the Standard Model conserve spin and angular momentum. The strong and electromagnetic interactions both conserve flavour, time reversal T, the charge conjugation quantum number C, the parity P and of course the combination of them (CP), while the weak interaction violates all of these symmetries (to some degree). CPT symmetry is the only symmetry obeyed by all three interactions.\\
\indent
Different states (particles) can be labelled using, for instance, the spectroscopic notation $n^{2S+1}L_J$ with $n$ being the main quantum number, \emph{S} the spin quantum number, \emph{L} the relative angular momentum quantum number and \emph{J} the total spin quantum number of the system. The total spin $J$ is expressed as the sum of \emph{L} and \emph{S}, $\mathrm{\bold{L}+\bold{S}}$. \\ 
\indent
Charge conjugation (C) is the operation where particles are replaced by their corresponding anti-particles in the same state. The C quantum number is given by \cite{shaw}

\begin{equation}
C_{\mathrm{boson}}=(-1)^{L},\mbox{    }
C_{\mathrm{fermion}}=(-1)^{L+S}.
\label{eq:C}
\end{equation}
The parity P for a meson and a baryon are expressed as \cite{shaw}

\begin{equation}
P_{\mathrm{meson}}=(-1)^{L+1},\mbox{    }
P_{\mathrm{baryon}}=(-1)^{L_{12}+L_3}
\label{eq:parity}
\end{equation}
where $\mathrm{L_{12}}$ and $\mathrm{L_3}$ are the internal angular momentum between two arbitrarily chosen quarks and the orbital angular momentum of the third quark about the center of mass of the pair.\\
\indent
In addition to the spectroscopic notation, one may add the quantum
numbers $J^{PC}$ of the configuration to more fully describe it.

\section{Physics of Interest to the PANDA Collaboration}
The PANDA experiment has many different physics objectives, mostly related to the strong interaction and some of them are mentioned below. The purpose of the PANDA hadron physics program is to study hadronic structures and hadronic interactions in the non-perturbative regime. New states will be searched for and possibilities for gluonic excitations such as hybrids and glueballs will be investigated \cite{rainerpwo}.

\subsection{Charmonium Spectroscopy}
Charmonium, the bound state of a charm quark and an anti-charm quark, is a very interesting configuration. The charm quark mass is relatively large, luckily heavy enough for non-relativistic calculations to be (barely) applicable \cite{qcd}. In addition, the strong coupling constant $\mathrm{\alpha_s}$ is fairly small for the system, $\mathrm{\approx 0.3}$, which makes it possible to use pertubative calculations \cite{qcd}. Charmonium states are also generally very narrow states, at least below the threshold of open charm production where the charmed quark pair must annihilate to create lighter quarks. Narrow states are easier to interpret, since the risk of having overlapping states is decreased and mixing effects between these states are generally small.\\
\indent
Charmonium studies started in $\mathrm{e^+e^-}$ collisions back in 1974. In these types of collisions, the quantum numbers of the intermediate photon, $J^{PC}$=$1^{--}$, dictates that only charmonium states with these quantum numbers can be directly created. However, if anti-protons are collided with protons, a whole new world of possibilities opens up. The initial system can have any quantum numbers that are available to a system comprising a fermion and an anti-fermion. The final state quantum numbers are given by the gluon(s) and quarks coming from the initial state. This makes it possible to end up with a broad range of allowed $J^{PC}$ quantum numbers. In the case of the created particle having a $J^{PC}$ that is ``forbidden'' according to the rules for the naive quark model mentioned in section \ref{config}, they are labelled ``exotic''\cite{shaw}. No such particles have so far been firmly established.

\subsection{Hybrids and Glueballs}
Hybrid and glueball configurations are thought to exist in parallel to the conventional hadrons. A hybrid is a meson state where gluonic excitations are present together with quarks, while a glueball is a state entirely built up by glue \cite{agnes}.\\
\indent
There are observed states which do not fully seem to fit into the naive quark model, where all hadrons can be described with three quarks or one quark and an anti-quark. For charmonium, this is the case e.g. for the recently observed so-called X, Y and Z states \cite{xyz}. Such states are candidates for being di-quarks, molecule states, exotic particles, hybrids or glueballs and the PANDA collaboration wishes to shed some light over this.

\subsection{Hyperons}
Hyperons are baryons with at least one s-quark. To conserve strangeness, they are always produced in a process where pairs of $\mathrm{\bar{s}s}$ quarks are created.\\
\indent
The proton and the $\mathrm{\Lambda}$ are assumed to have a di-quark-quark structure in the constituent quark model. The di-quark, being the ud-pair, is in an isospin and spin zero state and one may regard the di-quarks as spectators in the reaction $\mathrm{\bar{p}p\rightarrow \bar{\Lambda}\Lambda}$. This is important, since this implicates that the observables more directly reflect the dynamics of the underlying $\mathrm{\bar{u}u \rightarrow \bar{s}s}$-process \cite{trento}.\\
\indent
Studies have shown that $\mathrm{\bar{\Lambda}\Lambda}$ hyperon pairs are practically always produced with the $\mathrm{\bar{s}s}$ pair having parallel spins \cite{trento}. How this comes about is uncertain. Possibly, this could be a fundamental feature of the $\mathrm{\bar{s}s}$ production mechanism, or it could be related to a polarised $\mathrm{\bar{s}s}$-component inside the anti-proton/proton (polarisation meaning the direction, or orientation, of the spin). This intrinsic spin is however rather poorly known as it has been found that only a fraction of the spin is carried by the quarks \cite{trento}.\\ 
\indent
The different models give different predictions for the correlation between the initial proton spin and the final state $\mathrm{\Lambda}$ spin and it is still unclear how the polarisation arises and s-quarks are created \cite{trento}.

\subsection{Hypernuclei}
Hypernuclei are also of interest to PANDA. These are nuclei where (at least) one of the nucleons has been replaced by a hyperon. However, very different predictions for the spin-dependent contribution to the hyperon-nucleon interaction exist. A special $\mathrm{\gamma}$-ray detector will be available at PANDA for investigating excited hypernuclei by detecting the emitted photons from the de-excitation process with high resolution. With this technique, one will investigate the interactions between nucleons and hyperons. Also double hypernuclei and interactions between hyperons will be addressed \cite{tpr}.\\ 
\indent
Hyperatoms, where the atom contains a hyperon in an atomic orbit, are of interest for studies of hyperon properties. An especially interesting case is when the hyperon in the atomic orbit is a $\mathrm{\Omega^-}$-hyperon, because of its very long life time (82 ps) and its large spin of 3/2. A measurement of its electric quadrupole moment will give information on its shape, as well as the quark-quark interactions \cite{tpr}.

\chapter{FAIR and the PANDA Detector}
\label{panda}
\section{The GSI and FAIR Facilities}
Today the GSI facility includes a UNILNAC (heavy ion linear accelerator) delivering protons with an energy of up to 14~MeV/u, a heavy ion synchrotron (SIS) which accelerated particles to momenta of up to 2~GeV/u and an experimental storage ring (ESR) \cite{fair}. The future FAIR facility will be equipped with an additional double ring synchrotron (SIS100/300 for accelerations of heavy ion beams of up to 2.7~GeV/u and 34~GeV/u, respectively). The SIS100 ring will accelerate the protons which will be used to produce the secondary anti-proton beam. The ring has a circumference of 1100~meters and will be located 17~meters below ground. Three additional storage rings will be built: the CR (Collector Ring) where the anti-protons will be stochastically cooled, the NESR (New Experimental Storage Ring) and the HESR (High Energy Storage Ring). The HESR will store $10^{11}$ anti-protons with momenta between 1.5 and 15~GeV/c \cite{tpr}.\newpage

\begin{figure}[H]
\begin{center}
\includegraphics[bb=0 0 441 315,width=0.9\linewidth,angle=0]{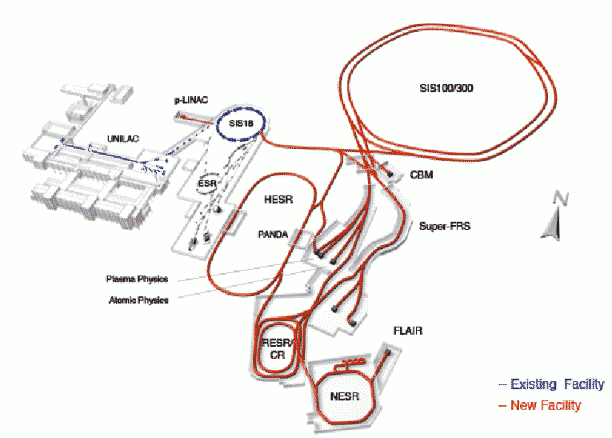}
\end{center}
\caption{The figure shows the existing GSI and future FAIR facilities, marked with blue and red respectively \cite{gsihp}.}
\label{fig:facility}
\end{figure}
\noindent
High intensity beams of anti-protons will be used for atomic-, nuclear- and particle physics at FLAIR, CBM will study relativistic heavy ion reactions \cite{gsidoc}. Radioactive nuclei beams having energies up to 1.5 GeV/nucleon will be available for Super FRS \cite{baseline}.\\
\indent
The cost of the new facility has been estimated to 1.2~billion Euros and it is planned to be completed in 2015 \cite{gsihp}.

\section{The PANDA Detector}
\begin{figure}[H]
\begin{center}
\includegraphics[bb=0 0 504 317,width=1.2\linewidth,angle=0]{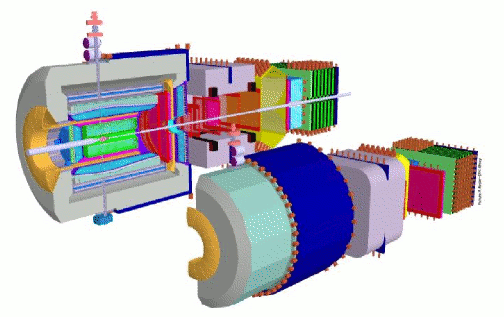}
\end{center}
\caption{The design of the complete PANDA detector, as it was
  described in the Conceptual Design Report in 2001 \cite{GSIconcept}.}
\label{fig:det}
\end{figure}

The PANDA detector, which is foreseen to be commissioned in 2014 or 2015, is one of the largest experiments at the new facility. It is designed to provide a nearly full coverage of the solid angle with excellent energy and angular resolution for neutral and charged decay particles. The detector layout can be seen in Figure~\ref{fig:det}.\\
\indent
The detector consists of two spectrometers: a target spectrometer (TS) with a superconducting solenoid and a forward dipole spectrometer (FS) for particles with opening angles of more than $\mathrm{\pm}$10$\mathrm{^{\circ}}$ in the horizontal and $\mathrm{\pm}$5$\mathrm{^{\circ}}$ in the vertical plane. The maximum opening angles in the FS are approximately 22$\mathrm{^{\circ}}$ in the vertical plane and slightly larger in the horizontal one.\\ 
\indent
More information on the topics in this chapter can be found in the PANDA Technical Design Report \cite{GSIconcept}.
\newpage

\subsection{The Target Spectrometer}
\begin{figure}[H]
\begin{center}
\includegraphics[bb=0 0 382 254,width=0.9\linewidth,angle=0]{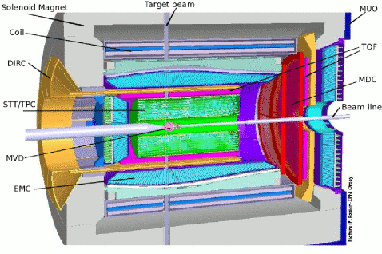}
\end{center}
\caption{The layout of the target spectrometer with
  its sub-detectors \cite{GSIconcept}.}
\label{fig:ts}
\end{figure}

The target spectrometer (TS) has a cylindrical geometry which surrounds the immediate interaction region and reaches out to a radius of about 2 meters. It can be seen in Figure~\ref{fig:ts} and includes the target system, a micro vertex detector (MVD), a straw tube tracker (STT) or alternatively a time projection chamber (TPC), a time-of-flight (TOF) detector, a detector for internally reflected Cherenkov light (DIRC) and an electromagnetic calorimeter (EMC). The coil of the solenoid magnet is placed outside of these sub-detectors. Muon detectors are placed outside the coil. 

\subsubsection{The Target}
The target system for PANDA must deliver a target thickness that gives a luminosity of $\mathrm{2\cdot 10^{32}}$ $\mathrm{/cm^2s}$. Assuming $\mathrm{10^{11}}$ stored anti-protons in the HESR, this translates into a target thickness of about $\mathrm{4\cdot 10^{15}}$~hydrogen atoms per $\mathrm{cm^2}$. Two alternatives, a cluster jet target and a pellet target, have been proposed.\\ 
\indent
The cluster jet target is an internal gas system which uses a continuous stream of hydrogen cluster gas that is being directed at the interaction region. A continuous flow can be delivered but the desired target density has not been reached yet.\\
\indent
The pellet target is an approach which uses frozen droplets of hydrogen (pellets). Hydrogen gas is liquefied and cooled down before being injected into a low pressure helium environment in form of a jet, which later breaks up to a uniform train of droplets. It is believed that this method can deliver the desired effective target thickness of $\mathrm{4\cdot 10^{15}}$~atoms/$\mathrm{cm^2}$. 

\subsubsection{The Micro Vertex Detector}
The micro vertex detector (MVD) is a radiation hard silicon detector, especially designed to detect secondary vertices of, for example, the decays of strange and charmed hadrons. Therefore it is of utmost importance that it is located close to the interaction point.\\
\indent
The detector features a barrel section, most likely consisting of four layers and six forward discs. The two innermost barrel layers will be made with pixel geometry and the forward discs will contain a mix of pixels and strips \cite{lars}. The pixel size will most likely be (100$\times$100)~$\mathrm{\mu m^2}$ to ensure good resolution  and radiation hardness close to the interaction point \cite{lars}. The two outermost barrel layers will consist of silicon strip detectors. The estimated spatial resolution of the detector is 100~$\mathrm{\mu}$m. 

\subsubsection{Tracking with the STT and the TPC}
The outer tracking system consists of two parts, one which will
be either of the straw tube tracker (STT) or the
time projection chamber (TPC) type, and a second one consisting of
Multi-wire Drift Chambers (MDCs) or Gas Electron Multipliers (GEMs).\\
\indent
The STT is a system of self-supporting gas filled straw detectors, arranged in 11 cylindrical and skewed double layers. The innermost layer has a radius of 16~cm and the outermost a radius of 42~cm. The total length of the detector will be 1.5~m. Charged particles entering the detector will produce electrons and positive ions that will drift in different directions in an electric field. Close to the wire, which is on positive voltage, avalanche amplification will occur and the electrons will be collected here while the ions drift towards the cathode. The resolution perpendicular to the beam line is about 150~$\mathrm{\mu}$m, depending on the drift distance \cite{stt}. The coordinates in the beam direction for this detector can be obtained in two ways. The first way is to use the charge division technique. The length-dependent wire resistivity affects the amplitude of the output signals and when reading out this at both straw ends, one can calculate where the interaction took place. The second way is to use the geometry of the skewed straws. The first technique is expected to give a resolution 0.5-1\% of the sensitive wire length (which translates into 7-15~mm), a value which is approximately 2-3 times larger than the resolution from the second method. The drift time in the detector depends on the gas mixture filling the straws, but varies between tens of nano seconds up to a few hundred nano seconds \cite{stt}.\\
\indent
The TPC is a much more complex detector than the STT and it is expected to give the best particle identification below momenta of 1~GeV/c. The detector itself is straight forward. However, the read-out electronics is very expensive and the online reconstruction is complicated. The TPC consists of two large gas filled cylindrical volumes with an electric field applied in the direction of the beam line. The field will separate electrons from positive gas ions created by traversing particles and the electrons will drift towards the readout anode end cap of the cylinder. Avalanche amplification will occur in Multi Wire Proportional Chambers (MWPCs), with the charge amplification most likely coming from GEMs. The read-out at the end cap will give two-dimensional information on the projection of the track. The third coordinate comes from drift time measurement of the primary electron clusters. The resolution for secondary vertices is foreseen to be 150~$\mathrm{\mu m}$ in $\mathrm{r \varphi}$-direction and 1~mm along the beam axis.\\
\indent
After the STT/TPC there will be either two MDCs or two GEM detectors in order not to lose information on charged particles in the gap after the STT/TPC which would otherwise exist in the detector.

\subsubsection{Charged Particle Identification}
Charged particle identification in the target spectrometer is done using information from many sub-detectors. For instance, energy loss per path length in a medium is a useful method for particle identification when the signal amplitude, as well as space coordinates, are known. This is not a problem for the TPC-option, but for the STT it poses a challenge since not as many measurements per track are performed and therefore fluctuations in $dE/dx$ can be large. Other identification techniques include time-of-flight measurements and
Detection of Internally Reflected Cherenkov (DIRC) light.\\
\indent
The PANDA time-of-flight (TOF) stop counters will provide a stop signal with respect to the start signal (given most likely by the MVD close to the interaction point) as a particle traverses the target spectrometer. Given that the particle is not too fast in relation to the time resolution, one can obtain velocity information for the particle. The TOF will consist of two parts, a barrel shape outside the tracker and an end cap in the forward spectrometer. Both consist of plastic scintillators with channel-plate photo multiplier read-out that can operate in magnetic
fields up to 2.2~T.\\
\indent
The DIRC identifies particles with momenta up to several GeV/c using totally internally reflecting Cherenkov photons and the best identification is done for momenta above 1~GeV/c. As particles enter the quarts bar, some of the radiated Cherenkov photons will always be internally reflected. These photons can be focused onto an array of photo multipliers or avalanche photo diodes where the Cherenkov angle is measured from the radius of the Cherenkov ring. This ring can be used to determine the velocity of the particle. The velocity is then used for particle identification, together with the momentum information from the drift chamber.

\subsubsection{The Electromagnetic Calorimeter}
The electromagnetic calorimeter is by far the single most expensive sub-detector. It must be able to detect photons with both high and low energy, meaning that it must give position and timing resolution over a wide dynamic range from tens of MeV up to several GeV. The proposed material for this is lead tungsten, $\mathrm{PbWO_4}$, a radiation hard and compact crystal which is a recently developed scintillator that has been chosen for other high-energy physics experiments such as CMS and ALICE at CERN.\\ 
\indent
The barrel part of the calorimeter will be 2.5~m long and filled with 11360 tapered crystals of 18 different shapes making sure there is a tilt towards the interaction point and as small gaps as possible between the individual crystals. The length of the crystals in the barrel part is expected to be 20~cm ($\approx$22~radiation lengths), while the 3864 crystals in the forward end cap may be longer \cite{mdr}. The backward end cap will contain 816 crystals.

\begin{figure}[H]
\centering
\includegraphics[bb=0 0 236 504,width=0.4\linewidth, angle=270]{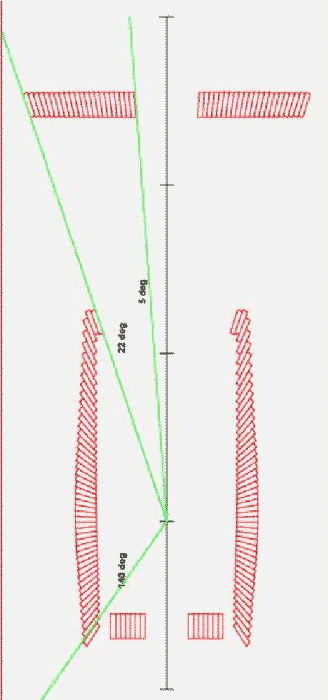}
\caption{The electromagnetic calorimeter in the target
  spectrometer with the barrel part, as well as the forward and
  backward end caps \cite{GSIconcept}.}
\label{fig:ecalparts}
\end{figure}
\noindent
Because the calorimeter will be located in of the solenoid, the
read-out has to be made using light sensors that are insensitive to
magnetic fields. This excludes the choice of photo multiplier
tubes and most likely the read-out will be made using Avalanche Photo Diodes (APDs) in the barrel and the end-cap. Vacuum triodes are considered for the forward end-cap due to the high count rate in this
region.\\ 
\indent
As the light yield of $\mathrm{PbWO_4}$ is relatively low compared to many other scintillators used in calorimeters, much effort goes into increasing the light yield. One way to do this is to cool the detector, as will be discussed in section \ref{pwo}. 

\subsubsection{The Magnet System}
Outside of the calorimeter there will be a superconducting coil with an inner radius of 90 cm and a length of 2.8~m, generating a field
strength of 2~T.

\subsubsection{Muon Detectors}
Muon detection will be done using one of three alternatives. The first is to use scintillator counters for time-of-flight measurements, the second is to use electromagnetic and hadronic calorimetry to measure $dE/dx$ and the third to use muon tracking. The muon tracking can be done either using Mini-Drift Tubes based on the Iarocci principle but operated in proportional mode, or drift tubes similar to those used for CMS at CERN. Also a combination of both types of mini-drift tubes is possible.

\subsection{The Forward Spectrometer}
The forward spectrometer consists of a large, normally conducting dipole magnet, six Multi-wire Drift Chambers (MDCs), possibly a Ring imaging Cherenkov Detector (RICH), a second electromagnetic calorimeter (F-EMC), a hadronic calorimeter (H-EMC) and a muon detector.

\begin{figure}[H]
\centering
\includegraphics[bb=0 0 390 250,width=0.9\linewidth, angle=0]{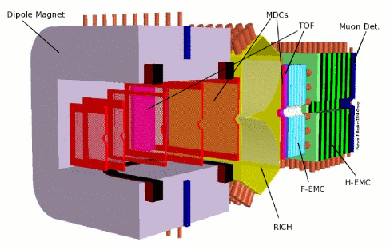}
\caption{The forward spectrometer with named sub-detectors \cite{GSIconcept}.}
\label{fig:fs}
\end{figure}

\subsubsection{The Magnet System}
The dipole magnet in the forward spectrometer will bend the charged particles to allow for a momentum analysis. The maximum bending power is 2~Tm, causing a bending of 2$\mathrm{^{\circ}}$ for the most energetic particles. The anti-proton beam will be deflected and bent back using a chicane to prevent interference.

\subsubsection{Tracking}
Particles emitted at angles lower than 22$\mathrm{^{\circ}}$ will not be fully covered by the central tracking and therefore it was initially suggested to put additional MDCs located 1.4 and 2~m downstream of the target, inside the magnet. Another pair of planar MDCs were discussed to be placed after the magnet to measure the deflections in the forward spectrometer dipole magnet, as well as a third pair located in the dipole magnet gap to trace low momentum particles.\\ 
\indent
The drift chambers are planned to be 1~cm thick and contain squared drift cells made up from cathode and sense wires mounted on self-supporting frames. The first two MDCs contain four pairs of octagonal detection planes in different angles, while the others are grouped in three double layers.

\subsubsection{Particle Identification}
The time-of-flight (TOF) wall will be located approximately 7~m from the interaction point. It is equipped with strips of plastic scintillators with photo multiplier read-out. The expected time resolution is 50~ps, which will be enough to distinguish pions from kaons at 2.8~GeV/c and pions from protons up to 4.7~GeV/c. A Ring Imaging Cherenkov Detector (RICH) will be probably be required for particle identification at higher momenta.

\subsubsection{The Forward Electromagnetic Calorimeter}
The forward electromagnetic calorimeter is planned to be a Shashlyk-type detector with alternating layers of lead and plastic scintillators for detecting photons and electrons. The scintillators are used for detection, while the lead layers act as energy absorbers and photon converters. The read-out will be done using wavelength shifting fibres and photo multipliers. 

\subsubsection{The Hadronic Calorimeter}
The second part of the forward calorimeter is the multi-purpose hadronic calorimeter. Firstly, it is designed to measure neutral hadrons like neutrons and anti-neutrons which are not detected anywhere else. Secondly, it will serve as a fast trigger for reactions with forward scattered hadrons. Thirdly, it will act as a muon filter for the muon detectors placed at the very end.\\
\indent
The calorimeter which will be used for this already exists. It comes from the WA80 experiment at CERN and has an electromagnetic and a hadronic section. The scintillator used in this detector is called PS-15A and it is based on polymethylmethacrylate (PMMA). 

\subsubsection{The Muon Detectors}
The final design for this detector part is not finished but it is under discussion to use the same principle as for the target spectrometer muon tracking.

\chapter{Energy Measurements with Crystals}
\label{egyres}
\section{Particle Interactions in Scintillators}
\subsection{Photon Interactions with Matter}
\begin{figure}[H]
\begin{center}
\includegraphics[bb=0 0 277 180,width=0.7\linewidth,angle=0]{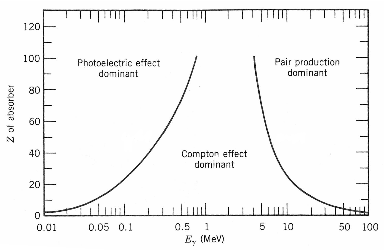}
\end{center}
\caption{Different photon interactions with matter and their
  dominating regions \cite{krane}.} 
\label{fig:interaction}
\end{figure}

There are three principal ways photons can interact with matter: via the
\emph{photoelectric effect}, \emph{Compton scattering} and
\emph{pair production}. The probability for the processes are strongly dependent on the energy and the atomic number of the material (Z), as can be seen in Figure~\ref{fig:interaction}. 

\emph{Photoelectric absorption} dominates for low energies, where the incoming photon ejects an electron from the material, resulting in a released electron with an energy equal to the energy of the photon minus its binding energy with which the electron was bound \cite{krane}. Experimental results have indicated a cross-section \cite{avhandling}

\begin{equation}
\sigma_{\mathrm{ph}}\propto \frac{Z^5}{E_{\gamma}^{7/2}}
\label{eq:phe1}
\end{equation}
in the low energy regime and \cite{avhandling}

\begin{equation}
\sigma_{\mathrm{ph}} \propto \frac{Z^5}{E_{\gamma}}
\label{eq:phe2}
\end{equation}
for $E_{\gamma}\gg m_e c^2$, where $Z$ is the atomic number of the material and $E_{\gamma}$ the photon energy.\\
\indent
\emph{Compton scattering} is a process in which the incoming photon scatters from a loosely bound atomic electron, which can be considered to be at rest. The result is a scattered photon and a scattered electron sharing the available energy. The cross-section for this process reduces to the Thomson scattering cross section at low energies. 

\begin{equation}
\sigma d\Omega=\frac{e^2}{4\pi m c^2}(\bold{\epsilon_i \cdot
  \epsilon_f})^2 d\Omega
\label{eq:thomson}
\end{equation}
with $\epsilon$ being the polarisation of the initial and final photon \cite{mandl}. This is a non-relativistic description of scattering of electromagnetic radiation. At energies where $E_{\gamma}\gg m_e c^2$, the cross-section for Compton scattering is proportional to \cite{avhandling}

\begin{equation}
\sigma_{\mathrm{Co}} \propto \frac{Z\ln{E_{\gamma}}}{E_{\gamma}}.
\label{eq:co}
\end{equation}
\indent
\emph{Pair production} is a process occurring in the neighbourhood of a nucleus (to conserve momentum), in which the photon converts into a electron-positron pair in the presence of an electromagnetic field. The threshold energy is twice the electron mass and the cross-section of the process can be approximated as

\begin{equation}
\sigma_{\mathrm{pair}}\propto Z^2 \ln{2E_{\gamma}}.
\label{eq:pair}
\end{equation}
Pair production is related to \emph{bremsstrahlung} where electromagnetic radiation is emitted as a result of an electrically charged particles being scattered in an electric field \cite{leo}. Since bremsstrahlung depends on the strength of the electric field, screening of the nucleus from the surrounding electrons is an important factor. Also for pair production this will be the case. The cross-section for pair production thus depends on the screening effect parameter $\xi$ given by \cite{leo} 

\begin{equation}
\xi=\frac{100m_ec^2 E_{\gamma}}{E_{e^+}E_{e^-} Z^{1/3}}.
\label{eq:xi}
\end{equation}
When $\xi$=0, there is complete screening and for $\xi$=1 there is
no screening. For photon energies 

\begin{equation}
E_{\gamma}\gg \frac{m_e c^2}{\alpha  Z^{1/3}}
\label{eq:Eg}
\end{equation}
$\mathrm{\xi}\longrightarrow0$, giving complete screening \cite{leo}. Here, $\alpha$ is the electromagnetic coupling constant. When there is no screening, one can calculate an energy-independent expression for the pair production cross-section \cite{leo}

\begin{equation}
\frac{1}{\lambda_{\mathrm{pair}}}=N\sigma_{\mathrm{pair}}\approx
\frac{7r_e^2}{9}
\label{eq:pair2}
\end{equation}
with $N$ being the density of atoms. It is related to the radiation
length $X_0$ (see section \ref{x0}) through \cite{leo}

\begin{equation}
\lambda_{pair}=\frac{9}{7} X_0.
\label{eq:pair3}
\end{equation}

An electromagnetic cascade with continuous pair production spreads in both transversal and longitudinal direction. A measure of the former is given by the so-called \emph{Moli{\`e}re radius} of the scintillator.

\subsection{Electron Interactions with Matter}
Electrons scatter via Coulomb interactions in the material and due
to their low mass, they will be largely deflected. Depending on how
they are scattered, they will travel different distances, or ranges, in the material. In addition, due to the scattering they will change the direction and magnitude of their velocity and therefore be subjected to accelerations and emit bremsstrahlung \cite{krane}.\\
\indent
The expressions for the energy losses per unit path length that the electron suffers is given by the Bethe-Bloch equation \cite{krane}, which has contributions from both collisional and radiative losses 

\begin{equation}
\frac{dE}{dx}={\left(\frac{dE}{dx}\right)}_{\mathrm{coll}}+{\left(\frac{dE}{dx}\right)}_{\mathrm{rad}},
\label{eq:bethe}
\end{equation}

\begin{multline}
{\left(\frac{dE}{dx}\right)}_{\mathrm{coll}} = \left(\frac{e^2}{4\pi
  \epsilon_0} \right)^2 \frac{2\pi N_0 Z \rho}{m c^2 {\beta}^2A} \\
\times \left(
  \ln{ \frac{T{(T+m c^2)}^2 {\beta}^2}{2 I^2 m c^2}} +(1-{\beta}^2) - 
(2\sqrt{1-{\beta}^2} -1+{\beta}^2)\ln{2}
  +\frac{1}{8}{(1-\sqrt{1-{\beta}^2})}^2 \right)
\end{multline}

\begin{equation}
{\left(\frac{dE}{dx}\right)}_{\mathrm{rad}}=\left(\frac{e^2}{4\pi \epsilon_0}
\right)^2 \frac{Z^2 N_0 (T+m c^2)\rho}{137 m^2 c^4 A}\left(
4ln\frac{2(T+m c^2)}{m c^2}-\frac{4}{3}\right)
\label{eq:bether}
\end{equation}
with \emph{T} being the kinetic energy of the electron, $N_0$ Avogadro\'s constant, \emph{Z} the atomic number, \emph{A} the atomic weight and $\rho$ the density of the material which the electron traverses. The electron mass is denoted \emph{m}.\\
The radiative term plays a larger role for high energies and heavy materials.

\subsubsection{Radiation Length}
\label{x0}
Radiation length is a concept frequently used in describing the characteristics of a detector material. It corresponds to the distance the electron has travelled when its energy has been reduced by a factor $1/e$, due to radiation losses only. For the high energy limit where collisional losses can be ignored to radiative ones, the radiation length becomes basically independent of the energy and is given by \cite{leo} 

\begin{equation}
\frac{1}{X_0}\approx \frac{4r_e^2\alpha \rho N_A
  Z(Z+1)}{A} \ln{\left(\frac{183}{Z^{1/3}}\right)}
\label{eq:rad}
\end{equation}
where $r_e$ is the classical radius of the electron, $N_A$
Avogadro\'s constant and A the atomic number.

\section{Energy Resolution}
When measuring a quantity (the incoming $\gamma$ energy in this case) there are always errors associated with the measurement, which makes the measured value fluctuate around an average value. In this particular case contributions come from statistical fluctuations, due to the Poisson statistics of the collected light in a scintillator, fluctuations associated with electronic noise and other instrumental effects. The relative influence of these different effects are generally not known in detail, but can be estimated from the energy dependence of the measured total fluctuation of the signal (the RMS-width $\sigma$ or the Full Width at Half Maximum, FWHM, of the peak in a measurement where the incoming photon energy is known). If we assume that the measured quantity x depends on many parameters u, v, ..., x=f(u,v,...), then the variance of x can be expressed as \cite{bevington}

\begin{equation}
{\sigma}^2_x = \lim_{N \rightarrow \infty} \frac{1}{N} \Sigma \left(
  (u_i-\bar{u})\left( \frac{\partial x}{\partial u}\right) +
  (v_i-\bar{v})\left( \frac{\partial x}{\partial v}\right) \right)^2. 
\label{eq:variance}
\end{equation}
For uncorrelated quantities, the above relation reduces to 

\begin{equation}
\sigma^2_x = \lim_{N \rightarrow \infty} \frac{1}{N} \Sigma \left(
  (u_i-\bar{u})^2\left( \frac{\partial x}{\partial u}\right)^2 +
  (v_i-\bar{v})^2\left( \frac{\partial x}{\partial v}\right)^2
  \right) = \sigma^2_u\left( \frac{\partial x}{\partial u}\right)^2+
  \sigma^2_v\left( \frac{\partial x}{\partial v}\right)^2.
\label{eq:uncorr}
\end{equation}
Thus we see that the variance can be written as a sum of individual contributions $\sigma_{x,u}$=$|\sigma_u \cdot dx/du|$, $\sigma_{x,v}$=$|\sigma_v \cdot dx/dv|$... For detecting photons from a scintillating crystal, one contribution is due to the Poisson statistics of the light collection process. Since the variance in the number of photo electrons at the cathode equals that number, the contribution $\sigma_{E,Poisson}$ to the uncertainty of the measured energy is proportional to the square root of the energy:

\begin{equation}
\sigma_{\mbox{\emph{E, Poisson}}}=a \cdot \sqrt{E}
\label{eq:poi}
\end{equation}
For scintillators having a high light yield this term is expected to only give a small contribution to the relative energy resolution, since the number of photons produced per incoming MeV is relatively large. For $\mathrm{PbWO_4}$ (see section \ref{pwo}) this is not the case, it is therefore very important to ensure a high efficiency in collecting the photons which are created. This can be done using a good reflective wrapping material and a good optical coupling between the PM tube and the crystal \cite{avhandling}.\\
\indent
The electronic noise describes the errors arising from the electrical set-up used for the measurements. The noise depends on the actual setting of the electronics such as high voltage etc, but does not depend on the signal strength and is thus independent of the energy:

\begin{equation}
\sigma_{\mbox{\emph{E, Noise}}}=b
\label{eq:noi}
\end{equation}

Lastly, one could in addition expect some fluctuations in the measured signal due to crystal properties such as non-uniformity of the produced light inside the crystals, temperature gradients, detector ageing, radiation damage etc. For a system of crystals errors in the inter-calibration will contribute. These fluctuations will be proportional to the signal strength, thus proportional to the energy:

\begin{equation}
\sigma_{\mbox{\emph{E, Crystal}}}=c \cdot E
\label{eq:sig}
\end{equation}
This term often dominates the energy resolution because the two other terms tend to be small \cite{avhandling}. Only for detectors where special care has been taken to prevent shower leakage and to inter-calibrational errors, this term can be manageable \cite{lyartikel}.\\
\indent
The energy resolution of scintillating crystals is thus often written as:

\begin{equation}
\sigma^2=\sigma_{\mbox{\emph{E, Poisson}}}^2 + \sigma_{\mbox{\emph{E, Noise}}}^2 + \sigma_{\mbox{\emph{E, Crystal}}}^2 =a^2 E + b^2  + c^2 E^2
\label{eq:step1}
\end{equation}
This can also be written as \cite{pdg}

\begin{equation}
\frac{\sigma}{E}=\frac{a}{\sqrt{E}} \oplus \frac{b}{E} \oplus c,
\label{eq:resolution}
\end{equation}
where the $\oplus$ sign indicates quadratic summing. 

\subsection{Energy Resolution for PANDA}
\begin{figure}[H]
\begin{center}
\includegraphics[bb=0 0 567 479,width=0.7\linewidth,angle=0]{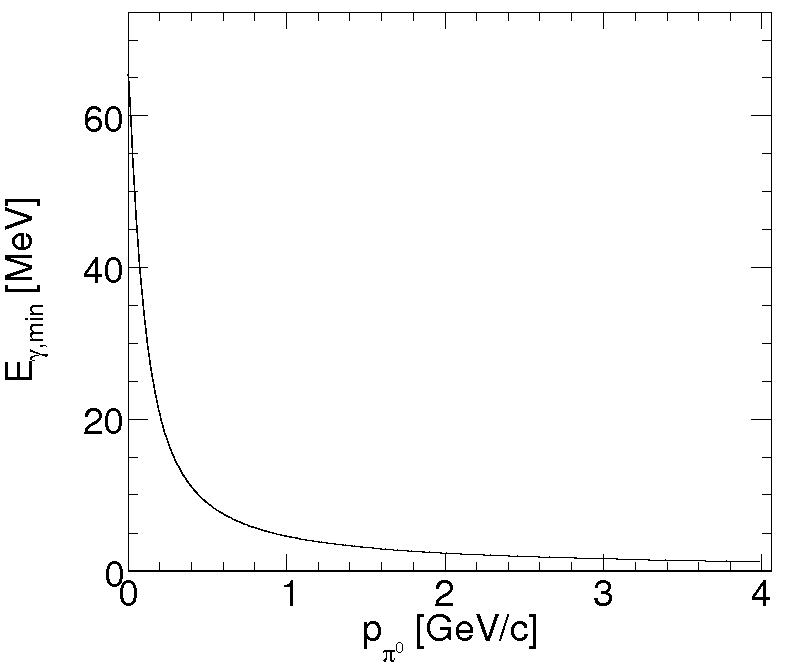}
\end{center}
\caption{The minimum energy of a decay photon as a function of the pion momentum \cite{jan}.}
\label{fig:janplot}
\end{figure}

The electromagnetic calorimeter plays a decisive role for most of the physics programs of PANDA and it must be able to cover a very large dynamic range (from tens of MeV to several GeV) of photons. Low energy thresholds are required for proper scans of mass and widths of channels with photons coming from isolated decays (photons from other decays but $\mathrm{\pi^0}$) such as $\mathrm{\bar{p}p \rightarrow \eta_c \rightarrow \gamma \gamma}$ and $\mathrm{\bar{p}p \rightarrow h_c \rightarrow \eta_c \gamma \rightarrow \gamma \gamma \gamma}$. The problematic backgrounds come from the high cross-section channels such as $\mathrm{\bar{p}p \rightarrow \pi^0 \gamma \rightarrow \gamma \gamma \gamma}$ and $\mathrm{\bar{p}p \rightarrow \pi^0 \pi^0 \rightarrow \gamma \gamma \gamma \gamma}$, where one photon is not detected \cite{jan}. These channels pose big challenges as the signatures look the same as for the true signal. For example, upper limits for the signal-to-background ratio for $\mathrm{\bar{p}p \rightarrow \eta_c \rightarrow \gamma \gamma}$ have been estimated for different energy thresholds, assuming 100\% detector efficiency\cite{alex}. For a threshold of 15~MeV the ratio was 1.75, for 10~MeV it was 2.82 and for 5~MeV it was 7.6. Corresponding Geant4 simulations have given signal-to-background ratios of 1.1 for 25~MeV and 0.7 for 50~MeV. The results were based on an energy resolution where $\sigma_{\mbox{\emph{E, noise}}}$=1.3~MeV. 50 photo electrons were assumed to be emitted per MeV at -25~$^{\circ}$C. \cite{tpr}.\\
\indent
Other problems come from the low mass of the pions and the forward boost of the system. This can cause very low energy photons to be emitted (for instance, a 1~GeV/c pion can emit a 4~MeV photon \cite{jan}) and if such a photon is lost it is not possible to distinguish the signal from the background. The dependence of the photon energy on the momentum of the pion is shown in Figure \ref{fig:janplot}.\newpage

\noindent
A third reason for the importance of a good calorimeter is to distinguish radiative (charmonium) decays from (for instance) charmed hybrids or glueballs that involve pions or etas ($\mathrm{\bar{p}p \rightarrow \chi_{c1,c2,c3} \rightarrow J/\psi \gamma}$ with a background of $\mathrm{\bar{p}p \rightarrow J/\psi \pi^0}$). Either the $\pi^0$ is needed to reconstruct the particle itself, or to reject the background. This is why the PANDA collaboration envisages a detector which can measure photon energies down to approximately 10~MeV. A high efficiency in detecting particles is crucial and a good energy resolution desired.\\
\indent
Excellent energy resolution is needed in the range of 100~MeV-1~GeV where many important channels decay to $\mathrm{\pi^0}$, and $\eta$, which then decay into photons (such as for instance $\mathrm{\psi \rightarrow J/\psi \rightarrow \pi^0 \pi^0}$, $\mathrm{X \rightarrow \chi_{c1} \rightarrow \pi^0 \pi^0}$, $\mathrm{X \rightarrow \eta_c \pi^0 \pi^0}$). The mass of a particle decaying to two photons is measured by the invariant mass $M^2$,

\begin{equation}
M^2=(E_1 + E_2)^2-(\textbf{p}_1 + \textbf{p}_2)^2=\frac{1}{c^2}\sqrt{2E_1 E_2 (1-\cos \alpha)}
\label{eq:pimass}
\end{equation}
where $E_1$ and $E_2$ are the energies of the decay particles, $p_1$ and $p_2$ the momentum vectors and $\alpha$ the angle between them. The mass resolution is dominated by the resolution of the lowest energy photon

\begin{multline}
\sigma_{m_{\gamma \gamma}}=\sqrt{ \left(\frac{\partial m_{\gamma \gamma}}{\partial E_1} \sigma_{E_1}\right)^2 + \left(\frac{\partial m_{\gamma \gamma}}{\partial E_2}\sigma_{E_1} \right)^2 + \left(\frac{\partial m_{\gamma \gamma}}{\partial \alpha}\sigma_{\alpha} \right)^2}\\
=\frac{m_{\gamma \gamma}}{2}\sqrt{ \left(\frac{\sigma_{E_1}}{E_1}\right)^2 + \left(\frac{\sigma_{E_2}}{E_2}\right)^2 + \left(\frac{\sin \alpha}{1-\cos \alpha}\sigma_{\alpha}\right)^2}
\end{multline}
\noindent
It is therefore important to ensure a good detection of the low energy photon so that the decay particle can be identified.\\
\indent
The granularity (position resolution) is given by geometrical constraints of the sub-detectors as well as the scintillator material, and it is important to have a good enough position resolution to reconstruct the opening angles of the $\mathrm{\pi^0}$. This is mainly a problem for high $\mathrm{\pi^0}$-momenta since it implies small opening angles. This effect is most important for the forward directions.

\section{$\mathrm{PbWO_4}$ Scintillator Characteristics}
\label{pwo}
\begin{figure}[H]
\begin{center}
\includegraphics[bb=0 0 395 230,width=0.7\linewidth,angle=0]{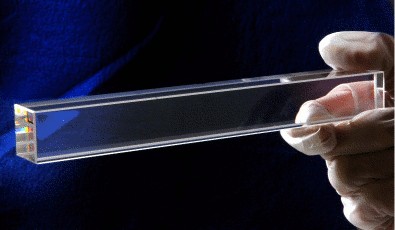}
\end{center}
\caption{One of the PWO crystals which has been delivered to Uppsala
  for future measurements of energy resolution and light yield uniformity. Photograph by \cite{teddy}.} 
\label{fig:pwo}
\end{figure}

Lead tungsten crystals, $\mathrm{PbWO_4}$ or PWO, were developed for the new
generation of high-energy physics experiments at LHC, CERN. Today it is
being used in the electromagnetic calorimeter of CMS, in PHOS and in
the photon spectrometer of ALICE. A photograph of a typical crystal can be
seen in Figure~\ref{fig:pwo}. 

The crystal development processes for these experiments have yielded high-quality and radiation hard crystals. More specifically, it seems the doping of the crystals is the key to limiting the reduction of the optical transmission to tolerable levels \cite{rainerpwo}.  Adding of trivalent rare earth ions (having atomic numbers between 58 and 70) to the crystal lattice makes inner shell transitions possible \cite{Rappe} and decreases cation and anion (i.e. positively and negatively charged ion) vacancies in the crystal. Unfortunately, addition of these ions also creates shallow electron centres which quench the scintillator light \cite{rainerpwo}. Some properties of lead tungstate are displayed in Table~\ref{tab:property}.

\begin{figure}[H]
\begin{center}
\includegraphics[bb=0 0 464 396,width=0.6\linewidth,angle=0]{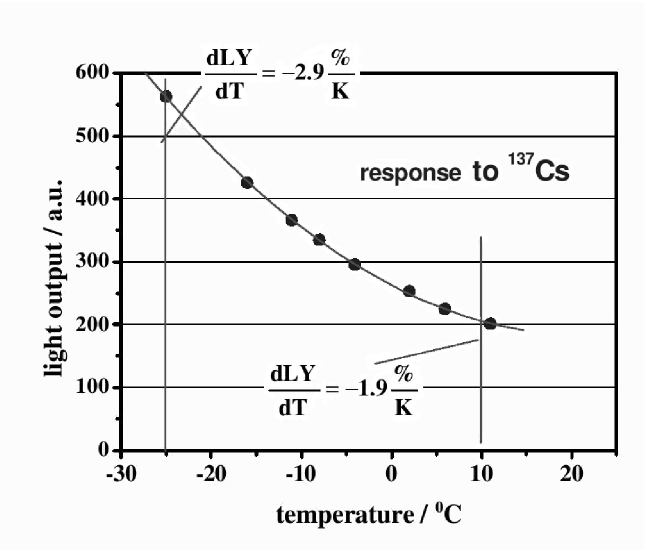}
\end{center}
\caption{The light yield of PWO as a function of the temperature \cite{loehner}.}
\label{fig:temp}
\end{figure}

\begin{table}[htb]
\centering
\begin{tabular}{|l|c|}
\hline
Property &PWO \\
\hline
\hline
Density [g/$\mathrm{cm^3}$] &8.28 \cite{pandadoc}\\
\hline
Radiation length [cm] &0.89 \cite{pandadoc}\\
\hline
Moli{\`e}re radius [cm] &2.2 \cite{lyartikel}\\
\hline
Refractive index  &2.3 \cite{pandadoc} \\
\hline
Decay time [ns]  &5/15/100 \cite{lyartikel}\\
\hline
Light Yield at 18 $^{\circ}C$ [phe/MeV] &20 \cite{pandadoc}\\
\hline
\end{tabular}
\caption{Some properties of lead tungsten. ``phe'' is short for ``photo   electrons'' and the three decay times correspond to the fast, medium and slow components.}
\label{tab:property}
\end{table}
\noindent
The very high density and short radiation length of PWO allows for a very compact detector. The high index of refraction is a very good quality since it reduces the risk of light scattering out of the crystal. The fast decay time allows for a high count rate.\\
\indent
The doping of PWO is essential to increase the low light yield, and so far PANDA has investigated crystals doped with impurities of Mo, La, Tb and Y \cite{pandadoc}. The light yield from PWO crystals has been measured to approximately 25 phe/MeV at room temperature \cite{Linda}. However, the light yield from PWO is very temperature dependent and increases with about 2\% per lowered degree C at 10~$^{\circ}$C, see Figure~\ref{fig:temp}. 

\chapter{Energy Resolution Measurements with PANDA Crystals}
\section{The Tagged Photon Facility at MAX-Lab in Lund}
\begin{figure}[H]
\begin{center}
\includegraphics[bb=0 0 468 325,width=1.0\linewidth,angle=0]{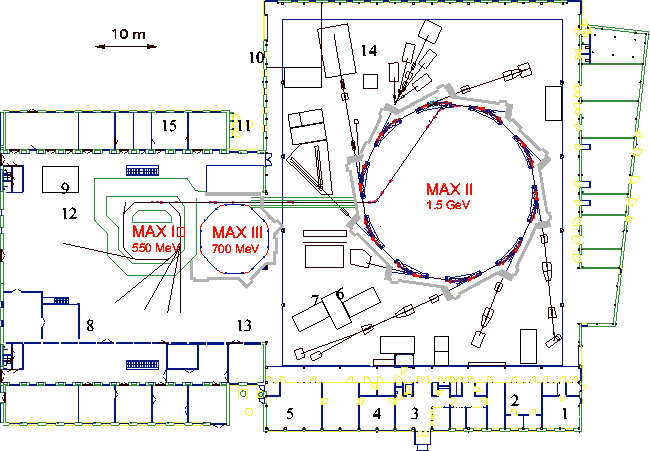}
\end{center}
\caption{Overview of the MAX-Lab facility \cite{maxanv}.}
\label{fig:maxlab}
\end{figure}

The electron accelerator facility MAX-Lab in Lund has been used to investigate the response of PWO crystals at low energies. The facility consists of three rings called MAX~I, MAX~II and MAX~III that are used for research with synchrotron radiation (electromagnetic radiation emitted when ultra-relativistic charged particles move through a magnetic field). An overview of MAX-Lab can be seen in Figure~\ref{fig:maxlab}.\\ 
\indent
The first step of the accelerator system is the pre-accelerator system. It consists of an electron gun, a linear accelerator and a recirculation system. After passing these three stages the electrons have reached an energy of 250-500~MeV. At this point they are injected into the storage rings where they are further accelerated. The energy of the electrons in the MAX I storage ring is approximately 550~MeV, about 1.5~GeV in the MAX~II ring and 700~MeV in the MAX~III ring \cite{maxlaborange}. \\
\indent
For nuclear physics applications, the electrons from MAX~I are extracted and transferred to the tagging spectrometer region. Here they will impinge on a radiator and photons will be emitted due to bremsstrahlung. The post-bremsstrahlung electrons are detected with a spectrometer \cite{taggingatmaxlab}.\newpage

\begin{figure}[H]
\begin{center}
\includegraphics[bb=0 0 858 293,width=1.0\linewidth,angle=0]{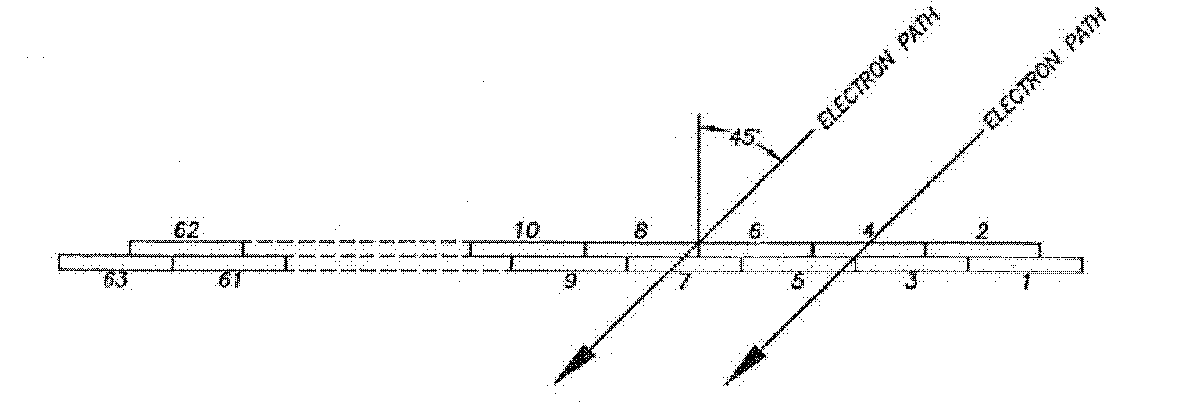}
\end{center}
\caption{The scintillator arrangement for the
  tagging system at MAX-Lab in Lund \cite{adler}.}
\label{fig:fp}
\end{figure}

\noindent
There are two tagging spectrometers, the first of which is a so-called end-point tagger capable of tagging photons close to the bremsstrahlung end point. The second tagger is the main tagger which can handle larger momentum values \cite{adler}.\\
\indent
The tagging system consists of two rows of overlapping plastic scintillators, 31 in the first row and 32 in the back row. All scintillators are 25~mm wide and they overlap to 50\% of their width  in the plane perpendicularly to the electron paths, see Figure~\ref{fig:fp}. The tagger signal is generated when a coincidence between two overlapping scintillators is registered. In total there are 62~tagged focal plane
 channels \cite{adler}.

\section{Measurement Set-up}
\begin{figure}[H]
\centering
\subfigure[The copper shell with the cooling pipes.]{\includegraphics[bb=0 0 238 317,width=0.35\linewidth,angle=0]{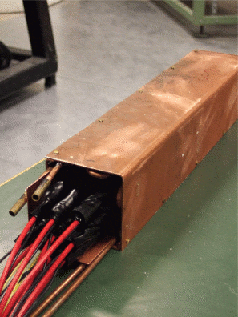}
\label{fig:copp}}
\subfigure[The set-up in the isolating box.]{\includegraphics[bb=0 0 238 317,width=0.35\linewidth,angle=0]{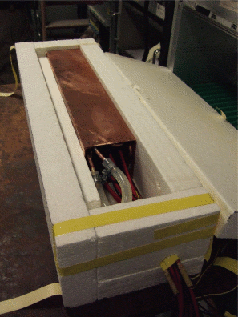}
\label{fig:copper}}
\caption{The crystal set-up used for the first measurements.}
\label{fig:setupboth}
\end{figure}

Two different sets of runs have been performed at MAX-Lab, one taking place in April 2007 and the other in September 2007. The purpose of the measurements was to investigate the energy resolution of PWO crystals between 19 and 55~MeV. Both measurements involved cooling, but the equipment used was more advanced for the September measurement. In addition to better and more stable cooling, the time information from the crystal read-out was saved during this run measurement and was later used during the analysis for background rejection.\\ 
\indent
The crystal set-up used for both experiments was a 3$\times$3-array of PWO crystals from Bogoroditsk in Russia, each with the dimension 2$\times$2$\times$20 $\mathrm{cm^3}$. Also a tenth PWO crystal was used and put on top of the set-up, perpendicular to the other nine crystals, to act as a detector for cosmic muons.\\ 
\indent
The signals were read out using Philips XP1911 Photo Multiplier Tubes (PM tubes). The polished crystal surfaces were wrapped with the mirror-like reflective foil VM2000 provided by 3M \cite{Rainer}. The crystals were attached to the PM tubes with VISCASIL silicon fluid (by General Electric) as an optical coupling, before being covered with black shrinking tape to prevent light leakage and to increase the stability.\\
\indent
For cooling, two different set-ups were used. For the April measurement, a small cooling machine with circulating cooling liquid was connected to a copper box surrounding the crystals. The copper block was then put inside an insulating box and kept with an over-pressure of nitrogen to prevent air from leaking in. The set-up is shown in Figure~\ref{fig:setupboth}. The temperature at which the measurement was performed was -15~$^{\circ}$C. Thermo elements were used to measure the temperature. The monitoring of the temperature was done with a web camera which was directed at the display of the thermo elements read-out.

For the September measurement, a climate chamber (V\"otsch~4021) was
available, in which it was possible to put the whole crystal array. It was
cooled to -25~$\mathrm{^\circ}$C, with an uncertainty of 0.1~$\mathrm{^\circ}$C. The climate chamber included a machine which dehumidified the air to ensure no ice would form on the cabling inside the chamber. The temperature inside the crystal array was not measured during the run, but from earlier investigations it
was known that temperature inside the array stabilised around the set
value after approximately 2h. 

\begin{figure}[H]
\centering
\subfigure[The crystal array inside the chamber.]{\includegraphics[bb=0 0 277 208,width=0.4\linewidth,angle=0]{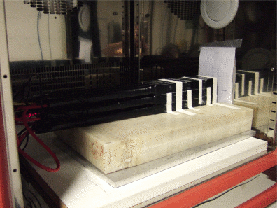}
\label{fig:arr}}
\subfigure[The climate chamber in place at MAX-Lab.]{\includegraphics[bb=0 0 285 214,width=0.4\linewidth,angle=0]{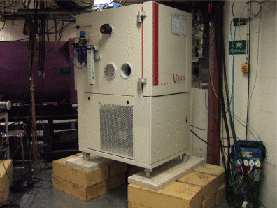}
\label{fig:clim}}
\caption{The crystal set-up for the September
  measurements with the crystal array  and the climate chamber}
\label{fig:setupbothsept}
\end{figure}

The position of the beam spot was mapped using a laser to make sure
the photons would go into the center crystal. For the second
measurement, the beam was let in through a hole in the side of the 
chamber which was covered with a rubber lid. The probability for
photon interaction in this material is very low and most photons will
pass right through it. Those few photons that do interact are most
likely scattered out of the direction of the beam and will not cause any
problems.
\newpage

\section{The Read-Out Electronics}
\label{elapril}
\begin{figure}[H]
\centering
\includegraphics[bb=0 0 763 536,width=1.0\linewidth, angle=0]{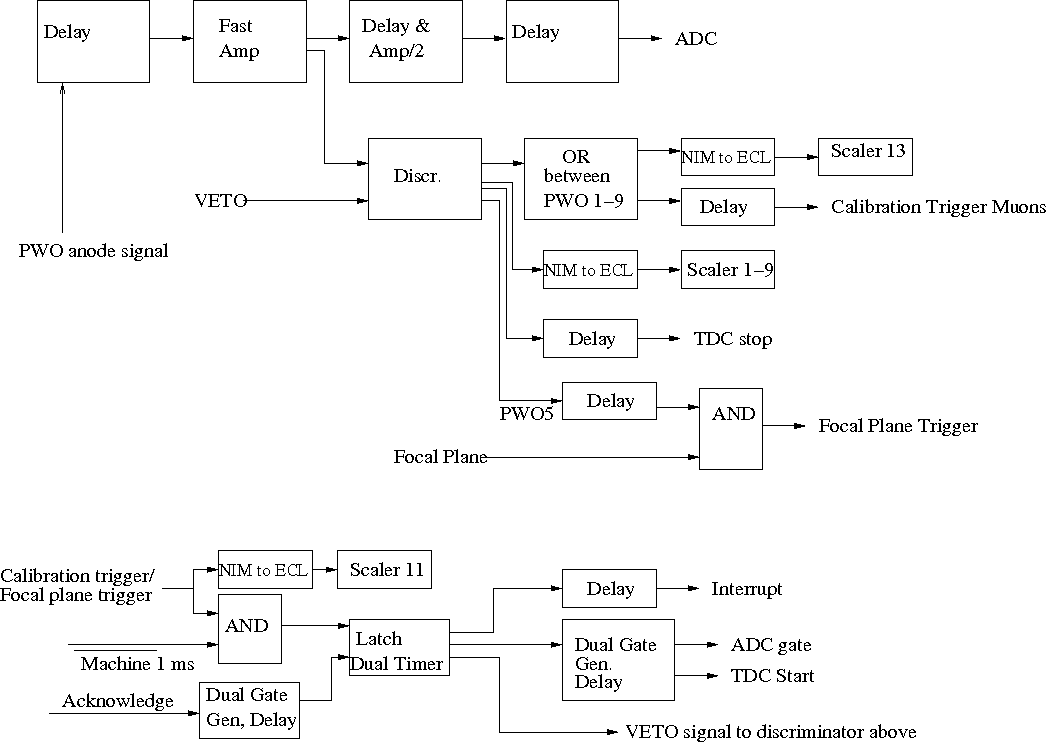}
\caption{The electronic set-up used for the September
  experiments.}
\label{fig:elsetup}
\end{figure}

The electronical set-up used in both measurements were basically
identical and can be seen in Figure~\ref{fig:elsetup}.
The signals from the nine crystals in the array were amplified and delayed in order to meet the timing requirements. Two different triggers could be used, one for triggering on cosmic muons for calibration purposes and one for triggering on the signal from the tagger focal plane in coincidence with the central crystal. The trigger took the data acquisition system into account by making sure that data was recorded when a detector had triggered and that no new events were processed while the system was busy. The $\mathrm{\overline{Machine}}$ corresponds to a signal from the accelerator, inhibiting any trigger generation during the first 1 ms of the machine cycle. The ``Acknowledge'' is a signal sent from the data acquisition system to mark that the information has been saved and the system is ready to treat new signals.\\ 
\indent
The only difference between the electronical set-ups used for the two measurements is that the timing information from the PWO signals was not recorded for the April measurement, but for the September measurement it was. In April, the timing was adjusted so that the true coincidences were recorded, but there was no TDC-information and therefore it was not possible to reduce the number of random coincidences by narrowing down the time interval.

\section{Analysis}
\label{analysis}
\begin{figure}[H]
\centering
\includegraphics[bb=0 0 567 491,width=1.0\linewidth, angle=0]{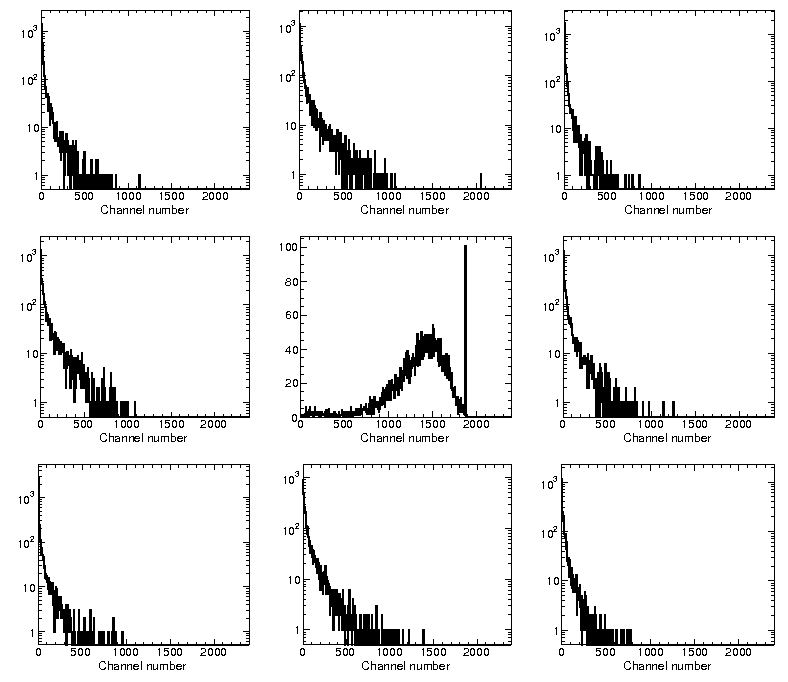}
\caption{A typical energy deposits in the nine PWO crystals from the
  September measurements.}
\label{fig:9pwo_sept}
\end{figure}

As soon as the photons reach the center crystal of the array, the shower process begins in both lateral and transversal directions, resulting in energy deposits in the central as well as in the surrounding crystals. The raw spectra for the September measurements can be seen in Figure~\ref{fig:9pwo_sept}. The spike in the central detector around channel number 1900 is an overflow peak, which collects signals with higher energies than the maximum value and puts then in a certain (or a few) bin(s).

\noindent
To obtain the energy resolution of the matrix, these energy contributions must be summed event wise. This is done using the CERN analysis program ROOT \cite{root}, but first all nine detectors must undergo a relative calibration using the zero point energy as well as another energy point. A threshold level for the addition of contributions was set to prevent noise from being added.\\
\indent
The timing information from the 62 focal planes detectors, and in the
case of the September measurement also the center PWO timing information,
were used to add the energy contributions for each event. The 
resulting peak which was obtained was then fitted with a Gaussian
distribution and the mean position as well as the sigma were used to
determine the energy resolution.

\section{Relative Calibration}
\begin{figure}[H]
\centering
\includegraphics[bb=0 0 568 479,width=1.0\linewidth, angle=0]{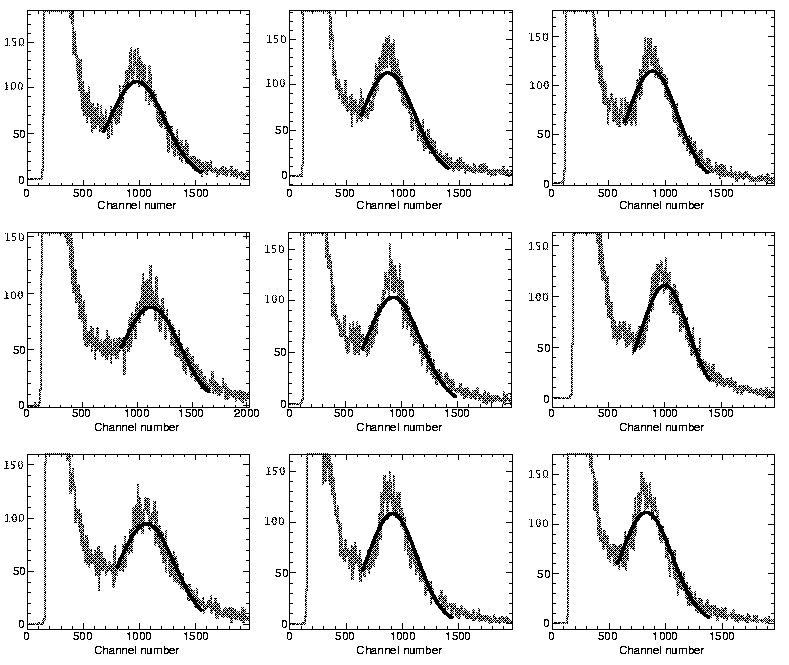}
\caption{The recorded muon spectrum from the September measurements.}
\label{fig:muon_sept}
\end{figure}

For the April measurement, a pedestal run was performed where the trigger signal came from the tenth crystal, located on top of the crystal array. The zero point energy could be extracted by fitting a Gaussian distribution to the noise peak. The second energy point was taken from the muon spectrum which was recorded during an over-night run. The threshold levels used in the analysis were chosen such that they were just above the energy at which the pedestal peak ended. The numerical values were between 0.3 and 0.9~MeV for the nine crystals, the large values stemming from some very wide pedestals.\\
\indent
Correspondingly, for the September measurement the two calibration points were taken from the zero point energy and a muon spectrum. The zero point energy was obtained from a pedestal run and the peaks were fitted with Gaussian distributions to obtain a mean value. The thresholds were determined in the same way as for the April measurements. The intervals for the thresholds were between 0.2 and 0.5~MeV. The second energy used for the calibration came from detected cosmic muons and the spectra can be seen in Figure~\ref{fig:muon_sept}. To get the position of the peak, Gaussian distributions were fitted around the muon peak.

\begin{figure}[H]
\centering
\includegraphics[bb=0 0 567 473,width=0.7\linewidth, angle=0]{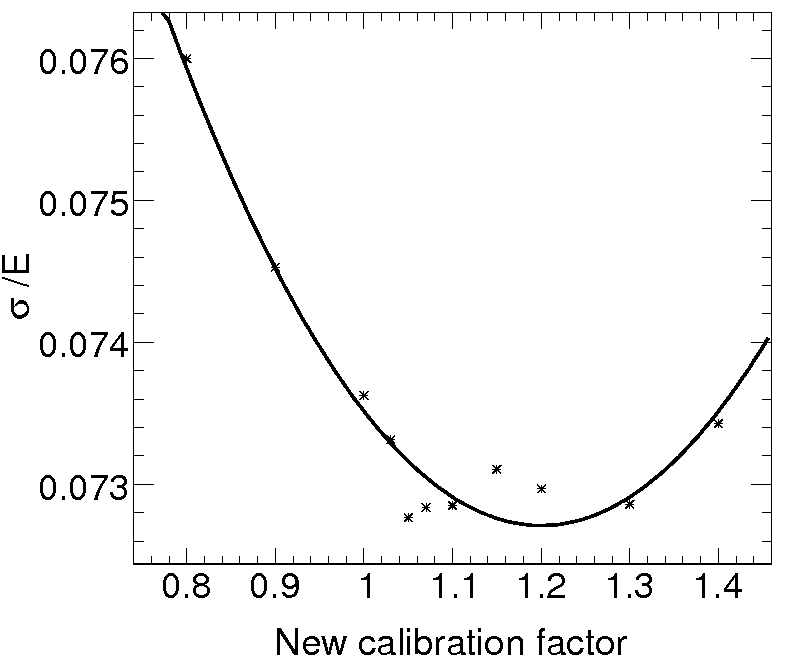}
\caption{A second degree polynomial fitted to energy resolution versus the new calibration factor for one of the eight surrounding crystals at an incoming photon energy of 51.6 MeV. The new calibration factor describes how much the old calibration factor for this specific crystal should change to give the lowest energy resolution.}
\label{fig:second}
\end{figure}

\noindent
The widths ($\sigma$) of the muon peaks vary between 6.5 and 7.6~MeV for the nine crystals, the center crystal having a muon peak with  $\sigma$=6.9~MeV. Depending on the interval chosen around the peak, the peak position changes by some hundred keV ($<$0.5~MeV).\\
\indent
Generally speaking, one may encounter some calibrational problems when using cosmic muons for calibration and a tagged low energy photon beam for measurement. The problem arises because the energy deposits inside the crystals from the muons and the photon beam take place at different locations. The cosmic muons will hit the crystal from above, along the whole length. The photons are directed to the front end side of the crystal array and will deposit their energy in that part. If the light yield along the crystal is uniform, this is not a problem. However, in chapter \ref{homo} where light yield uniformity is investigated, one clearly sees a dependence of the light yield on the distance between the incoming photon and the PM tube. If, however, the light non-uniformity is identical for all detectors, the relative calibration is not affected.\\
\indent
For one of the non-tapered crystals (crystal label $\mathrm{20\_216}$) wrapped in VM2000, the average number of emitted photo electrons per incoming MeV (phe/MeV) over the whole crystal length is 40.1. If one only considers the two data points which are located the farthest away from the PM tube, this number changes to 38.2 ($\approx$95.2\% of the light yield of the whole crystal). The corresponding numbers for the second VM2000-wrapped non-tapered crystal are 38.9 and 37.6~phe/MeV ($\approx$96.6\% of the light yield of the whole crystal). The difference between the two crystals is 1.4\%, which is not very much. However, as this study has not been done for the crystals in the array we do not know for sure if this effect is negigible. To be on the safe side though, it would be better to use a source which irradiates the crystals from the front end side for future calibrations. Alternatively, one could demand, by a coincidence arrangement, that the muons pass the relevant parts of the crystals.\\
\indent
An investigation was performed for the September measurement to study if the calibration could be improved. For each crystal, a new calibration factor in the range 0.80 to 1.40 of the old one was tried in order to search for a minimum in the relative energy resolution. This was done for photon energies 24.5 and 51.6~MeV. In Figure~\ref{fig:second} the result of such an optimisation for the detector below the central one is shown. A second order polynomial fit yields an additional calibration factor of 1.2 to optimise the resolution. The final calibration factor for each crystal was taken as the average of the two calibration factors obtained for the low and the high energy. The new calibration factors ranged between 1.0 from 1.2 times the old factor, with six of them being in the interval of 1.0-1.1. The energy resolution was improved (from 0.0127 to 0.0126 at 18.9~MeV and from 0.074 to 0.072 at 51.6~MeV). 

\noindent

\section{Results from Measurements below 60 MeV, April 2007}
\begin{figure}[H]
\centering
\includegraphics[bb=0 0 567 498,width=0.7\linewidth, angle=0]{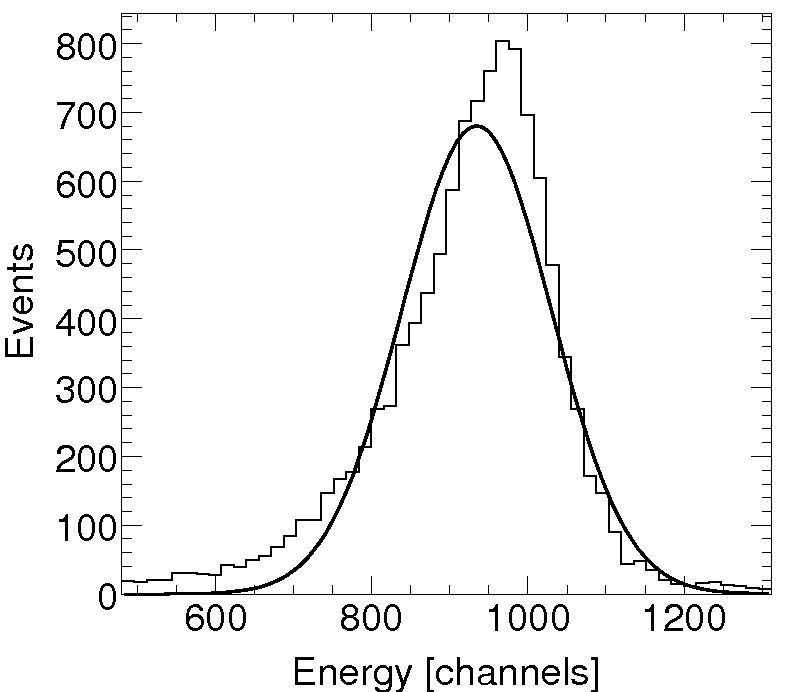}
\caption{A typical summed event spectrum for an incoming photon energy $\mathrm{E_{\gamma}}$=35~MeV, from the run in April. A Gaussian distribution (full drawn line) was fitted to the data.}
\label{fig:gauss_april_fp1}
\end{figure}

The 61 working taggers corresponded to photon energies ranging from 19.0
to 55.6~MeV. The relative calibration was performed and the contributions from the nine crystals were added as described in section \ref{analysis} and the resulting energy peaks were fitted with Gaussian distributions as shown in Figure~\ref{fig:gauss_april_fp1}. \newpage

\begin{figure}[H]
\centering
\includegraphics[bb=0 0 567 489,width=0.7\linewidth, angle=0]{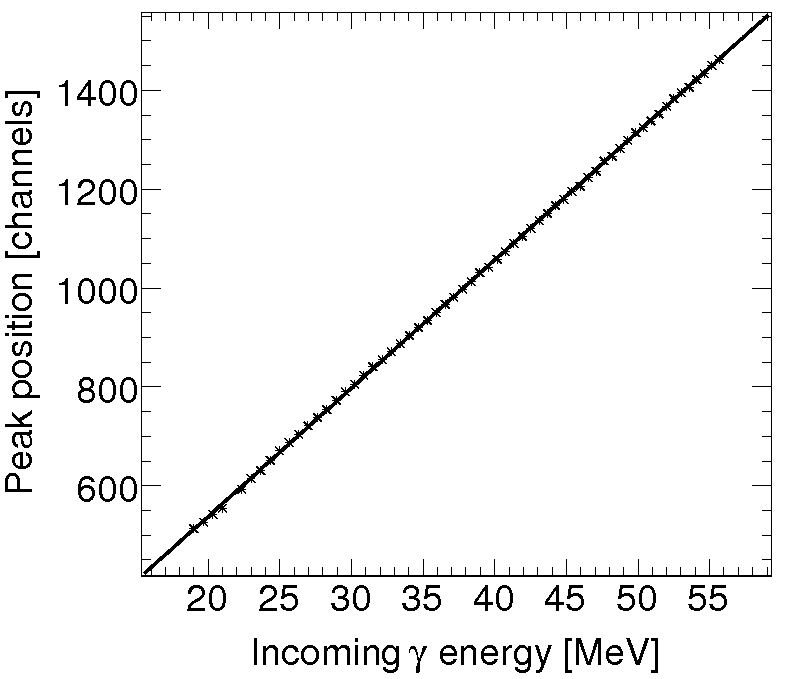}
\caption{The channel number for the Gaussian centroids as a function of incoming photon energy. A straight line fit is also shown, $\mbox{peak position} = 19.6 + 25.9[MeV^{-1}] \cdot E_{\gamma}$. The $\chi^2$/d.o.f. of the fit is 11.7.}
\label{fig:calib_april}
\end{figure}

The Gaussian distribution was used to given a simple description of the system. As the fit is not perfect, one may imagine two contributions (one from the signal and one from the leakage out of the crystal array) to the peak shape. The signal information was obtained from fitting the region corresponding to half of the height of the left hand side and the full right hand side of the peak. The relative energy resolution $\sigma /E$ decreases with 16\% for $E_{\gamma}$=21.0~MeV and with 19\% for $E_{\gamma}$=53.0~MeV when doing this.\\
\indent
In Figure~\ref{fig:calib_april} the fitted peak position are shown as a function of the incoming photon energy. As expected, there is a clear linear dependence.\newpage
 
\begin{figure}[H]
\centering
\includegraphics[bb=0 0 567 486,width=0.7\linewidth, angle=0]{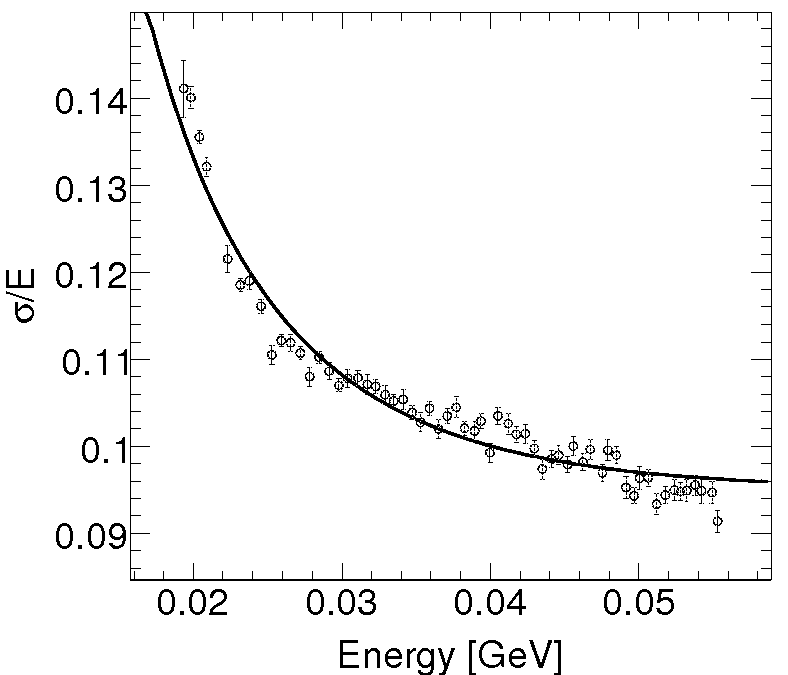}
\caption{The measured relative energy resolution (open circles) as a
  function of tagged photon energy. The full drawn line marks the fit of the standard energy resolution expression in Equation \ref{eq:resolution}. The value for the square of the $a$-parameter becomes negative and therefore all parameters from the fit are presented squared in Table~\ref{tab:res_apr}. The $\chi^2$/d.o.f. of the fit is 5.09.} 
\label{fig:resol_april}
\end{figure}

To obtain the energy resolution, the mean value of the fitted Gaussian distribution of the summed energy peak was assumed to correspond to the incoming tagged photon energy. The width ($\sigma$) was given by the fit. The relative resolution, $\mathrm{\sigma /E}$, as a function of the incoming photon energy is shown in Figure~\ref{fig:resol_april}. The value E was taken from the tagged photon energy and assumed to correspond to the mean value of the Gaussian distribution. A full drawn line fitted to the data is also shown in the same figure. The function describing the data is

\begin{equation}
\frac{\sigma}{E} = \sqrt{ \frac{a^2}{E}+\frac{b^2}{E^2}+c^2}
\label{eq:fiteq}
\end{equation}
where the parameters $a^2$, $b^2$ and $c^2$ are determined by minimising the $\chi^2$-value of the fit. The reason for using this fit function instead of one where $a$, $b$ and $c$ are fitted, is that forcing the square of the parameters to positive puts $a$ to zero.\newpage

\begin{table}[htb]
\centering
\begin{tabular}{|l|c|}
\hline
  &Value\\
\hline
\hline
$a^2$ [$GeV$] &(-1.64$\pm$0.24)$\cdot 10^{-4}$ \\
\hline
$b^2$ [$GeV^2$] &(6.30$\pm$0.38)$\cdot 10^{-6}$ \\
\hline
$c^2$ &(1.016$\pm$0.036)$\cdot 10^{-2}$ \\
\hline
\end{tabular}
\caption{Squared values for the parameters $a$, $b$ and $c$, taken from the fit in Figure~\ref{fig:resol_april}.} 
\label{tab:res_apr}
\end{table}
\noindent
The covariance matrix for this fit is:

\[
\emph{$COV_{April}$}=\left(
\begin{array}{lll}
5.7\cdot 10^{-10}   &-8.9\cdot 10^{-12}  &-8.4\cdot 10^{-9} \\
-8.9\cdot 10^{-12}  &1.4\cdot 10^{-13}   &1.3\cdot 10^{-10} \\
-8.4\cdot 10^{-9}   &1.3\cdot 10^{-10}   &1.3\cdot 10^{-7} \\
\end{array}
\right)
\]
\noindent
The correlations between parameters are calculated according to Equation \ref{eq:corr}, 

\begin{equation}
\mathrm{corr_{xy}}=\frac{COV(X,Y)}{\sigma_X \sigma_Y}
\label{eq:corr}
\end{equation}
where X and Y denote the parameters in question. Inserting numbers gives the correlations that are presented in Table~\ref{tab:corrtabapril}. Very large correlations (or anti-correlations) between the parameters are observed. The conclusion is that the three parameters cannot be independently determined by fitting in this limited energy interval.

\begin{table}[htb]
\centering
\begin{tabular}{|l|c|}
\hline
  &Correlation\\
\hline
\hline
corr$_{a^2 b^2}$ &-0.99\\
\hline
corr$_{a^2 c^2}$ &-0.99\\
\hline
corr$_{b^2 c^2}$ &0.96\\
\hline
\end{tabular}
\caption{Correlations between the squared energy resolution fit parameters for the April measurement.} 
\label{tab:corrtabapril}
\end{table}

\begin{figure}[H]
\centering
\includegraphics[bb=0 0 567 478,width=0.7\linewidth, angle=0]{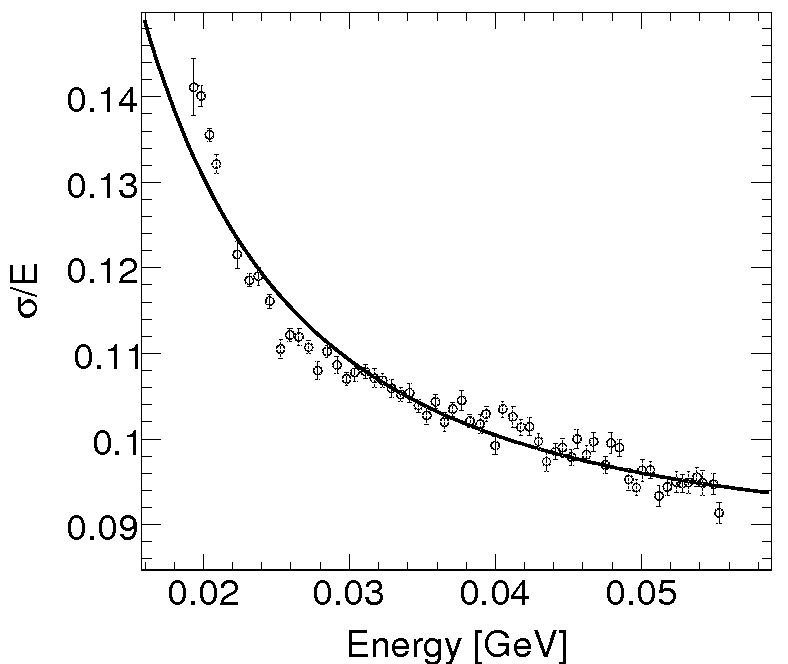}
\caption{The fit to the measured relative energy resolution (open circles) as the$a$ parameter has been given a fixed value corresponding to 50~phe/MeV. The fit parameters are presented in Table~\ref{tab:res_apr_fixed}. The $\chi^2$/d.o.f. of the fit is 6.02.} 
\label{fig:resol_april_fixed}
\end{figure}

One can easily understand that the imaginary value given for the $a$-parameter which should describe the Poisson statistics, is not reasonable as it does not have a physical interpretation. Considering that we know the approximate value of this parameter from previous measurements (the number of emitted photo electrons should be close to 50 per MeV at this temperature and about 30 per MeV at room temperature \cite{Linda}), the value of $a$ can be calculated according. At $E$=1~GeV the number of photo electrons $N$ is:

\begin{equation}
N=50000=\left( \frac{E}{\sigma_{\mbox{\emph{E, Poisson}}}}\right)^2=\left( \frac{1}{\emph{a}} \right)^2E
\label{eq:poisson}
\end{equation}
The value for $a$ becomes 1/$\sqrt{50000}$~$\mathrm{\sqrt{GeV}}$. With this input, the other two parameters can be fitted again. The resulting fit can be seen in Figure~\ref{fig:resol_april_fixed}.

\begin{table}[htb]
\centering
\begin{tabular}{|l|c|}
\hline
  &Value\\
\hline
\hline
b [$GeV$] &(1.85$\pm$0.23)$\cdot 10^{-3}$ \\
\hline
c &(8.63$\pm$0.72)$\cdot 10^{-2}$ \\
\hline
\end{tabular}
\caption{Values for the parameters $b$ and $c$, taken from the fit in Figure~\ref{fig:resol_april_fixed}. $a$ has been fixed to 1/$\sqrt{50000}$ $\mathrm{\sqrt{GeV}}$.} 
\label{tab:res_apr_fixed}
\end{table}
\noindent
The covariance matrix for this fit is:

\[
\emph{$COV_{April, fixed}$}=\left(
\begin{array}{ll}
2.8\cdot 10^{-15}   &-2.3\cdot 10^{-12}  \\
-2.3\cdot 10^{-12}  &2.6\cdot 10^{-9}   \\
\end{array}
\right)
\]
\noindent
giving a correlation between the two fitted parameters of -0.86.\newpage

\section{Results from Measurements below 60 MeV, September 2007}
The tagged photon energies for this measurement ranged from 18.9 to 51.6~MeV. Only 47 tagged energies were used as the high amplification resulted in overflow for the highest energies. The summed energy peaks were fitted with Gaussian distributions to obtain the mean value of the peak and the $\sigma$, in order to calculate the energy resolution. When using the timing information, the $\sigma$ of the fitted Gaussian distribution decreases at the same time as the mean value increases, compared to when not using this information. The change in $\mathrm{\sigma /E}$ for $\mathrm{E_{\gamma}}$=35~MeV is 6.8\%.

\begin{figure}[H]
\begin{center}
\includegraphics[bb=0 0 568 492,width=0.7\linewidth,angle=0]{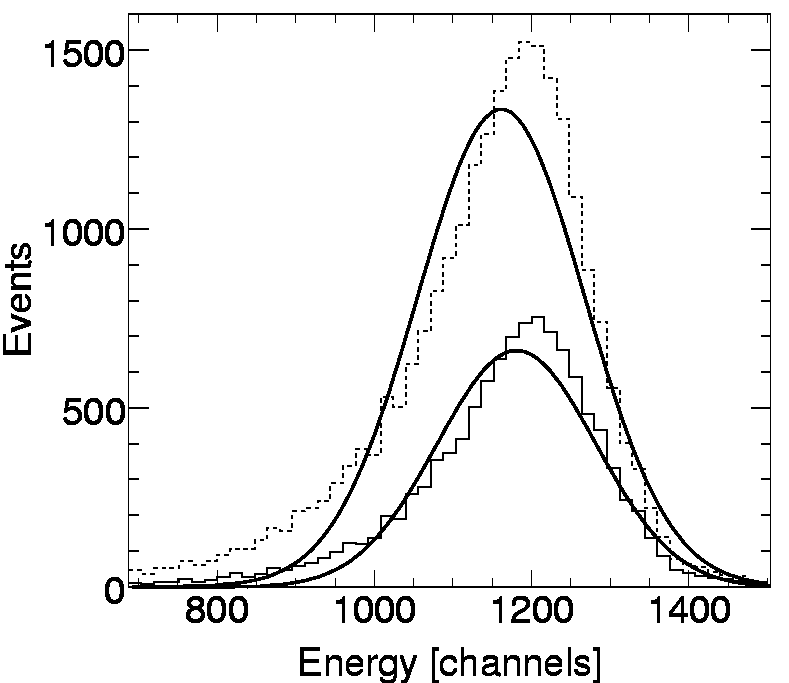}
\end{center}
\caption{A typical event summed spectrum for an incoming photon energy of $E_{\gamma}$=35 MeV from the run in September. The full drawn peak is obtained when the TDC-information is used, while the dashed curve corresponds to the peak one gets without using this information.}
\label{fig:twofits}
\end{figure}
\begin{figure}[H]
\begin{center}
\includegraphics[bb=0 0 567 485,width=0.7\linewidth,angle=0]{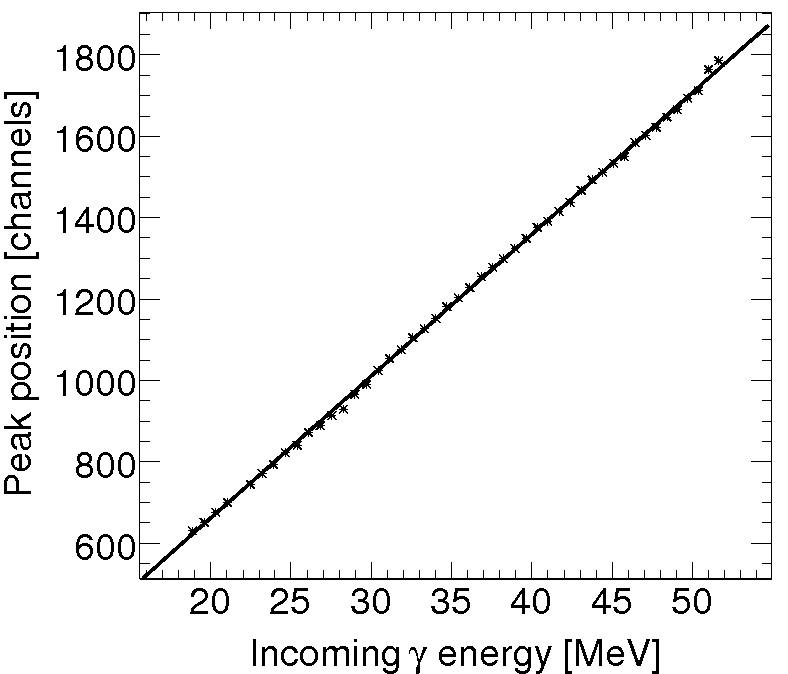}
\end{center}
\caption{Centroid channel number for the Gaussian distributions as a function of incoming photon energy. The straight line fit is described by $\mbox{peak position} = -35.1 + 34.9[MeV^{-1}] \cdot E_{\gamma}$. The $\chi^2$/d.o.f. of the fit is 20.5.}
\label{fig:kalchek}
\end{figure}

\noindent
If only half of the left hand side of the peak is fitted, the relative energy resolution $\sigma /E$ decreases with 18.0\% for both $E_{\gamma}$=21.0~MeV and $E_{\gamma}$=49.0~MeV.\\
\indent
By plotting the peak positions from the fit of the summed energy peaks as a function of the incoming photon energies, one can see a clear linear dependence, see Figure \ref{fig:kalchek}. The reason for the large $\chi^2$/d.o.f. value is that the errors in peak position taken from the fit are very small.

\begin{figure}[H]
\centering
\includegraphics[bb=0 0 567 492,width=0.7\linewidth, angle=0]{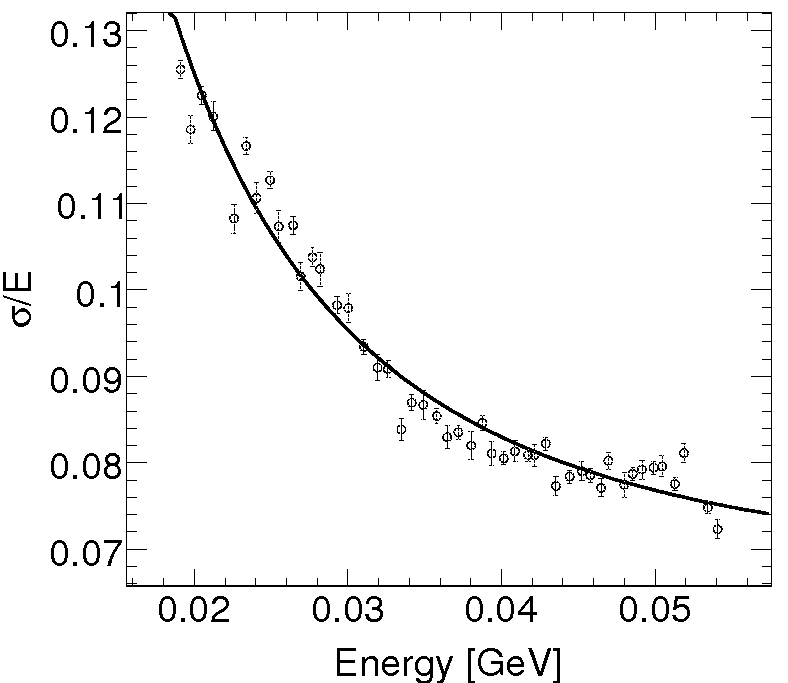}
\caption{The measured energy resolution of the September data. As the fit gives a negative value for the square of the Poisson parameter $a$, all three fit parameters are presented squared in Table~\ref{tab:res_sep_newcal}. The $\chi^2$/d.o.f. of the fit is 6.69. }
\label{fig:res_sept_newcal}
\end{figure}

\noindent
As for the April measurement, the mean values of the fitted Gaussian distributions were taken to correspond to the incoming photon energies, and the widths of the peaks were given by the fit. The relative energy resolution, $\mathrm{\sigma /E}$, as a function of the incoming photon energy can be seen in Figure~\ref{fig:res_sept_newcal} together with a full drawn line, showing the fit of Equation~\ref{eq:fiteq}.\newpage

\begin{table}[htb]
\centering
\begin{tabular}{|l|c|}
\hline
  &Value \\
\hline
\hline
$a^2$ [$\mathrm{GeV}$] &(-3.1$\pm$2.4)$\cdot 10^{-5}$ \\
\hline
$b^2$ [$\mathrm{GeV^2}$] &(5.07$\pm$0.38)$\cdot 10^{-6}$ \\
\hline
$c^2$ &(4.45$\pm$0.35)$\cdot 10^{-3}$ \\
\hline
\end{tabular}
\caption{Values for the square of the parameters $a$, $b$ and $c$, obtained from the fit in Figure~\ref{fig:res_sept_newcal}. The energy resolution is expressed as $\sigma/E=a/\sqrt{E} \oplus b/E \oplus c$.} 
\label{tab:res_sep_newcal}
\end{table}
\noindent
Again, large correlations are obtained in the fit, cf. Table~\ref{tab:corrtabnew}. 

\begin{table}[htb]
\centering
\begin{tabular}{|l|c|}
\hline
  &Correlation\\
\hline
\hline
corr$_{a^2 b^2}$ &-0.99\\
\hline
corr$_{a^2 c^2}$ &-0.99\\
\hline
corr$_{b^2 c^2}$ &0.96\\
\hline
\end{tabular}
\caption{The correlations between the squared parameters $a$, $b$ and $c$ obtained from the fitted September data.} 
\label{tab:corrtabnew}
\end{table}

\begin{figure}[H]
\centering
\includegraphics[bb=0 0 567 500,width=0.7\linewidth, angle=0]{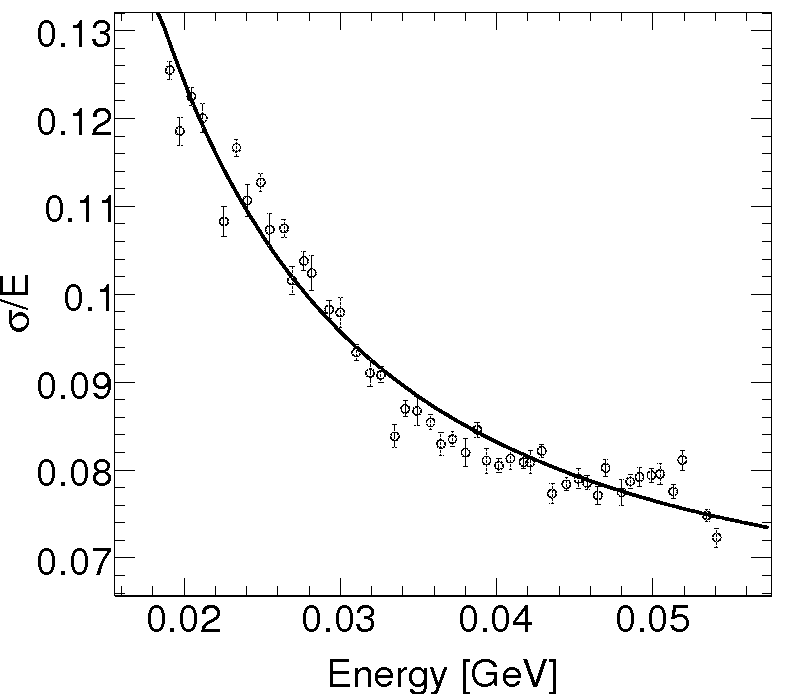}
\caption{A fit to the measured relative energy resolution (open circles) with the $a$ parameter fixed to a value that corresponds to 50 emitted phe/MeV. The fit parameters are presented in Table~\ref{tab:res_sept_fixed}. The $\chi^2$/d.o.f. of the fit is 6.64.} 
\label{fig:resol_sept_fixed}
\end{figure}

Because of the negative $a$-parameter obtained from the fit, it would again be interesting to fix its value to something reasonable and look at the new fit. $a$ was set to correspond to 50~phe/MeV, like for the April measurements, and a new fit was performed. The result is seen in Figure~\ref{fig:resol_sept_fixed}.

\noindent
The values of the parameters become

\begin{table}[htb]
\centering
\begin{tabular}{|l|c|}
\hline
  &Fitted value \\
\hline
\hline
b [$GeV$] &(2.07$\pm$0.24)$\cdot 10^{-3}$ \\
\hline
c &(6.12$\pm$0.69)$\cdot 10^{-2}$ \\
\hline
\end{tabular}
\caption{Values for the parameters $b$ and $c$, taken from the fit in Figure~\ref{fig:resol_sept_fixed}. $a$ has been fixed to 1/$\sqrt{50000}$ $\mathrm{\sqrt{GeV}}$.} 
\label{tab:res_sept_fixed}
\end{table}
\noindent
The resulting correlation between the two fitted parameters is -0.84.

\section{Comparison with Previous Data}
\begin{figure}[H]
\centering
\includegraphics[bb=0 0 567 484,width=0.7\linewidth, angle=0]{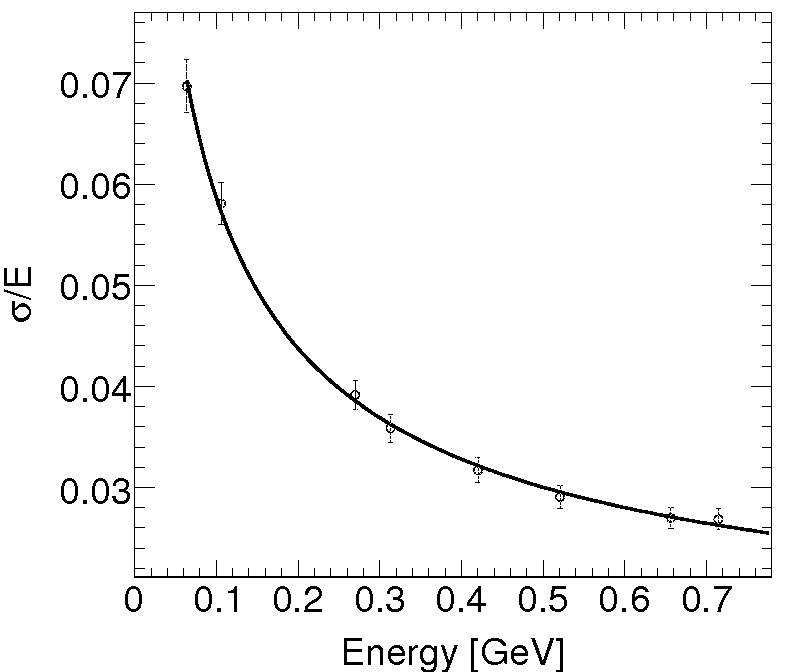}
\caption{The measured relative energy (open circles) and a fit to the data (full drawn line) given in Equation \ref{eq:resolution}. The data was taken from measurements in Mainz \cite{exjobb} and the fit parameters are presented in Table~\ref{tab:res_mainz}. The $\chi^2$/d.o.f. of the fit is 0.21.}
\label{fig:mainz}
\end{figure}

The PANDA detector will need to cover a wide range of energies in the electromagnetic calorimeter and detecting low-energy photons will be as important as detecting high-energy ones. Here we compare the present results with results obtained for higher photon energies.\\
\indent
In my Master thesis\footnote{Published as S.~Ohlsson before my name change in 2007}, I presented results from similar measurements at MAMI, Mainz, with PWO crystals at energies between 64 and 715~MeV for an array of 3$\times$3~crystals. The crystals were 15~cm long, non-tapered and cooled to -24~$\mathrm{^\circ}$C during the measurements \cite{exjobb}. The results from the thesis were used for a new analysis and the relative energy resolution and a fit to the data is shown in Figure~\ref{fig:mainz}.

\begin{table}[htb]
\centering
\begin{tabular}{|l|c|}
\hline
  &Value\\
\hline
\hline
$a^2$ [$\mathrm{GeV}$] &(3.67$\pm$0.52)$\cdot 10^{-4}$ \\
\hline
$b^2$ [$\mathrm{GeV^2}$] &(-4.1$\pm$3.8)$\cdot 10^{-6}$ \\
\hline
$c^2$ &(1.84$\pm$0.93)$\cdot 10^{-4}$ \\
\hline
\end{tabular}
\caption{Values for the squared parameters $a$, $b$ and $c$, obtained from the fit in Figure~\ref{fig:mainz}. The energy resolution is expressed as $\sigma/E=a/\sqrt{E} \oplus b/E \oplus c$.} 
\label{tab:res_mainz}
\end{table}
\noindent
Here, the square of the $b$-parameter (which describes the noise contribution to the energy resolution) becomes negative but within two standard deviations consistent with the noise contribution measured in the September data ($b \approx$2~MeV). The correlation between the parameters is somewhat smaller in this case, cf. Table~\ref{tab:corrtabmainz}. 

\begin{figure}[H]
\centering
\includegraphics[bb=0 0 567 492,width=0.7\linewidth, angle=0]{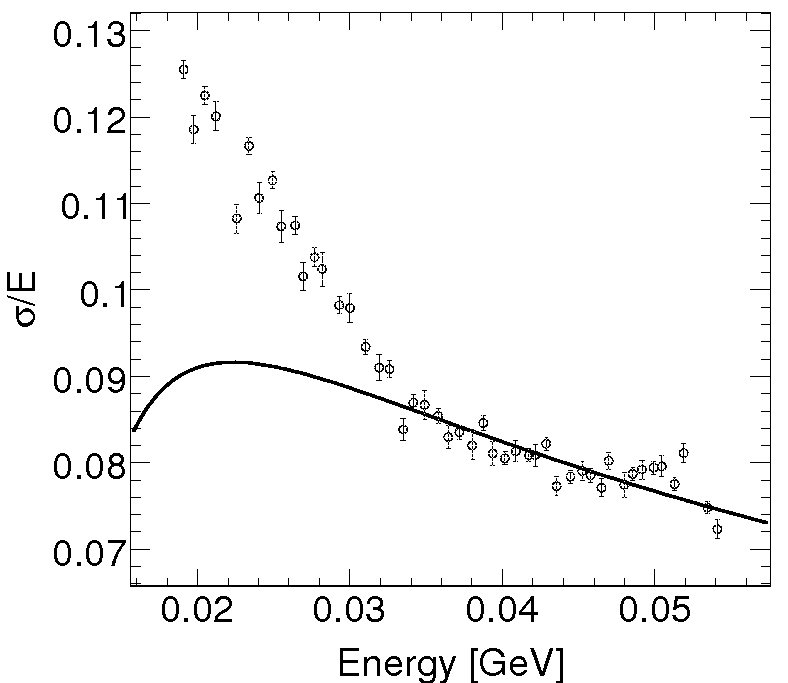}
\caption{The September data points (open circles) with the Mainz fit shown with a full drawn line.}
\label{fig:septmainzfit}
\end{figure}

\begin{table}[htb]
\centering
\begin{tabular}{|l|c|}
\hline
  &Correlation\\
\hline
\hline
corr$_{a^2 b^2}$ &-0.93\\
\hline
corr$_{a^2 c^2}$ &-0.94\\
\hline
corr$_{b^2 c^2}$ &0.85\\
\hline
\end{tabular}
\caption{The correlations between the squared parameters $a$, $b$ and $c$ obtained from the fit to the Mainz data.} 
\label{tab:corrtabmainz}
\end{table}
\noindent
The parameters for the energy resolution in the interval 64 to 715~MeV are very different from those obtained from the Lund measurements. This is clearly seen in Figure~\ref{fig:septmainzfit} where the September data is shown together with the fit to the Mainz data. The Mainz fit predicts that the energy resolution curve should turn downwards at low energies, in contrast to what has been measured. This feature comes from the sign of the squared $b$ parameter.

\begin{figure}[H]
\centering
\includegraphics[bb=0 0 567 488,width=0.7\linewidth, angle=0]{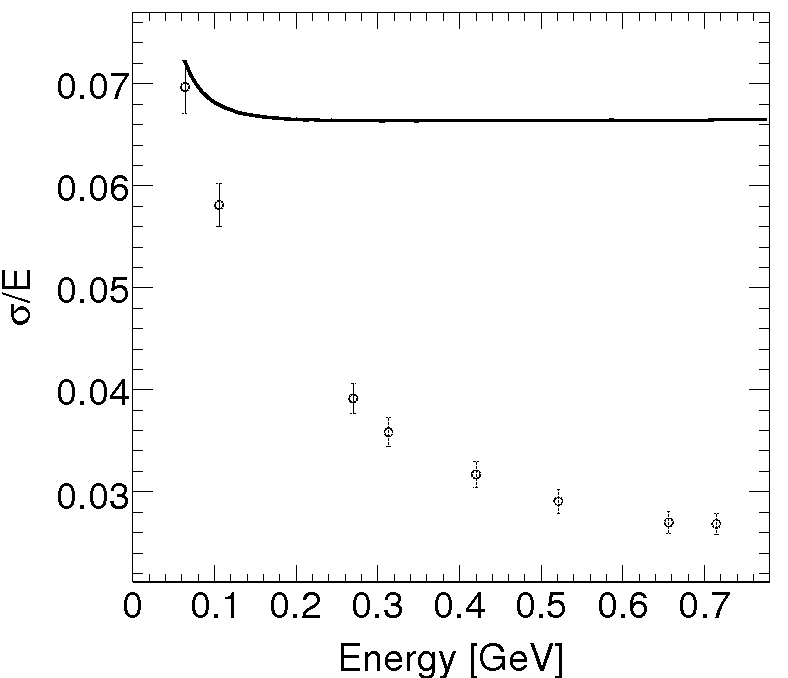}
\caption{The Mainz data points are shown with open circles and the fit from the September data is shown with a full drawn line.}
\label{fig:mainzWsept}
\end{figure}

\noindent
At the same time the fit to the September data does not agree with the Mainz data, cf. Figure~\ref{fig:mainzWsept}.\newpage

\begin{figure}[H]
\centering
\includegraphics[bb=0 0 567 504,width=0.7\linewidth, angle=0]{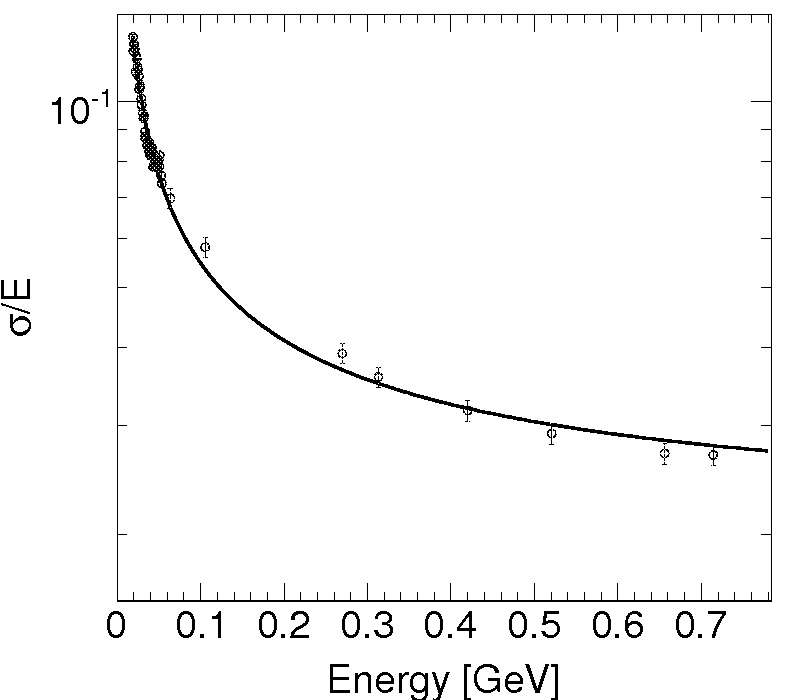}
\caption{The data points from the September measurements shown together with the Mainz data points (open circles). The fit is marked with a full drawn line. The fit parameters can be found in Table~\ref{tab:res_tot}. The $\chi^2$/d.o.f. of the fit is 6.78.}
\label{fig:mainzlund}
\end{figure}

In Figure~\ref{fig:mainzlund}, the September data points are shown together with the Mainz data points. A full drawn line corresponds to the fitted energy resolution over the entire energy interval. The parameters from the fit are presented in Table~\ref{tab:res_tot}.

\begin{table}[htb]
\centering
\begin{tabular}{|l|c|}
\hline
  &Value \\
\hline
\hline
a [$\mathrm{\sqrt{GeV}}$] &(1.55$\pm$0.21)$\cdot 10^{-2}$ \\
\hline
b [$\mathrm{GeV}$] &(9.50$\pm$3.3)$\cdot 10^{-4}$ \\
\hline
c &(2.11$\pm$0.60)$\cdot 10^{-2}$ \\
\hline
\end{tabular}
\caption{Values for the parameters $a$, $b$ and $c$, obtained by fitting Equation \ref{eq:fiteq} to the the September and the Mainz data simultaneously.} 
\label{tab:res_tot}
\end{table}
\noindent
The correlations of the parameters are clearly smaller in this fit to data in a larger energy interval, cf. Table~\ref{tab:total}. 

\begin{table}[htb]
\centering
\begin{tabular}{|l|c|}
\hline
  &Correlation\\
\hline
\hline
corr$_{a^2 b^2}$ &-0.93\\
\hline
corr$_{a^2 c^2}$ &-0.69\\
\hline
corr$_{b^2 c^2}$ &0.48\\
\hline
\end{tabular}
\caption{The correlations between the squared parameters $a$, $b$ and $c$ obtained from the fit to the combination of the September data and the Mainz data.} 
\label{tab:total}
\end{table}
To understand if the values in Table~\ref{tab:res_tot} are reasonable, one can look closer at the $a$-parameter corresponding to the Poisson statistics from the light collection. Here, it corresponds to about 3.9~phe/MeV (at a temperature of -25$^{\circ}$C). It is not a reasonable number. As the Poisson term $a$ is far smaller in reality than given by the fit suggests that this term includes other effects as well. This means that the interpretation of the energy resolution terms is more difficult. The correlations between parameters are still large but they are considerable smaller between $b^2$ and $c^2$ as well as between $a^2$ and $b^2$.\\

\section{Discussion and Conclusion from the Energy Resolution Measurements}
The fit to the April data as well as the September data yielded results with a negative square of the $a$-parameter (the Poisson parameter) in the expression for the energy resolution. If all parameters were forced to be larger than or equal to zero, the negative parameters were set to 0. That the two results gave similar results was not surprising as the measurements were in many aspects identical, except for the cooling which was better for the September measurements. The differences regarding the analysis were the optimisation of the calibration for the September measurement and the timing information that could be used to reduce the number of random coincidences. The better equipment and analysis result in a smaller relative energy resolution, $\sigma /E$, for the September data. After imposing the demand that the Poisson parameter should correspond to 50~phe/MeV, the fits became more in line with what one would expect. The fit to the eight high energy data points resulted in a squared $b$-parameter (the noise parameter) which was negative. The fit to the combination of the improved September data points together with Mainz data points gave however an energy resolution with only positive terms. In terms of the standard parameterisation (given in Equation \ref{eq:resolution}), the statistical term appears to be reasonable, but could probably be made smaller using longer crystals (especially for high energies), less wrapping material and performing a better calibration. The small constant term $c$, is given by the Mainz data points which force the asymptotic value of $\sigma/E$ down to approximately 0.02. However, one should keep in mind that the measurement in Mainz was different to those in Lund regarding crystal geometry and electronical set-ups.\\
\indent
The correlation between the three fit parameters were calculated for every energy resolution fit and very large values were obtained for both the low energies and the high energies. For the case when a fit was made to the relative energy resolutions in both the low and the high energy interval, the correlations became smaller. The values of the correlations are a clear indication of the inability to simultaneously determine the values of all three fit parameters. It means that changing the value of one parameter does not necessarily result in a different fit, since it translates into changes also in the other parameters so that a similar result can be obtained. This is clearly not desired, since each parameter should describe individual and independent contributions of both statistical fluctuations, noise and crystal properties. Due to this feature, it is not possible to discuss the energy resolution in terms of this standard expression, at least not unless the energy region over which the fit has been performed is very large.\\
\indent
Regarding the calibration, it was shown that using a pedestal peak and cosmic muons did not yield the best possible calibration. Lower values of $\sigma/E$ were obtained by slightly adjusting the calibraton constants for the eight crystals surrounding the central one. In conclusion, for future measurements it is important to ensure a good calibration. Even if the effect of the light yield uniformity of the crystals along the crystal length did not seem to play an important role here, one should use a source placed at the front end side so that the energy deposits from the source will be similar to those from the beam. Perhaps a more careful calibration can in addition lower the contribution to the energy resolution of the conventional ``constant term'' (which was very large for the low energy fits), or the intrinsic crystal properties. One could argue that this contribution has already been lowered in the September measurement, compared to the April measurement, by using more advanced and efficient cooling which lowers the temperature gradients inside the crystal.\\
\indent
From the presented figures, it is clear that the energy resolutions from the measurements in Lund and Mainz agree very well in the region around 50-70~MeV. The measured value for $\sigma /E$ is 0.072 for $E_{\gamma}$=51.6~MeV at Lund and 0.07 for $E_{\gamma}$=64~MeV in Mainz. It is also evident from Figures \ref{fig:septmainzfit} and \ref{fig:mainzWsept}, that an energy resolution parameterisation in one interval can not be applied to another energy interval. The low energy regime where the energy resolution is varying much with energy, needs to be carefully mapped since every data point gives an important contribution to describing the overall shape. Also, just fitting data points in the low energy region is not sufficient to describe the asymptotic behaviour at higher energies. The measured energy resolution at low energies is totally different from the extrapolated resolution from the Mainz data points. Also the energy resolution of the Mainz data is far better than the extrapolation of the September data suggests.\\
\indent
In addition, the energy resolution also depends on, for instance, the shower leakage out of the crystals, the light yield uniformity along the crystal, the absorption of light inside the crystal and developments of electromagnetic cascades in the material before the scintillator. These contributions may have energy dependences not described by the conventional formula of Equation \ref{eq:resolution}. 

\chapter{Light Yield Uniformity Tests of PANDA Crystals}
\label{homo}
A very desirable feature of the calorimeter is a uniform light output from the crystals. The light sensor is located at one end and thus no position sensitivity is possible. Light yield uniformity means that, given the same energy deposition, the same number of photons should reach the light sensor, irrespectively of where they are produced. All measurements described in this section were performed at room temperature.

\section{Set-Up for Uniformity Tests}
\begin{figure}[H]
\centering
\includegraphics[bb=0 0 685 345,width=0.8\linewidth,angle=0]{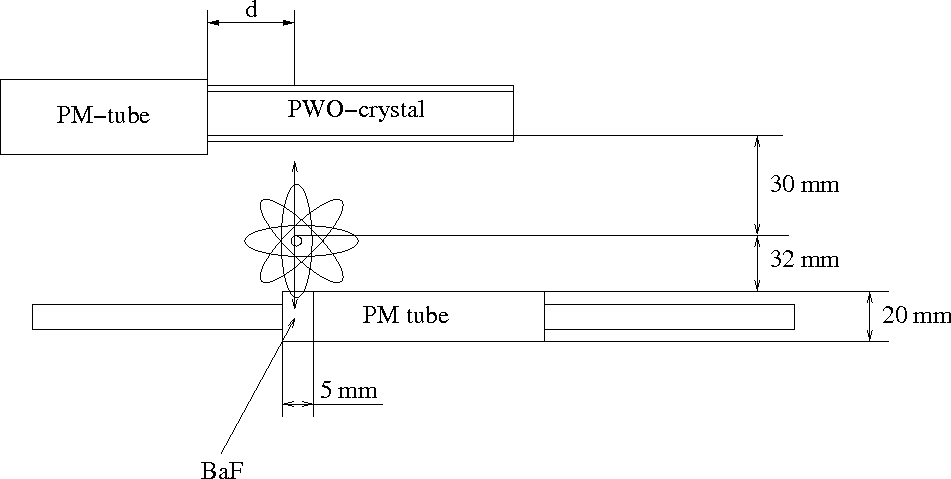}
\caption{The principle of the uniformity measurements. The $^{22}$Na source is located between the two scintillators and therefore it is possible to know where (distance d from the PM tube) the annihilation photons strike the PWO-crystal.}
\label{fig:sthlmsetup2}
\end{figure}

For light yield uniformity investigations, it is of utmost importance to know where in the crystal the radiation enters. For this study, a $^{22}\mathrm{Na}$ source was used. This source decays via $\beta^+$ radiation and the emitted positron very quickly annihilates with an electron from the surroundings, causing two photons, each with an energy of 511~keV, to be emitted back to back. If one uses a reasonably small scintillator to detect one of these photons and simultaneously records a signal from the main detector, the position of the incoming photon is known. This principle is shown in Figure~\ref{fig:sthlmsetup2}.\newpage

\begin{figure}[H]
\centering
\includegraphics[bb=0 0 290 218,width=0.6\linewidth,angle=0]{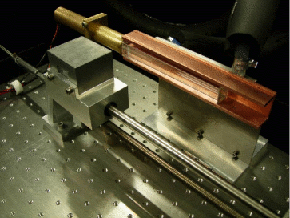}
\caption{The set-up for the
  uniformity tests without the second scintillator and its PM tube, which were later attached where the lead collimator is located in this photograph.}
\label{fig:sthlmsetup1}
\end{figure}

\noindent
The small BaF-crystal and its PM tube were attached on a metal block which could be slided on a rod, using a handle outside the box in which the set-up was placed, see Figure~\ref{fig:sthlmsetup1}.

Two different PWO crystal shapes were studied, three tapered and two non-tapered ones. Their dimensions are shown in Table~\ref{tab:homodim} below. 

\begin{table}[htb]
\centering
\begin{tabular}{|l|c|c|}
\hline
Shape  &Tapered &Non-tapered \\
\hline
\hline
Front-end dim. [mm$^2$]  &28$\times$28 &20$\times$20\\
\hline
Back-end dim. [mm$^2$] &22$\times$22  &20$\times$20\\
\hline
Length [mm] &200 &200 \\
\hline
\end{tabular}
\caption{The dimensions of the PWO crystals used
  for the uniformity measurements. The front-end is the one attached
  to the PM tube.}
\label{tab:homodim}
\end{table}
\noindent
The PM tube was of the type Hamamatsu R2083 and had a diameter of 51~mm, thus fully covering the end face of the crystal. Two different wrapping materials were tried, firstly white reflective Teflon surrounded by aluminium foil and secondly VM2000 \cite{Rainer}. In order to understand the contribution to the light yield uniformity, the light yield was firstly investigated along the crystal. Then, the reflective properties were changed in some regions and it was studied how this affected the light yield uniformity. The electronics used for the studies were a high voltage supply, a pre-amplifier, an amplifier with a 1 $\mathrm{\mu}$s shaping time and a Multi Channel Analyser.

\section{Statistics}
In order to translate the measured results to the number of photo electrons emitted from the photo cathode, the energy resolution must be carefully investigated.\\
\indent
For a Poisson distribution with a mean value N, the standard deviation $\mathrm{\sigma}$ is given by \cite{bevington}

\begin{equation}
\sigma =\sqrt{N}
\label{eq:gauss}
\end{equation}
Here, \emph{N} is the true value of what is being measured, i.e the number of photo electrons. The pulse height $S$ of the measured peak is proportional to the number of photo electrons produced according to $S=k N$, with $k$ being a proportionality factor. Assuming that only the statistical fluctuations contribute to the width of the peak, the standard deviation of the pulse height $\sigma_S$, is given by $\sigma_S = k \sigma = k \sqrt{N}$. 

\begin{equation}
\frac{\sigma_S}{S} \propto
  \sqrt{\frac{k \sqrt{N}}{kN}} = \frac{1}{\sqrt{N}}
\label{eq:sigma2}
\end{equation}
and

\begin{equation}
N = \left( \frac{S}{\sigma_S} \right)^2.
\label{eq:bgphe}
\end{equation}
$k^{-1}$ was determined separately as the average of the number of photo electrons per channel for all measurements,

\begin{equation}
\frac{N}{S}= \left( \frac{S}{\sigma_S} \right) ^2 /S = \frac{S}{\sigma_S ^2}
\label{eq:sgimadet}
\end{equation}
 and N=$k^{-1}$S was used for the calculations.\\
\indent
The assumption that only statistical fluctuations contribute to the width of the peak is of course a very crude approximation, as the terms $b$ and $c$ in Equation~\ref{eq:resolution} are put to zero. However, it can be justified by the fact that it gives a maximum Poisson width, or a lower bound for the number of photo electrons and any other contributions would just improve the situation. Also, taking other contributions into account would be very difficult, as the expression (meaning the individual contributions) for the energy resolution is not known in this low energy regime.

\section{Analysis}
Different positions along the crystals were investigated and at
each one, pulse height spectra were recorded and fitted with a
Gaussian function. The peak positions and widths were used to
calculate the number of emitted phe/MeV.

\section{Results}
\begin{figure}[H]
\centering
\includegraphics[bb=0 0 644 568,width=0.7\linewidth,angle=0]{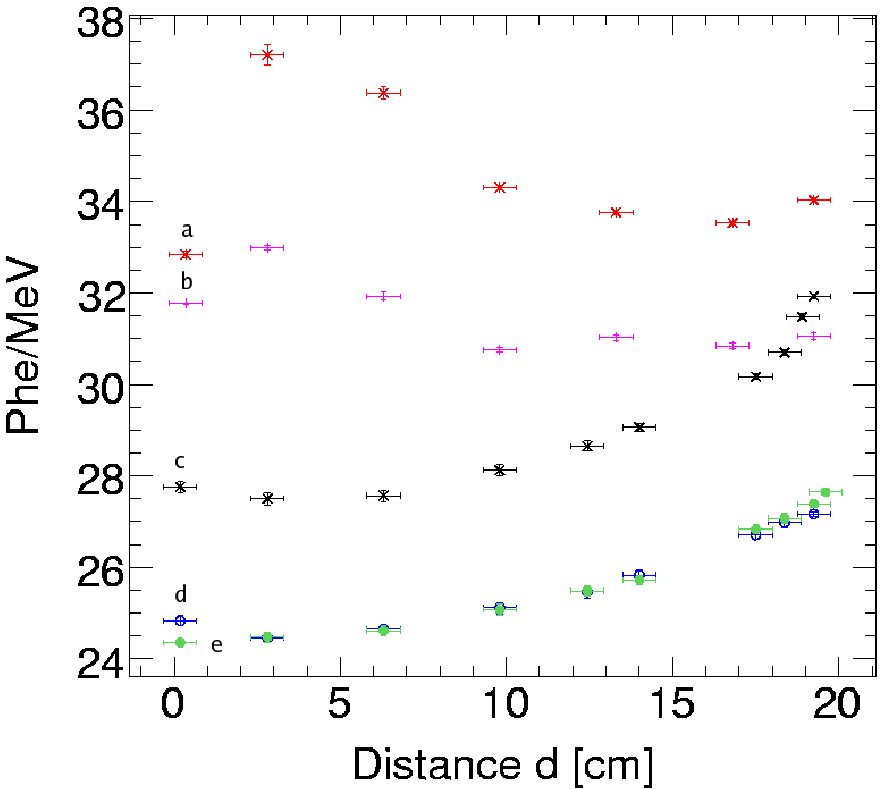}
\caption{Measured light collection yield as a function of the distance d defined in Figure~\ref{fig:sthlmsetup2} for two non-tapered (a and b) and three tapered (c, d and e) PWO crystals. The crystals were wrapped with white Teflon.}
\label{fig:teflon}
\end{figure}

\begin{figure}[H]
\centering
\includegraphics[bb=0 0 644 568,width=0.7\linewidth,angle=0]{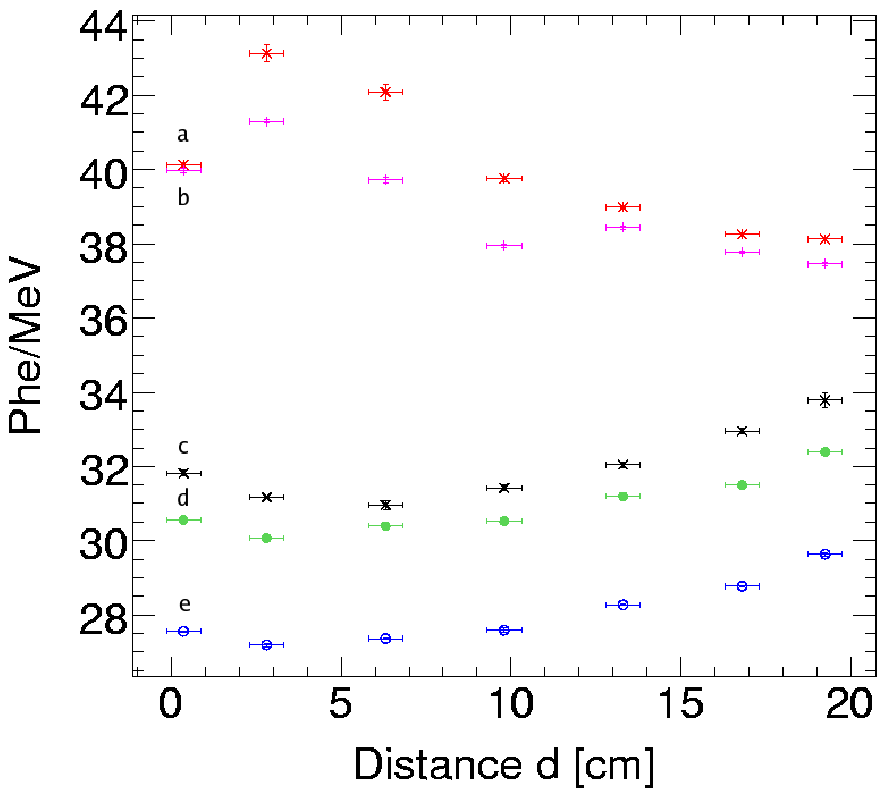}
\caption{Measured light collection yield as a function of the distance d defined in Figure~\ref{fig:sthlmsetup2} for two non-tapered (a and b) and three tapered (c, d and e) PWO crystals. The crystals were wrapped with VM2000.} 
\label{fig:vm}
\end{figure}

The plotted light yield as a function of the distance between the point of interaction and the PM tube for the measurements with Teflon wrapping is shown in Figure~\ref{fig:teflon}. There is a clear dependence of the light yield on the shape of the crystals. The tapered crystals deliver more light when the source is as far away from the PM tube as possible, while the non-tapered crystal light yields seem to be maximum close to the PM tube.\\
\indent
Using instead the mirror-like wrapping VM2000 the shapes of the non-uniformity is similar, but the light output is approximately 17\% larger than with Teflon wrapping (see Figure~\ref{fig:vm}).

\newpage

We quantified the uniformity by calculating, for each crystal, the ratio
$\alpha_{NU}$ defined as:

\begin{equation}
\alpha_{NU}=\frac{\sqrt{\frac{1}{n-1}\sum_{i=1}^n{(N_i-\bar{N})^2}}}{\bar{N}}
\label{eq:alfanu}
\end{equation}
where $N_i$ is the measured number of photo electrons at the i:th interaction point and n the number of data points. This parameter describes the spread of number of phe/MeV and for a totally uniform light yield it should be 0. Using 5 data points from each crystal, $\alpha_{NU}$ has been calculated. To avoid end effects, the five data points were chosen correspondingly to the distances for the second to the sixth data point in the a-curve in Figure~\ref{fig:teflon}. In the cases where the exact distances were not measured, they were interpolated. The uncertainties in $\alpha_{NU}$ were calculated with the error propagation formula, assuming that the spread in the number of photo electrons (in Figure~\ref{fig:teflon}) was the the only uncertainty. The crystal identification information and the average value of the number of emitted phe/MeV are presented in Table~\ref{tab:ratioTef1}, while the $\alpha_{NU}$-parameters for both wrapping materials as well as the uncertainties in this value are displayed in Table~\ref{tab:ratioTef2}. 

\begin{table}[htb]
\centering
\begin{tabular}{|l|c|c|c|c|}
\hline
PWO  &Crystal label &$\bar{N}_{Teflon}$ &$\bar{N}_{VM}$ \\
\hline
\hline
a &20\_016 &35.035  &40.44 \\
\hline
b &20\_017 &31.51  &39.035 \\
\hline
c &27\_Left &28.44  &31.71 \\
\hline
d &26\_Left &25.24  &27.84 \\
\hline
e &28\_Left &25.23  &30.74 \\
\hline
\end{tabular}
\caption{Number of photo electrons per MeV, N, averaged over the point of interaction for each of the five crystals and for the two wrapping materials. Crystal a and b are non-tapered, c-e are tapered.} 
\label{tab:ratioTef1}
\end{table}

\begin{table}[htb]
\centering
\begin{tabular}{|l|c|c|}
\hline
PWO   &$\alpha_{NU}$ (Teflon) &$\alpha_{NU}$ (VM2000) \\
\hline
\hline
a &(4.718$\pm$0.041)$\cdot 10^{-2}$ &(5.152$\pm$0.046)$\cdot 10^{-2}$ \\
\hline
b &(3.017$\pm$0.067)$\cdot 10^{-2}$ &(3.787$\pm$0.045)$\cdot 10^{-2}$ \\
\hline
c &(3.73$\pm$0.23)$\cdot 10^{-2}$  &(2.55$\pm$0.20)$\cdot 10^{-2}$ \\ 
\hline
d &(3.21$\pm$0.38)$\cdot 10^{-2}$ &(2.395$\pm$0.041)$\cdot 10^{-2}$ \\ 
\hline
e &(3.23$\pm$0.31)$\cdot 10^{-2}$ &(1.92$\pm$0.20)$\cdot 10^{-2}$ \\ 
\hline
\end{tabular}
\caption{he measured non-uniformity, $\alpha_{NU}$, for each of the five crystals and for the two wrapping materials.} 
\label{tab:ratioTef2}
\end{table}
As can be seen in Table~\ref{tab:ratioTef2}, $\alpha_{NU}$ grows for the non-tapered crystals when the Teflon wrapping is substituted to VM2000, while the opposite is true for tapered crystals.

\section{Light Yield Uniformity Improvements}
\begin{figure}[H]
\centering
\includegraphics[bb=0 0 486 108,width=0.7\linewidth,angle=0]{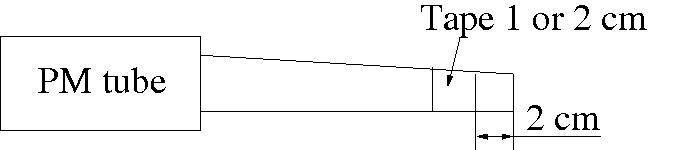}
\caption{The location of the tape for the study of light yield uniformity improvements. The tape is put 2~cm from the end side of the crystal and it is either 1 cm or 2~cm wide. For the 2$\times$2~cm of tape option, there is an equal amount of tape on the opposite side of the crystal.}
\label{fig:tape}
\end{figure}

Attempts to investigate the possibilities to make the light yield more uniform were done using one tapered crystal (crystal label $\mathrm{28\_Left}$, seen in Figure~\ref{fig:vm} as data set e). In order to investigate the importance of photon reflections at different parts of the crystal surface, four different ways were tried. They were: 1) no reflective wrapping on the crystal side opposite of the PM tube, 2) black tape (1~cm wide) put 2~cm from the end of the crystal, 3) black tape (2~cm wide) put 2~cm from the end of the crystal and finally 4) two stripes of 2~cm wide tape put at two opposite sides of the crystal about 2~cm from the end side, cf. Figure \ref{fig:tape}. 

\begin{figure}[H]
\centering
\includegraphics[bb=0 0 647 568,width=0.7\linewidth,angle=0]{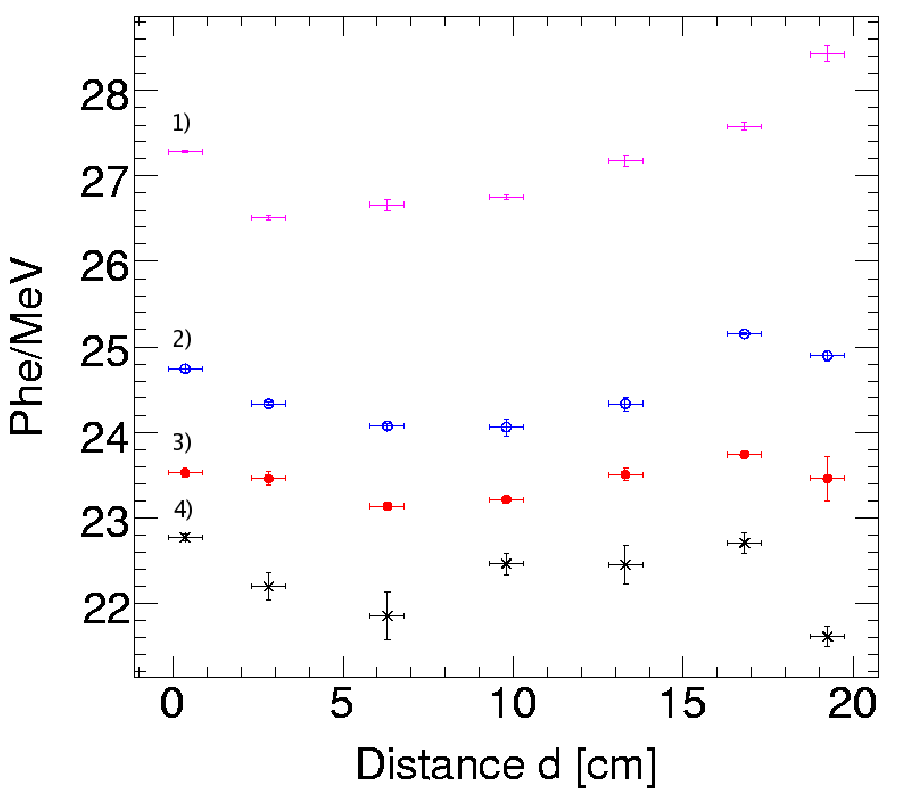}
\caption{Measured light collection yield as a function of interaction point in the crystal for four different ways of making the light yield more uniform: 1) no VM2000 at the back-end, 2) 1~cm of tape, 3) 2~cm of tape and 4) 2$\times$2~cm of tape.}
\label{fig:sthlm3}
\end{figure}

The expectation was that without reflective wrapping on the short end of the crystal, the light would scatter out, thereby decreasing the overall light yield and perhaps affecting the shape of the non-uniformity. For using black tape, it was expected that the scattered photons would not be reflected back into the crystal, but instead be absorbed and thereby decreasing the light yield in that specific region. The results from the uniformity improvement measurements are displayed in Figure~\ref{fig:sthlm3}. 
\newline

$\alpha_{NU}$ was calculated for the four measurements and
the result can be seen in Table~\ref{tab:tejprat}.

\begin{table}[htb]
\centering
\begin{tabular}{|l|c|c|c|}
\hline
Modification &Crystal label &$\bar{N_{VM}}$  &$\alpha_{NU}$ \\
\hline
\hline
Unmodified &$\mathrm{28\_Left}$ &30.74 &(1.92$\pm$0.20)$\cdot 10^{-2}$ \\
\hline
1) Back free from VM2000 &$\mathrm{28\_Left}$ &26.94 &(1.62$\pm$0.12)$\cdot 10^{-2}$ \\
\hline
2) 1 cm tape &$\mathrm{28\_Left}$ &24.39 &(1.828$\pm$0.052)$\cdot 10^{-2}$ \\
\hline
3) 2 cm tape &$\mathrm{28\_Left}$ &23.41 &(1.030$\pm$0.083)$\cdot 10^{-2}$ \\
\hline
4) 2$\times$2 cm tape &$\mathrm{28\_Left}$ &22.33 &(1.457$\pm$0.032)$\cdot 10^{-2}$ \\
\hline
\end{tabular}
\caption{The uniformity ratio $\alpha_{NU}$ and the average light collection yield $\bar{N}_{VM}$, calculated for modified VM2000 wrappings. $\alpha_{NU}$ for the unmodified case is included for comparison.} 
\label{tab:tejprat}
\end{table}

\newpage
\section{Discussion and Conclusions from the Uniformity Results}
\begin{figure}[H]
\centering
\includegraphics[bb=0 0 824 440,width=0.8\linewidth,angle=0]{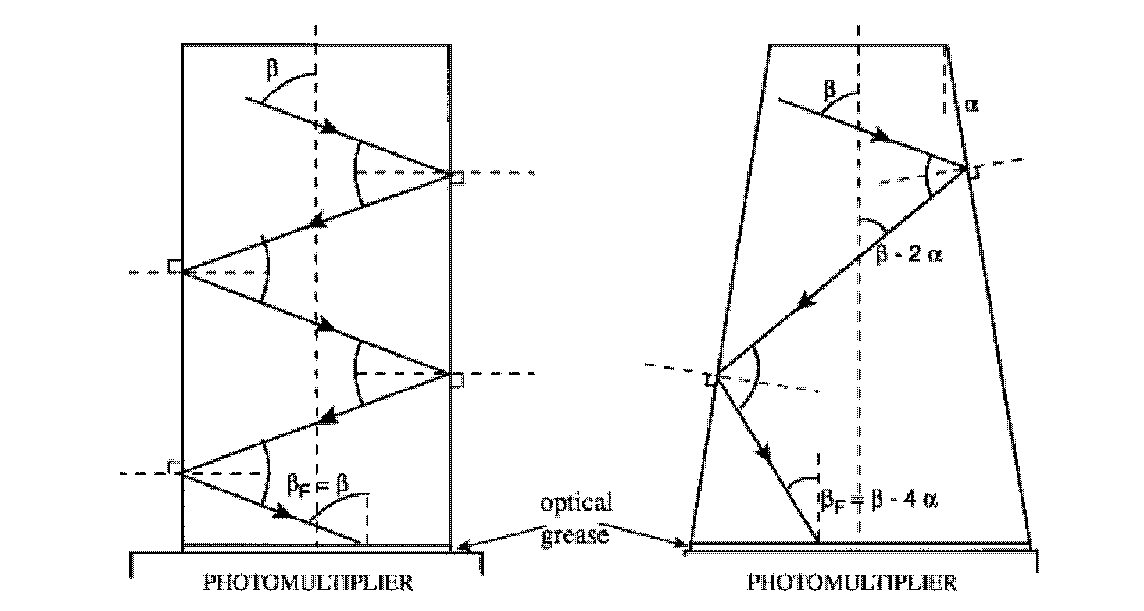}
\caption{The geometrical focusing effect where
  the tapered geometry to the right is seen to result in more
  favourable angels, through which the light can escape the
  scintillator \cite{lyartikel}.}
\label{fig:geomeff}
\end{figure}

It is of utmost importance to keep a stable temperature while investigating the uniformity of the detector response from the PWO crystals due to the temperature dependent light yield. For these measurements, the variation in temperature was not larger than 0.1 $\mathrm{^\circ}$C. Using the dependence of the light yield on the temperature mentioned in section \ref{pwo}, the corresponding change in the light output was smaller than 0.2\% and hence, the measured variation in the light collection over the crystal length (cf. Figures \ref{fig:teflon} and \ref{fig:vm}) is not a temperature effect.\\
\indent
As can be seen from the results in Figure~\ref{fig:teflon} and \ref{fig:vm}, there are clear results that the VM2000 wrapping is superior to Teflon when it comes to reflecting scattered photons back into the crystals. Further, the light collected from the crystals is far from uniform and it also seems to be very dependent on the shape of the scintillator. From Figure~\ref{fig:teflon} and \ref{fig:vm} one can see that the overall shape of the light yield uniformity profile with the increase at large distances does not seem to depend on the wrapping material as long as the same material covers the whole crystal. According to $\alpha_{NU}$, the uniformity improves when using VM2000 over Teflon for the tapered crystals, while the opposite is true for the non-tapered crystals. When tape is put on the crystals and the same non-uniformity quantity $\alpha_{NU}$ is calculated, it is seen that $\alpha_{NU}$ is decreased by a large amount. When applying 2~cm of black tape, $\alpha_{NU}$ is approximately 50\% lower compared with the case of normal VM2000 wrapping. Even using no tape at all, but only leaving the short end opposite to the PM tube free from VM2000, lowers $\alpha_{NU}$ with about 15\%.\\ 
\indent
Contributions to the non-uniformity for both non-tapered and
tapered crystals come from light attenuation along the crystal due to intrinsic absorption inside the material, reflective properties of the crystal surface, transmission through the surface, the wrapping material as well as from diffusion on impurities and bubbles.\\
 \indent
For the tapered crystals, the path the photons travel inside the scintillator is in general longer than for non-tapered crystals due to purely geometrical reasons. Also the number of reflections inside the crystals are larger here. Both effects increase risk of losing photons either due to internal absorption or scattering out of the scintillator thereby decreasing the number of phe/MeV.\\
\indent
For tapered crystals, the so-called focusing effect of the tapered shape is important. This effects favours light produced far from the PM tube (in the small end) because the reflections yield angles which are more favourable for transmitting the light into the PM tube, see Figure~\ref{fig:geomeff} \cite{lyartikel}.\newpage

Total internal reflection occurs above a certain critical angle in the medium. For the PWO-to-air surface the critical angle is 25.8 degrees. At the PM tube the critical angle is increased by using an optical coupling grease. As the angles ($\beta_F$ in Figure \ref{fig:geomeff}) of the photons striking the window of the PM tube are in general larger for non-tapered compared to tapered crystals, photons exit with a larger probability in the latter case. If the photon instead moves away from the PM tube with a small angle, it will for both crystal shapes suffer more reflections, but for the tapered geometry the number of reflections can grow very quickly. One should also keep in mind that it is very likely that neither wrapping material can be put close enough to the crystal so that an air gap is avoided. This air gap influences the critical angle above which total internal reflection occurs. As the difference in refraction index is larger between air and PWO than for either wrapping and PWO, this critical angle will be smaller with an air gap present and the chance for having total internal reflection and keeping the photons inside the crystal is larger. For smaller angles some photons will however escape and if there is no reflective foil beyond the air gap, these photons will disappear.\\ 
\indent
The results presented in this licenciate thesis show that the number of phe/MeV is larger for the non-tapered compared to the tapered crystals, despite the obvious geometrical disadvantages for such crystals mentioned in this section. The explanation is probably different crystal properties such as the intrinsic light yield.\\
\indent
It is possible to compensate for the non-uniformity of the light yield but to the cost of a decreased overall light yield. This is not an ideal solution for this type of crystal where the light yield is low to start with.\\ 
\indent
Since the electromagnetic shower occurs mainly at a depth of 4-12~$\mathrm{X_0}$ into the crystal\cite{lyartikel}, one could experiment with putting tape at different positions. It is suggested that correcting for the light yield in the region 4-12~$\mathrm{X_0}$ would have a critical effect while changes in the region 12-25~$\mathrm{X_0}$ only would effect late developing showers\cite{lyartikel}. Corrections in the region 0-4~$\mathrm{X_0}$ would not affect the non-uniformity profile very much.\\
\indent
According to \cite{lyartikel}, diffuse wrappings like TYVEK and Millipore seem to give higher light yields than Teflon does for PWO crystals, but a comparison with VM2000 was not made. It would be interesting to compare the results from such wrappings with the ones used in the investigations described here. Since the NU profile of the light yield is not affected by the different wrapping materials as long as the same type of wrapping is used for the entire crystal, these studies can be made independently of each other. 

\chapter{Simulation Studies}
\label{sim}
\section{Introduction}
The challenging physics program of PANDA and the complexity of the detector require that substantial effort is devoted to simulations of the physics channels of interest to the collaboration. One of these topics concerns hyperon physics. The $\mathrm{\Lambda}$ particle is the lightest of the hyperons and it also frequently occurs as a decay product from excited hyperons, often together with photons (detected in the calorimeter). This makes it an obvious starting point for simulations of hyperon channels.\\
\indent
High quality data for the $\mathrm{\bar{p}p\rightarrow\bar{\Lambda}\Lambda}$ reaction up to momenta of 2~GeV/c exist from the PS185 experiment at LEAR, CERN. These data can serve as a benchmark for simulations of the same reaction for PANDA. 

\section{About the $\mathrm{\Lambda}$ State}
All ground state hyperons, described as baryons with strange quarks,
decay only via the weak interaction, except for the $\mathrm{\Sigma^0}$
which decays electromagnetically. Because of this, the life times for
hyperons are reasonably long. For example, a produced $\Lambda$ has an
average flight path of several centimetres (c$\mathrm{\tau}$=7.89~cm
\cite{pdg}). This typically means that the decay point is located inside, or in
the vicinity of, the MVD-detector for PANDA.\\
\indent
Parity is not conserved in weak decays. Because of this, an asymmetry in the directions of the decay particles may be observed. If so, one can measure the hyperon polarisation, and in case of anti-hyperon-hyperon pairs also spin correlations \cite{trento}.\\ 
\indent
If viewing the proton and the $\mathrm{\Lambda}$ in the constituent quark model, one can group u- and d-quarks inside the proton and the $\mathrm{\Lambda}$ into an isospin and spin zero di-quark state and and let the remaining u- and s-quarks reflect the spin and isospin properties of the particles.

\section{The Coordinate System}
The center of mass system for the $\bar{p}p \rightarrow \bar{\Lambda}\Lambda$ is shown in Figure~\ref{fig:lamframe}. In addition, we define a coordinate system for the decay particles in the rest frame of the anti-hyperon/hyperon. Here, the z-axis is given by the direction of the  outgoing $\mathrm{\Lambda}$ ($\mathrm{\bar{\Lambda}}$). The y-axis points in a direction perpendicular to the scattering plane of the $\mathrm{\bar{p}p}$-system and the outgoing hyperons, and the x-axis is chosen such that the system is right-handed. Formally, the axis are defined as \cite{trento}:

\begin{figure}[t]
\begin{center}
\includegraphics[width=\linewidth,bb=0 0 1041 454]{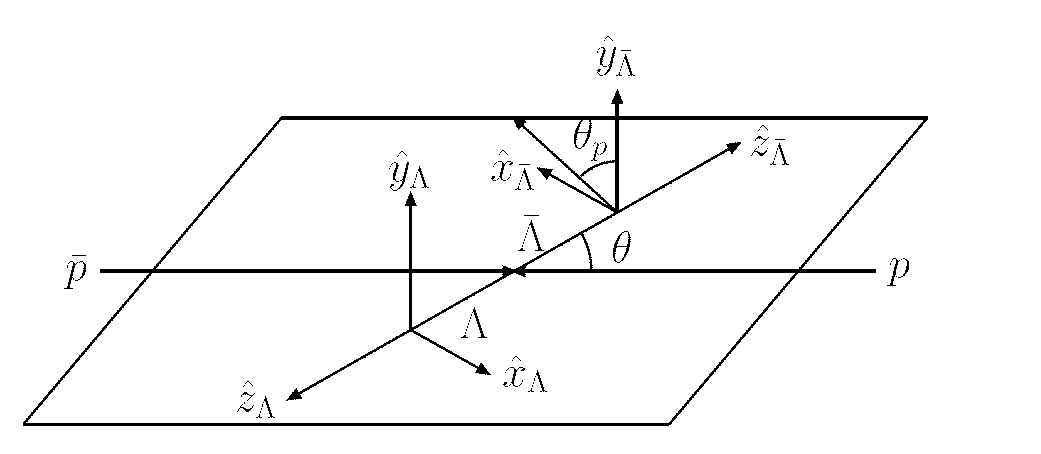}
\end{center}
\caption{The center-of-mass frame for the $\mathrm{\bar{p}p}$ collision, in which the coordinate systems for the rest frames of the $\mathrm{\Lambda}$ and $\mathrm{\bar{\Lambda}}$ are shown. These coordinate systems are different for each event.}
\label{fig:lamframe}
\end{figure}    

\begin{equation}
\hat{x}=\hat{y} \times \hat{z}
\label{eq:zhat}
\end{equation}

\begin{equation}
\hat{y}=\frac{\bar{k}_i \times \bar{k}_j}{|\bar{k}_i \times
  \bar{k_j}|}=\hat{n}
\label{eq:yhat}
\end{equation}

\begin{equation}
\hat{z}=\hat{k}_j
\label{eq:xhat}
\end{equation}
where the $\bar{k_i}$ is the momentum vector of the initial beam. Together with the momentum vector of the outgoing hyperon, $k_j$, they define the interaction plane and $\hat{n}$ is the normal vector to this plane. The index j refers to either the $\mathrm{\Lambda}$ or the $\mathrm{\bar{\Lambda}}$ particle. As can be seen, the axis perpendicular to the interaction plane, $\hat{y}$ or $\hat{n}$, is the same for the two particles. 

\section{The Angular Distributions of the $\mathrm{\Lambda}$}
The PS185 experiment collected data on the cross-sections for $\mathrm{\bar{p}p}$ going to different single strangeness hyperon states up to a maximum $\mathrm{\bar{p}}$-momentum of 2~GeV/c. Of special interest to our simulations is the reconstruction of the differential cross-sections for the $\mathrm{\Lambda}$, in the reaction $\mathrm{\bar{p}p \rightarrow \bar{\Lambda}\Lambda}$, and the polarisation as a function of the scattering angle in the center of mass (CM) system. 

\begin{figure}[H]
\centering
\includegraphics[bb=0 0 507 476,width=0.8\linewidth,angle=0]{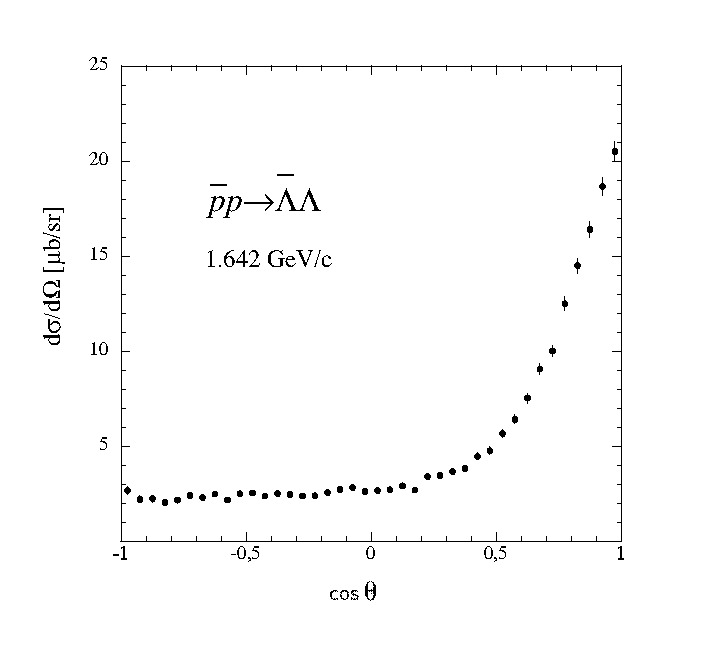}
\caption{The CM differential cross-section measured by PS185
  for the $\bar{\Lambda}$ in $\bar{p}p\rightarrow
  \bar{\Lambda}\Lambda$ reaction at the beam momentum 1.642~GeV/c \cite{trentobild}.} 
\label{fig:angdist_ps}
\end{figure}
To fully describe the distribution over the angular interval \cite{fmm}, Legendre polynomials of up to degree eight, multiplied with energy dependent coefficients $\mathrm{A_n(\epsilon)}$ expanded in third order polynomials of $\epsilon$, were used \cite{tord},

\begin{equation}
f(\cos\theta)=A_0(\epsilon)\left(
P_0(\cos\theta)+A_1(\epsilon)P_1(\cos\theta)+...+A_8(\epsilon)P_8(\cos\theta)\right). 
\label{eq:fit}
\end{equation}
The general expression for the orthogonal Legendre polynomials
$P_n(x)$ is \cite{beta}

\begin{equation}
P_n(x)=\frac{1}{2^n n!}\frac{d^n}{dx^n}(x^2-1)^n.
\label{eq:leg}
\end{equation}
The coefficients $A_n(\epsilon)$ are polynomials depending on the excess energy
$\epsilon$ 

\begin{equation}
\epsilon=\sqrt{s}-\sum m_f=\sqrt{s}-2m_{\Lambda}
\label{eq:epsilon}
\end{equation}
with s being the squared total energy in the CM system. Using the four-momentum vectors $p$ of the particles, the total energy is

\begin{equation}
s=(p_1+p_2)^2=p_1^2-2p_1p_2+p_2^2=m_1^2-2p_1p_2+m_2^2
\label{eq:s}
\end{equation} 
The $A_n$ coefficients are given by \cite{tord}

\begin{equation}
\label{eq:A}
A_n(\epsilon)=a_{n0}+a_{n1}\epsilon+a_{n2}\epsilon^2+a_{n3}\epsilon^3
\end{equation}
where the coefficients $a_n$ are fitted to the experimental data. These have been implemented in the event generator for the PANDA simulations to describe the angular distributions in the hyperon energy region of PS185 ($<$2 GeV/c).

\section{$\Lambda$ Polarisation}
To generally describe the influence of the spin on the $\bar{p}p
\rightarrow \bar{\Lambda} \Lambda$ reaction, a spin observable
$\chi_{jk\mu\nu}$ can be used. It is defined as \cite{trento}

\begin{equation}
\chi_{jk\mu\nu}=\frac{\mathrm{Tr}(\sigma^c_{\mu} \sigma^d_{\nu} M \sigma^a_j
  \sigma^b_k M^{\dag})}{Tr(MM^{\dag})} =
  \frac{ \frac{1}{4} \mathrm{Tr}(\sigma^c_{\mu} \sigma^d_{\nu} M \sigma^a_j 
  \sigma^b_k M^{\dag}) }{I_0}
\label{eq:spin}
\end{equation}
where the $j$, $k$, $\mu$ and $\nu$ indices refer to the spin projection of the
beam, target, scattered and recoil particles while the indices $a$-$d$
represent the different particles in the reaction
$a+b\rightarrow c+d$. The $\mathrm{\sigma}$ matrices are the
three Pauli matrices $\mathrm{\sigma_1}$, $\mathrm{\sigma_2}$ 
and $\mathrm{\sigma_3}$, plus the identity matrix
$\mathrm{I_0={\sigma}_0}$. M is the transition operator for
$\bar{p}p$ to $\bar{\Lambda}\Lambda$. The
expression is derived from a relation that describes the
initial-state density matrix for spin 1/2 particles 
with a certain polarisation. Using expressions for the density
matrices before and after interaction and the transition operator M from
the $\mathrm{\bar{p}p}$ to $\mathrm{\bar{\Lambda}\Lambda}$, one arrives at
Equation \ref{eq:spin}. More details on this can be found in \cite{theory}.\\
\indent
Fortunately, the $\mathrm{4^4=256}$ spin observables above can be reduced using
symmetries of parity and charge conjugation and geometrical
identities. The parity conservation, applied to strong interaction
processes such as $\mathrm{\bar{p}p \rightarrow \bar{\Lambda} \Lambda}$, states that the reaction probability should be unaffected by a spatial inversion of the coordinates. This gives that the polarisation of the observables in $\hat{x}$- and $\mathrm{\hat{z}}$-directions must be zero, since they otherwise would change
sign. Furthermore, charge 
conjugation symmetry applied to the self-conjugated $\mathrm{\bar{p}p}$ and
$\mathrm{\bar{\Lambda}\Lambda}$ systems ensures that the polarisations for
$\mathrm{\bar{\Lambda}}$ and $\mathrm{\Lambda}$ must be
equal. Finally, for polarised 
beam or target, invariance under rotation of the scattering plane
reduces the number of spin observables additionally. These symmetries
decrease the number of spin observables to the more manageable number of 40.\\
\indent
The $\mathrm{\Lambda}$ decays to either $\mathrm{p\pi^-}$ or
$\mathrm{n\pi^0}$ with branching ratios of 
64\% and 36\% respectively. In this thesis, only investigations of the
charged decay mode have been done. The decay distribution for the
$\mathrm{\Lambda}$ with polarisation P can be expressed as

\begin{equation}
I(\theta_{\mathrm{p}})=\frac{1}{4\pi}\left( 1 + \alpha P \cos
\theta_{\mathrm{p}} \right)
\label{eq:decdist}
\end{equation}
where $\mathrm{\theta_p}$ is the proton emission angle projected on the
spin projection direction of the $\Lambda$ ($\hat{y}$ in the case of unpolarised
beam and target) in the rest frame of the $\Lambda$,
while $\mathrm{\alpha}$ is the so-called asymmetry parameter and is
related to the probability for the decay baryon to be emitted in the
spin direction of the decaying hyperon. This value has been
experimentally measured to be 0.64 for the $\mathrm{\Lambda}$ (-0.64 for the $\mathrm{\bar{\Lambda}}$) \cite{trento}.

\subsection{How to Reconstruct the $\mathrm{\Lambda}$ Polarisation}
\label{howtoreco}
The polarisation is extracted from the reconstructed data by
considering the distribution of the decay particles according to Equation
\ref{eq:decdist}. The probability for a particle to fall within this
distribution over the entire interval is 100\%,
meaning that the probability density function should be 

\begin{equation}
A\int \frac{1}{4\pi}(1+\alpha P \cos\theta) d\cos\theta=1
\label{eq:dist}
\end{equation}
with A being a normalisation constant. Substituting $\cos\theta$ with
$x$ and performing the integration gives
A=$\mathrm{2\pi}$. Hence, the normalised distribution function becomes

\begin{equation}
f(x)=\frac{1}{2}(1+\alpha Px).
\label{eq:norm}
\end{equation}
Calculating the mean value of $x$ according to \cite{bevington} gives

\begin{equation}
<x>=\int_{-1}^1 xf(x)dx=\frac{1}{2}\int_{-1}^1 (x+\alpha P x^2)dx=\frac{\alpha P}{3}
\end{equation}
and substituting $x$ with $\cos\theta$ gives the relation 

\begin{equation}
P=\frac{3}{\alpha} <\cos\theta> \approx \frac{3}{\alpha}
\overline{\cos\theta} =\frac{3}{\alpha N}\sum_{i=1}^N \cos\theta
\label{eq:sum}
\end{equation}
with N being the number of reconstructed events. To get an estimate of the uncertainty in the reconstructed polarisation, the square root of the variance is calculated according to \cite{frodesen}

\begin{equation}
\frac{3}{\alpha}V(\overline{\cos\theta})=\left(\frac{1}{N}\right) ^2
V\left(\sum_{i=1}^N\cos\theta_i 
\right)=\frac{3}{\alpha}\frac{1}{N}V(\cos\theta)
\label{eq:mean}
\end{equation}
where $V(cos\theta)$ is the estimated value for the variance of the polarisation
distribution 

\begin{equation}
V(\cos\theta)=\frac{1}{N-1}\sum_{i=1}^N
\left(\cos\theta_i-\overline{\cos\theta}  \right)^2.
\label{eq:varpol}
\end{equation}

\section{The PANDA Software Frameworks}
PANDA has today two different software frameworks, one which is referred to as
the BaBar-like framework and the other referred to as the PandaROOT
framework. Both are C++ based software for Monte Carlo simulations,
event reconstruction and physics analysis for the PANDA experiment.\\
\indent
The BaBar-like framework was the first one available to the
collaboration and it was to a large extent inherited from the 
BaBar collaboration. It contains many well debugged libraries and
tools and has been successfully used in the BaBar
experiment for many years. It has been decided to use this
framework until the PANDA physics book is completed in 2008. After
that, PandaROOT will be the framework for Monte Carlo simulations and data
reconstruction.\\
\indent
The PandaROOT framework is an object-oriented ROOT-based framework
inspired by the CMS collaboration. It is fully
based on the use of C++ class libraries from the ROOT data analysis
framework \cite{root}. ROOT is a widely used software package in
nuclear and particle physics and thus has a large community of
developers which the PANDA community can profit from. Through the use
of ROOT, the PandaROOT framework implements the concept of Virtual
Monte Carlo (VMC) \cite{vmc}. VMC is a set of library classes which enables
the user to implement a particle physics detector simulation without
beforehand defining the particle transport code, such as Geant3\cite{geant3},
Geant4\cite{geant4} or Fluka\cite{fluka}, to be used.

\section{Reconstruction}
\subsection{Generation of Particles}
Events for the simulations have been produced by a modified generator
which was originally used at the PS185 experiment at LEAR. As of now,
it is an integrated part of the software framework and the information
on the differential cross-sections for $\mathrm{\bar{p}p\rightarrow
  \bar{\Lambda}\Lambda}$ production has, as well as polarisation and
spin correlations, been added.\\
\indent
The generator makes a call to the CERNLIB routine FOWL which then returns
the momentum vectors for the generated $\mathrm{\bar{\Lambda}\Lambda}$
in the CM system, according to a isotropic distribution. This
distribution is then adapted to the experimental differential cross
section shown in Figure~\ref{fig:lamframe}. The selection process is done by keeping $\Lambda$($\bar{\Lambda}$)-events with a probability given by the value of the angular distribution for this certain $\mathrm{\cos\theta}$, divided by the maximum value of the angular distribution.\\
\indent
The momentum of the generated anti-protons has been set to 1.64~GeV/c
as this was one of the energies used also for the PS185
experiment. The beam spread is set to 0.01\%.
The interaction point is set to (0,~0,~0), with a smearing in x- and
y-direction according to a Gaussian distribution centred around zero
and with a standard deviation of 1~mm, given by the pellet target. The smearing in the z-direction is confined to a circular region with a radius of 1~mm. The branching fraction of $\Lambda\rightarrow p\pi^-$ as well
as the polarisation have been set to 100\%. The transport code used for the p,
$\mathrm{\bar{p}}$, $\mathrm{\pi^+}$ and $\mathrm{\pi^-}$ to simulate
the interactions with the detector materials is Geant4.\\
\indent
The sub-detectors used for the reconstruction is the MVD, the STT, the
two MDCs (each having 8 layers) in the target spectrometer and the six
MDCs (each having 6 layers) in the forward spectrometer \cite{jan}. 198~000 pairs of $\mathrm{\bar{\Lambda}\Lambda}$ were generated.

\subsection{Angular Distribution}
\label{angdist}
\begin{figure}[H]
\centering
\includegraphics[bb=0 0 567 385,width=0.6\linewidth,angle=0]{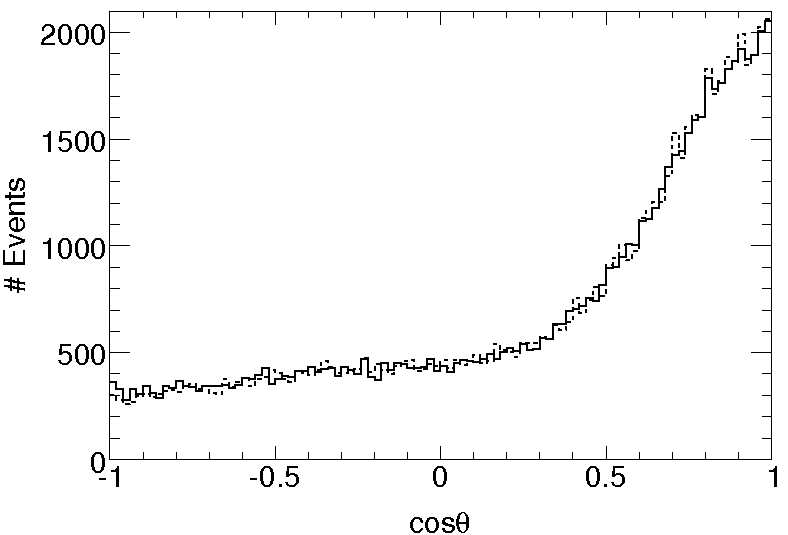}
\caption{The reconstructed $\cos\mathrm{\theta}$ distribution
  for the $\mathrm{\bar{\Lambda}}$ is shown with a full drawn line and
  the Monte Carlo truth for those particles with a dashed line.}
\label{fig:lamcos}
\end{figure}

\begin{figure}[H]
\centering
\includegraphics[bb=0 0 567 376,width=0.6\linewidth,angle=0]{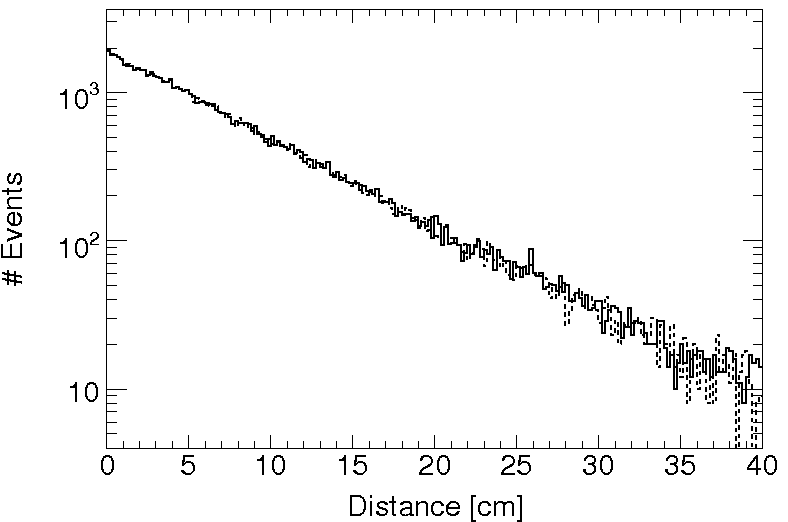}
\caption{The reconstructed c$\tau$ of the simulated
$\mathrm{\bar{\Lambda}}$ is shown with a full drawn line and the
  Monte Carlo truth of those events with a dashed line.}
\label{fig:ctau}
\end{figure}

The reconstructed distribution for $\cos\theta$ for the
$\mathrm{\bar{\Lambda}}$ in the CM frame together with the Monte Carlo truth for those events are shown in Figure~\ref{fig:lamcos}. There is an excellent agreement between the two curves.\\
\indent
The c$\mathrm{\tau}$ of the $\mathrm{\bar{\Lambda}}$ was also simulated and reconstructed. The result can be seen in Figure~\ref{fig:ctau}. For the reconstructed events, c$\mathrm{\tau}$=7.419$\pm$0.029~cm, while the Monte Carlo truth of those events gives c$\mathrm{\tau}$=7.396$\pm$0.029~cm. The Monte Carlo truth of all generated $\bar{\Lambda}$ yields a decay length of 7.862$\pm$0.018~cm. The established experimental value is 7.89~cm \cite{pdg}. It should be investigated why there is a 5~mm bias on the decay length of the $\mathrm{\bar{\Lambda}}$. \newpage

\subsection{Detector and Detection Efficiency}
\begin{figure}[H]
\centering
\subfigure{\includegraphics[bb=0 0 567 382,width=0.6\linewidth,angle=0]{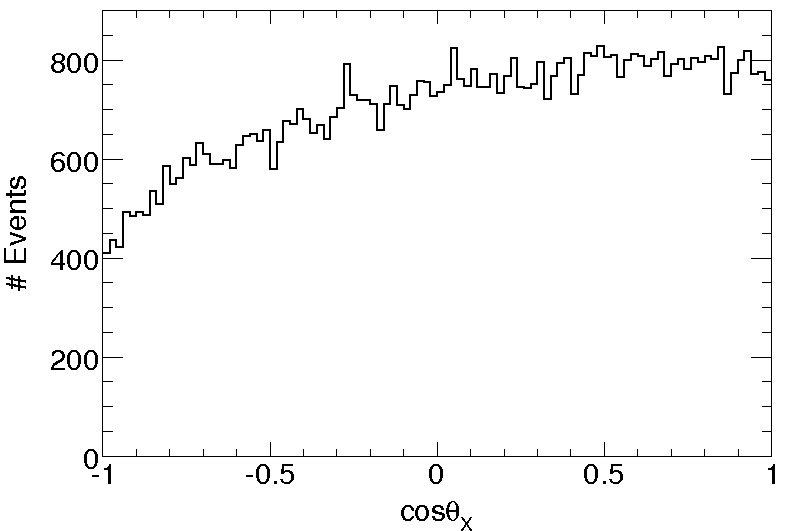}
\label{fig:xnon}}
\subfigure{\includegraphics[bb=0 0 567 383,width=0.6\linewidth,angle=0]{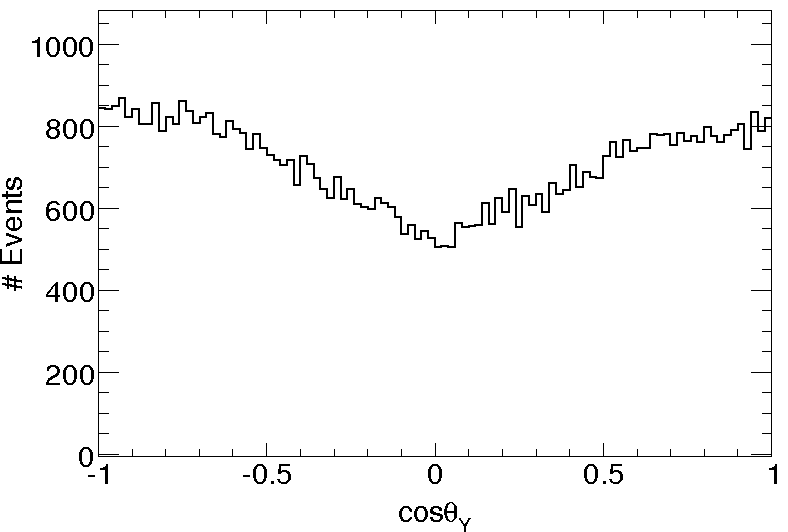}
\label{fig:ynon}}
\subfigure{\includegraphics[bb=0 0 567 383,width=0.6\linewidth,angle=0]{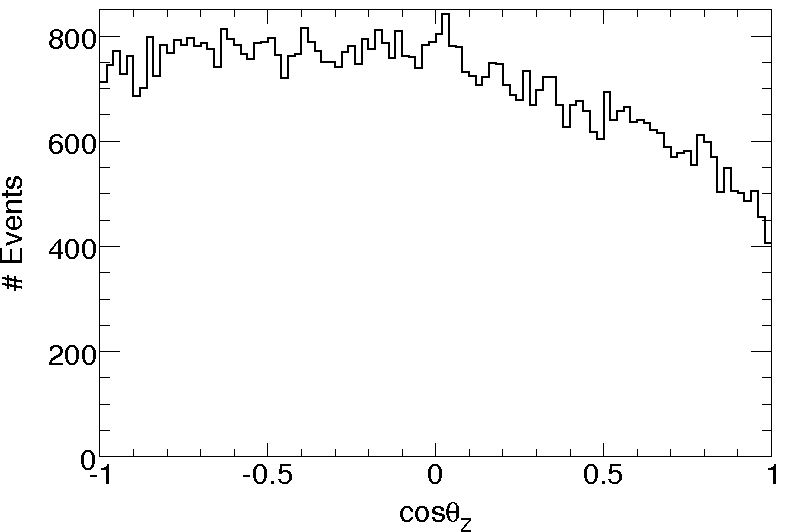}
\label{fig:znon}}
\caption{Reconstructed distributions for
  $\cos\theta_{\bar{p}}$ for the decay anti-proton with respect to the
  x-axis \ref{fig:xnon}, y-axis \ref{fig:ynon} and z-axis
  \ref{fig:znon} in Figure~\ref{fig:lamframe} for unpolarised $\mathrm{\bar{\Lambda}}$
  particles.} 
\label{fig:nonpol}
\end{figure}

We define the \emph{detection efficiency} by the ratio of the number of 
reconstructed events to the number of generated particles. Out of the
198 000 generated $\mathrm{\bar{\Lambda}\Lambda}$
pairs, 70 267 are reconstructed, translating into roughly 35\%.\\
\indent
The \emph{detector efficiency} reflects the ability of the detector to
correctly detect particles over the angular interval. From reconstructed
unpolarised $\mathrm{\Lambda}$ and $\mathrm{\bar{\Lambda}}$
particles, one can investigate the $\cos\theta_{p}$ distributions of the
daughter proton and anti-proton in x-, y-
and z directions defined earlier over the interval (-1,~1) in the
$\mathrm{\bar{\Lambda}}$ rest frame. For full detector efficiency,
these plots should be isotropic if there is no polarisation in any
direction. However, this is not the case, as can be seen in Figure
\ref{fig:nonpol}. 

\begin{figure}[H]
\centering
\subfigure{\includegraphics[bb=0 0 567 407,width=0.6\linewidth,angle=0]{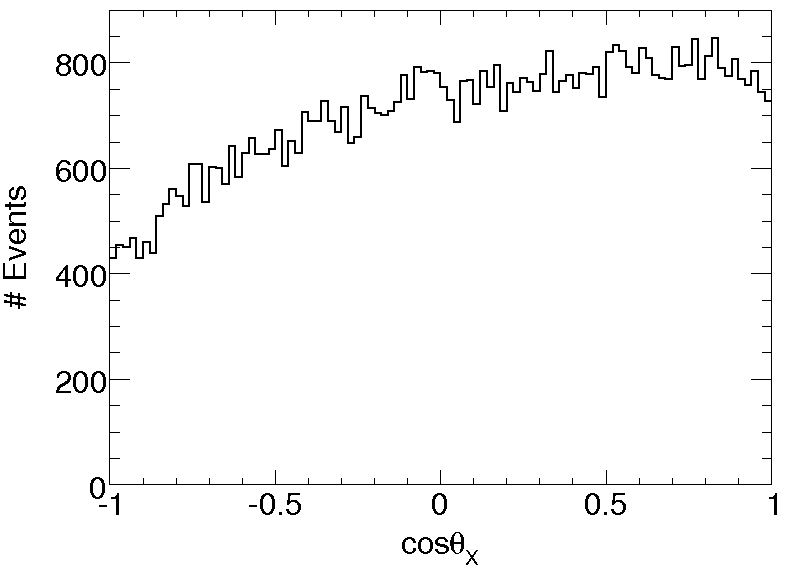}
  \label{fig:xpol}} 
\subfigure{\includegraphics[bb=0 0 567 382,width=0.6\linewidth,angle=0]{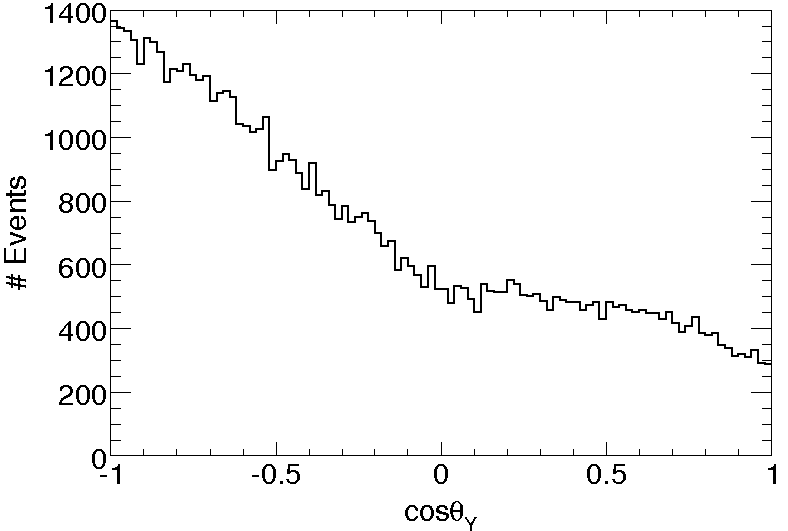}
  \label{fig:ypol}} 
\subfigure{\includegraphics[bb=0 0 567 389,width=0.6\linewidth,angle=0]{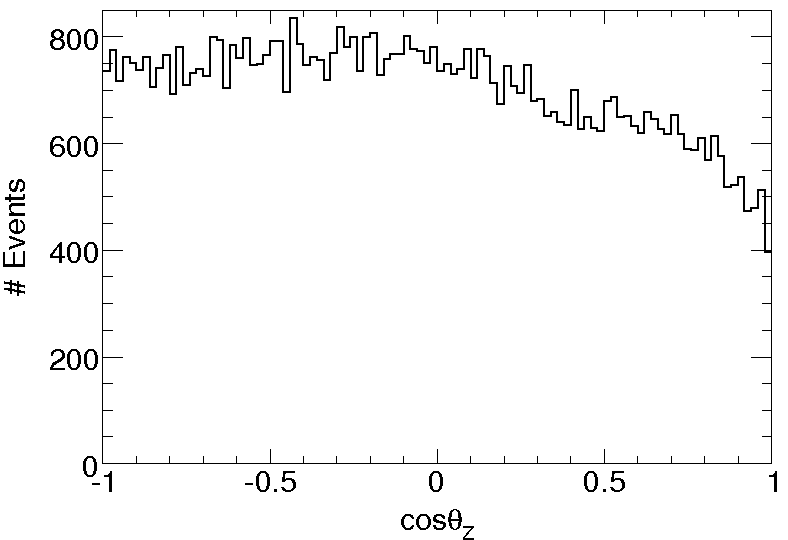}
  \label{fig:zpol}} 
\caption{Reconstructed distributions for
  $\cos\theta$ for the daughter anti-proton with respect
  to the x-axis \ref{fig:xnon}, y-axis \ref{fig:ynon} 
  and z-axis \ref{fig:znon} for polarised $\mathrm{\bar{\Lambda}}$ particles.}
\label{fig:pol}
\end{figure}

There is a lack of anti-protons in the backward x-direction in Figure
\ref{fig:xnon} and there are anti-protons missing 
in the $\cos\theta_y=0$ region of Figure~\ref{fig:ynon}, meaning
perpendicular to the y-axis. For the outgoing
$\mathrm{\bar{\Lambda}}$, the x-axis tends to 
point in the backward direction of the laboratory system (in the
direction of the negative z-values rather than the positive), as can
be seen in Figure~\ref{fig:lamframe}, even if it varies with every
event.\\ 
\indent
The missing anti-protons in Figure~\ref{fig:xnon} indicate that the
corresponding missing $\pi^+$ will go in the direction of
$\cos\theta_x$ being close to 1, meaning backwards in the laboratory
system. In the same way, from Figure~\ref{fig:ynon} one understands
that the missing $\pi^+$ may go either along the positive x-axis or in
the direction of the negative z-axis. In cases when the pions do go in these
directions, they will become slow because of the strong forward
boost due to the incoming anti-protons. If so, one would expect a dip
in the region of $\cos\theta$ approaching +1 in
Figure~\ref{fig:znon} and this is exactly what is seen.\\
 \indent
The corresponding distributions of reconstructed
$\cos\theta_{\mathrm{\bar{p}}}$ (see Figure~\ref{fig:lamframe}) in the case of polarised $\mathrm{\bar{\Lambda}}$ is seen in Figure~\ref{fig:pol}. 

\begin{figure}[H]
\centering
\includegraphics[bb=0 0 567 544,width=0.6\linewidth,angle=0]{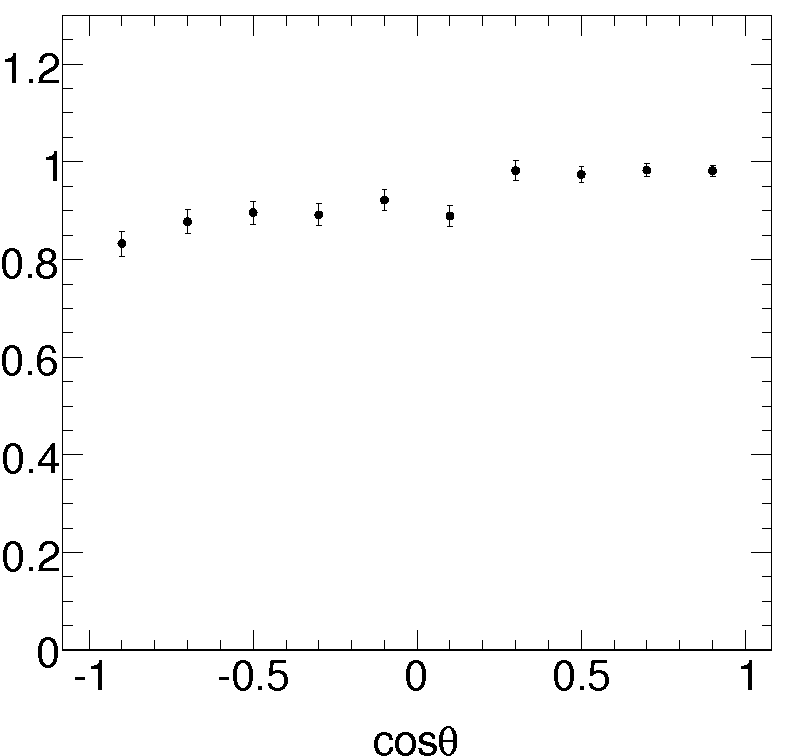}
\caption{The calculated polarisation as a function of
  $\cos\theta_{\bar{\Lambda}}$ in the CM system. }
\label{fig:Polbin}
\end{figure}

In Figure~\ref{fig:xpol} and Figure~\ref{fig:zpol} one would once again
expect a flat distribution and for Figure~\ref{fig:ypol} a slope
corresponding to the polarisation times the asymmetry parameter
$\mathrm{\alpha /2}$. However, one notices a lack of particles in the same regions
as in Figure~\ref{fig:nonpol}.\\
\indent
To further investigate the detector response in different angular
regions of the detector, one can plot the polarisation as a function of
$\cos\theta_{\bar{\Lambda}}$ in the CM system. For an isotropic
detector response, one would expect a constant polarisation independent of the $\mathrm{\theta_{\bar{\Lambda}}}$ angle as the polarisation is set to 100\%. As can be seen in Figure~\ref{fig:Polbin}, this is not quite the case.

\noindent
The reason why the distribution is not constant over the entire angular is not known and should be investigated further.

\subsubsection{Slow Pions}
\begin{figure}[H]
\centering
\includegraphics[bb=0 0 567 374,width=0.6\linewidth,angle=0]{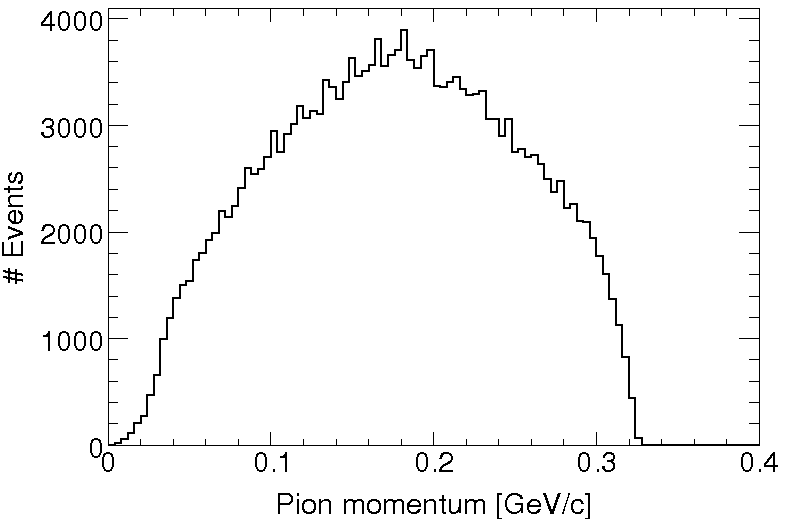}
\caption{Magnitude of the $\mathrm{\pi^+}$ momentum for
  the Monte Carlo truth of the generated particles.}
\label{fig:pimommc}
\end{figure}

\begin{figure}[H]
\centering
\includegraphics[bb=0 0 567 375,width=0.6\linewidth,angle=0]{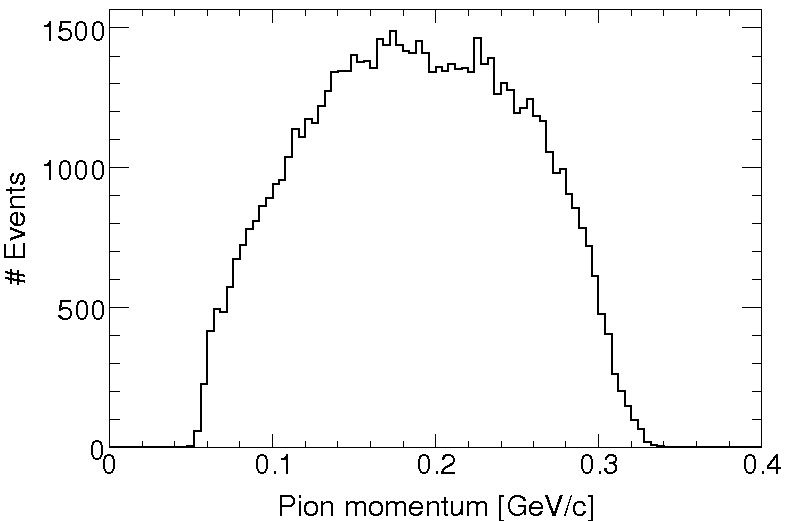}
\caption{Magnitude of the $\mathrm{\pi^+}$ momentum for
  the reconstructed particles.}
\label{fig:pimomreco}
\end{figure}

For a correct reconstruction of the $\mathrm{\Lambda}$ and
$\mathrm{\bar{\Lambda}}$, it
is important to correctly reconstruct the decay particles - the
proton and the pion. The pion is the most critical one, since its much
lower mass generally causes it to be emitted with a larger velocity
than the proton. If the pion goes backwards in the detector, seen in the laboratory system, it will become very slow because of the boost in the opposite direction and may even spiral in the detector due to the solenoid field, without
undergoing detection. There may also be difficulties with the software regarding handling of spiralling particles, resulting in fewer pions being reconstructed.\\
\indent
Figures \ref{fig:pimommc} and \ref{fig:pimomreco} show
histograms of the pion momentum from the generated Monte Carlo pions as well as of the reconstructed ones. The decay pions have momenta between 0 and approximately 0.3~GeV/c. However, only pions with momentum above 0.05~GeV/c are reconstructed. Consequently, events with such slow pions will be missed in Figure~\ref{fig:nonpol} and \ref{fig:pol}.
\newpage

Another way to study where the reconstruction of the pion fails
is to look at the the $\mathrm{\pi^+}$ angles in the laboratory system
and compare them to the reconstructed ones. Also here, one can see
that the angular distributions differ. 
\newpage

\begin{figure}[H]
\centering
\subfigure[Laboratory emission angle of the Monte Carlo $\pi^+$.]{\includegraphics[bb=0 0 568 381,width=0.6\linewidth,angle=0]{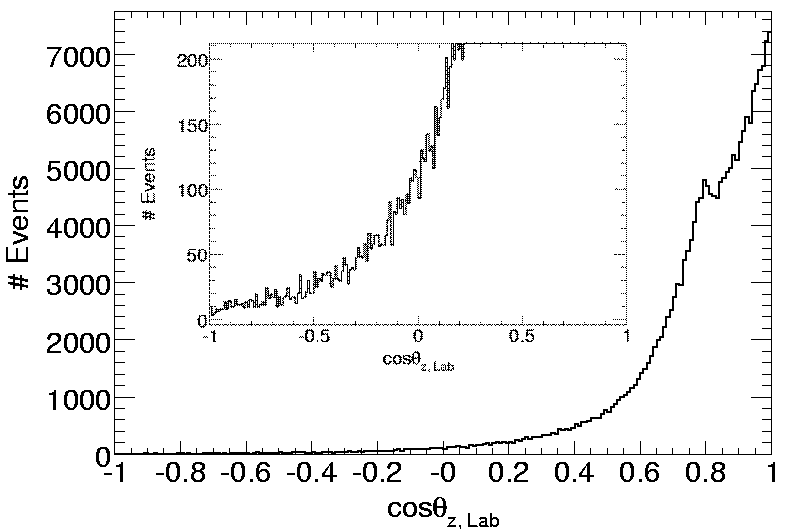}
  \label{fig:pilabmc}} 
\subfigure[Laboratory emission angle of the reconstructed $\pi^+$.]{\includegraphics[bb=0 0 568 383,width=0.6\linewidth,angle=0]{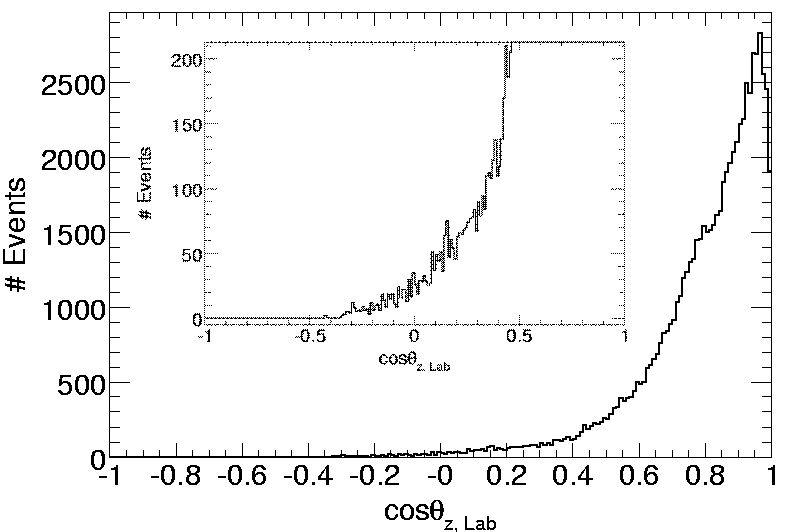}
  \label{fig:pilabreco}} 
\caption{Angle of the Monte Carlo truth and the reconstructed $\mathrm{\pi^+}$
  in the detector. The angle $\mathrm{\theta_{\bar{\Lambda}}}$ is measured in the
  laboratory system.} 
\label{fig:pi}
\end{figure}
\noindent
Zooming in in Figure~\ref{fig:pilabmc} and \ref{fig:pilabreco}, one
can see from Figure~\ref{fig:pi} that pions going backward
in the detector are not reconstructed, probably because they are
simply too slow to undergo reconstruction in the solenoid field. This
is also what was suggested in connection to figures \ref{fig:nonpol} and
\ref{fig:pol}. The peak close to $\cos\theta$=0.7 in \ref{fig:pilabmc}
comes from the case when the $\bar{\Lambda}$ goes in the direction of
$\cos\theta_{\bar{\Lambda}}$ close to 1. In this case, the pion can not
go backwards in the laboratory system. If it is emitted ``backwards''
in the $\bar{\Lambda}$ rest system, with respect to the flight
direction of the $\bar{\Lambda}$ in the laboratory system, there is a maximum
in its laboratory emission angle $\theta_{Lab}$. This is what is seen.

\subsection{The Polarisation}
\label{polarisation}
The polarisation and its errors were calculated using the formulae
given in section \ref{howtoreco}. The results on the reconstructed
polarisation are presented in the Table~\ref{tab:polreco}, while the Table~\ref{tab:polmc} contains the average polarisation calculated from the Monte Carlo truth of the reconstructed particles. The polarisation should be 100\% in y-direction and 0\% in x- and z-direction.

\begin{table}[htb]
\centering
\begin{tabular}{|l|c|c|c|}
\hline
  &x-direction &y-direction &z-direction \\
\hline
\hline
100\% pol  &-0.3511$\pm$9.9$\cdot10^{-3}$  &1.151$\pm$1.1$\cdot10^{-2}$
  &0.2546$\pm$9.9$\cdot10^{-3}$\\ 
\hline
0\% pol &-0.3445$\pm$9.9$\cdot10^{-3}$ &-0.066$\pm$1.1$\cdot10^{-2}$
  &0.2846$\pm$9.8$\cdot10^{-3}$\\ 
\hline
\end{tabular}
\caption{Reconstructed average polarisation for
  non-polarised and polarised $\mathrm{\bar{\Lambda}}$-particles, as
  well as calculated standard deviation for these values.}
\label{tab:polreco}
\end{table}

\begin{figure}[H]
\centering
\subfigure{\includegraphics[bb=0 0 567 544,width=0.5\linewidth,angle=0]{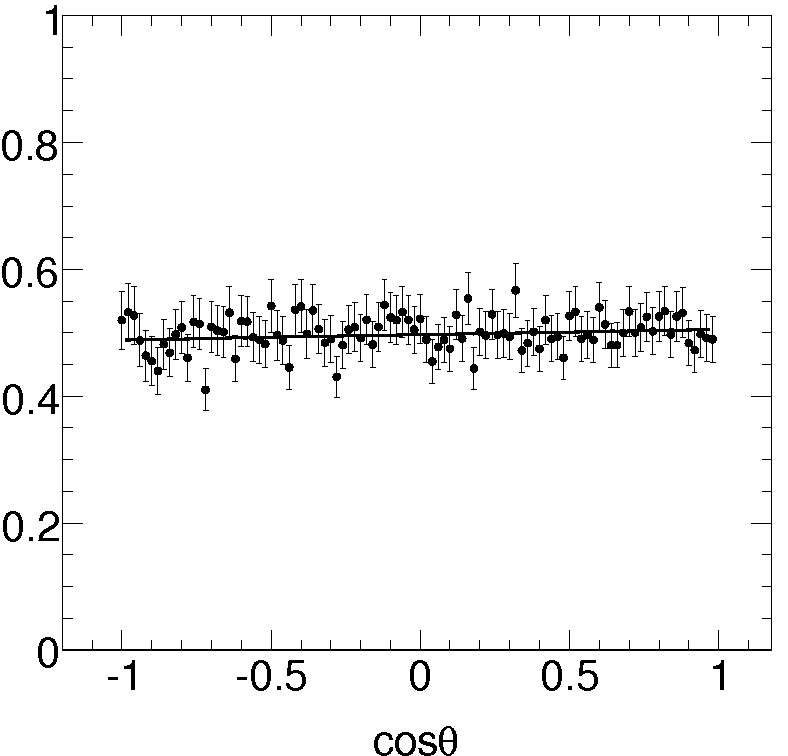}
\label{fig:effx}}
\subfigure{\includegraphics[bb=0 0 567 544,width=0.5\linewidth,angle=0]{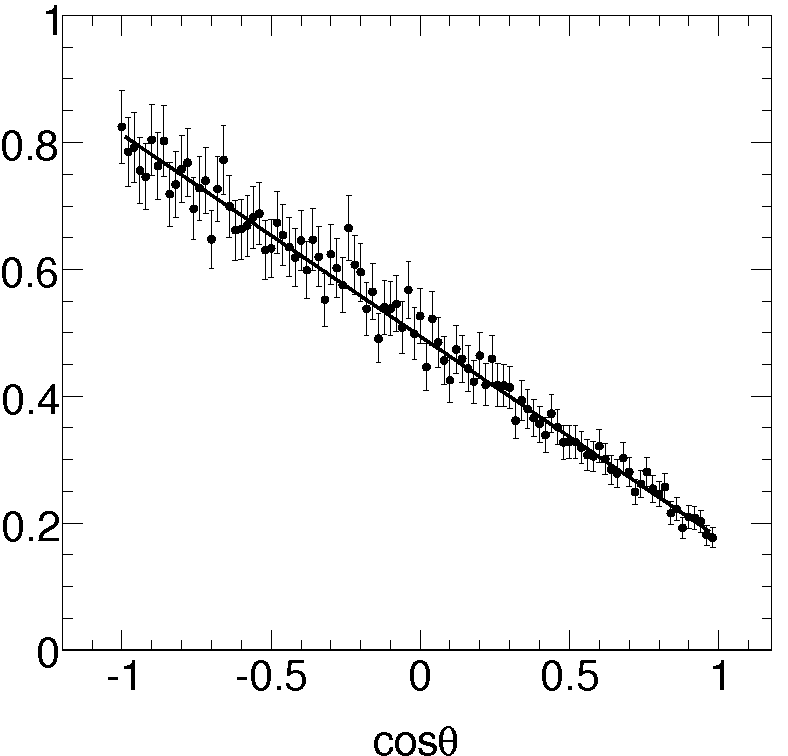}
\label{fig:effy}}
\subfigure{\includegraphics[bb=0 0 567 544,width=0.5\linewidth,angle=0]{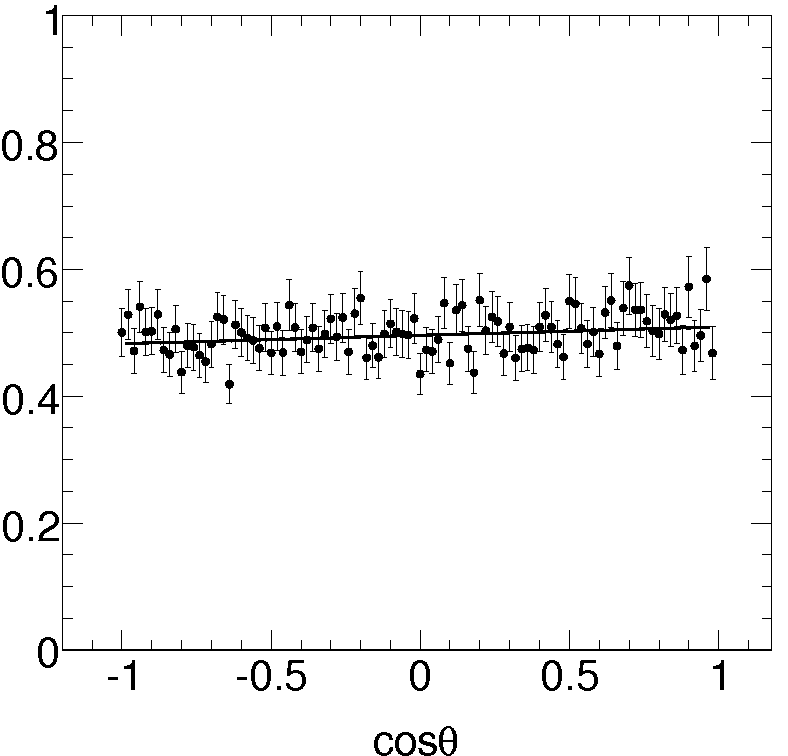}
\label{fig:effz}}
\caption{Efficiency calibrated $\cos\theta$
  distributions. The slopes of the fitted functions correspond to $\alpha P/2$.}
\label{fig:pol1}
\end{figure}

\begin{table}[htb]
\centering
\begin{tabular}{|l|c|c|c|}
\hline
  &x-direction &y-direction &z-direction \\
\hline
\hline
100\% pol  &-0.3368$\pm$9.9$\cdot10^{-3}$  &1.156$\pm$1.1$\cdot10^{-2}$
  &0.2546$\pm$9.9$\cdot10^{-3}$\\ 
\hline
0\% pol &-0.3311$\pm$9.9$\cdot10^{-3}$ &-0.068$\pm$1.1$\cdot10^{-2}$
  &0.2944$\pm$9.8$\cdot10^{-3}$\\ 
\hline
\end{tabular}
\caption{Polarisation of the Monte
  Carlo truth of the reconstructed particles in the cases of no and
  full polarisation, respectively. The calculated standard deviation
  for these values are also shown.}
\label{tab:polmc}
\end{table}
\noindent
One can see that the reconstructed polarisations agree quite well with the Monte Carlo truth for those particles, however it is far from being (0,~1,~0) in the x-, y- and z-directions. The polarisation even takes on unphysical values of having a magnitude larger than 1. Taking the statistical uncertainties into account does not reduce the discrepancy between the generated and reconstructed events enough, and the reason must therefore be caused by the detector response and reconstruction inefficiencies.\\
\indent
The polarisations can also be extracted from the plots of the
$\cos\theta$-distributions for the decay protons,
Figure~\ref{fig:pol}. A straight line fit over the interval should
have a slope of $\alpha P/2$ with $P$ being the polarisation
(here equal to 1 since it has been set to 100\% in the event generator). To get a realistic measure of the 
polarisation, the histograms should however first be
compensated for the acceptance of the detector and the reconstruction ability,
Figure~\ref{fig:nonpol}. Using the Monte Carlo truth of the
reconstructed non-polarised $\mathrm{\bar{\Lambda}}$ to calibrate for
the acceptance and efficiency gives the result seen in Figure~\ref{fig:pol1}. The uncertainties have been calculated from the error propagation formula, using an uncertainty in each bin of the histograms proportional to the square root of the number of events in that bin.\newpage

\noindent
The fitted values for the slopes obtained from the three plots in Figure
\ref{tab:effpol} are shown in Table~\ref{tab:effpol}.

\begin{table}[htb]
\centering
\begin{tabular}{|l|c|c|c|}
\hline
Slope, x-direction &Slope, y-direction &Slope, z-direction \\
\hline
\hline
0.0084$\pm$0.0066 &-0.3175$\pm$0.0058 &0.0133$\pm$0.0067\\
\hline
\end{tabular}
\caption{Calculated reconstructed average polarisation for
  polarised $\mathrm{\Lambda}$-particles. The values have been
  corrected for the detector response.}
\label{tab:effpol}
\end{table}
\noindent
The values from the slopes should be multiplied with 2/$\mathrm{\alpha}$ to get the polarisation, due to the factor 1/2 in Equation \ref{eq:norm}. Hence, one gets the polarisation multiplied with the asymmetry factor along the y-axis to be 0.63, leading to a polarisation of (99$\pm$1.8)\%. The polarisation along the x and the z-axis are (2.6$\pm$2.0)\% and (4.1$\pm$2.1)\% respectively. The values including errors are displayed in Table~\ref{tab:effpol}.

\subsection{Momentum and Vertex Reconstruction}
\begin{figure}[H]
\centering
\subfigure{\includegraphics[bb=0 0 567 383,width=0.6\linewidth,angle=0]{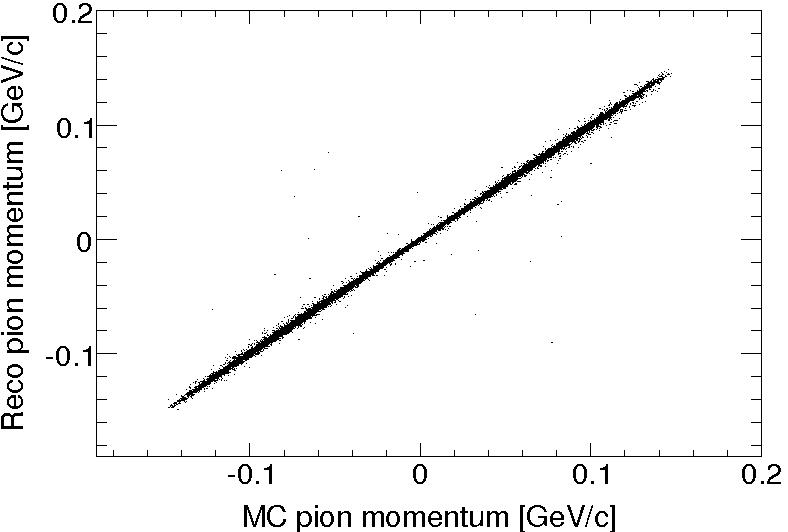}
  \label{fig:prx}} 
\subfigure{\includegraphics[bb=0 0 567 383,width=0.6\linewidth,angle=0]{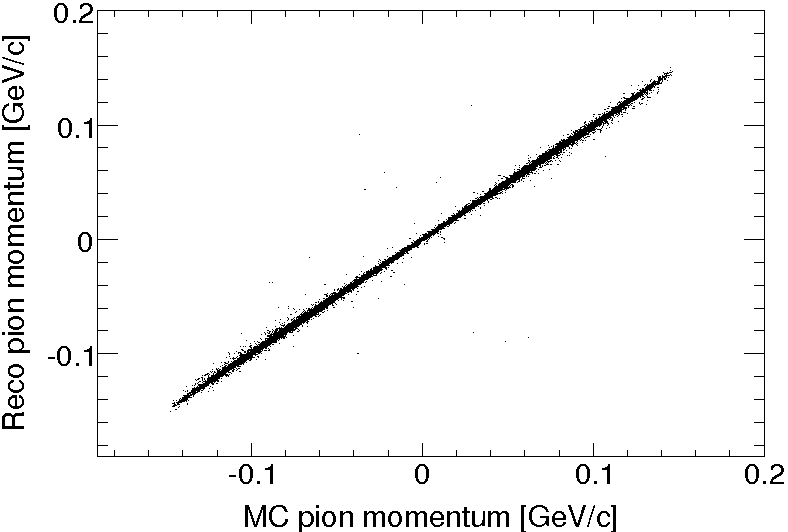}
  \label{fig:pry}} 
\subfigure{\includegraphics[bb=0 0 567 384,width=0.6\linewidth,angle=0]{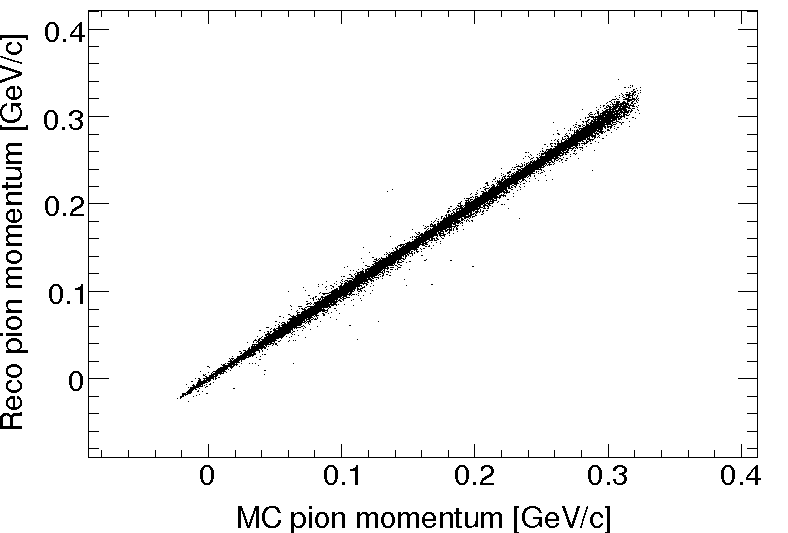}
  \label{fig:prz}} 
\caption{Reconstructed $\mathrm{\pi^+}$ momenta
  versus Monte Carlo momenta in the x-direction \ref{fig:pix},
  y-direction \ref{fig:piy} and z-direction \ref{fig:piz}.}
\label{fig:pirecvsmc}
\end{figure}

\begin{figure}[H]
\centering
\subfigure{\includegraphics[bb=0 0 567 376,width=0.6\linewidth,angle=0]{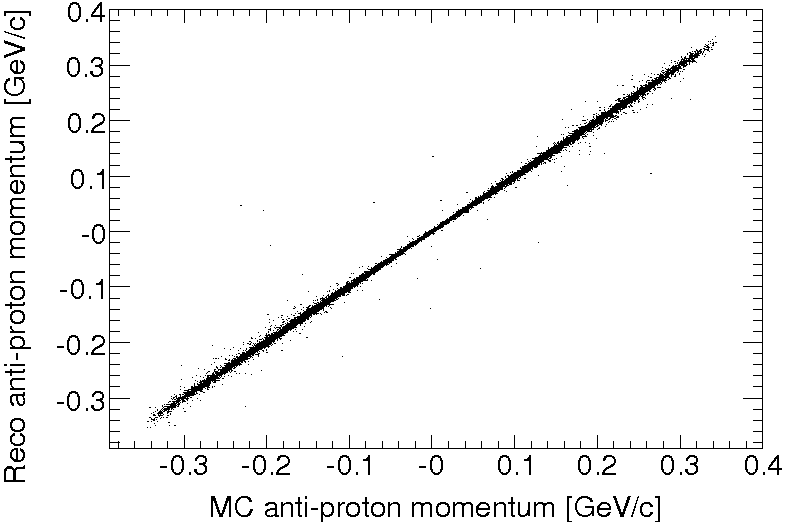}
  \label{fig:pix}} 
\subfigure{\includegraphics[bb=0 0 567 383,width=0.6\linewidth,angle=0]{recopionvsy.png}
  \label{fig:piy}} 
\subfigure{\includegraphics[bb=0 0 567 384,width=0.6\linewidth,angle=0]{recopionvsz.png}
  \label{fig:piz}} 
\caption{Reconstructed anti-proton momenta versus Monte Carlo momenta in
  the x-direction \ref{fig:prx}, y-direction \ref{fig:pry} and z-direction \ref{fig:prz}.}
\label{fig:prrecvsmc}
\end{figure}

\begin{figure}[H]
\centering
\subfigure{\includegraphics[bb=0 0 567 386,width=0.6\linewidth,angle=0]{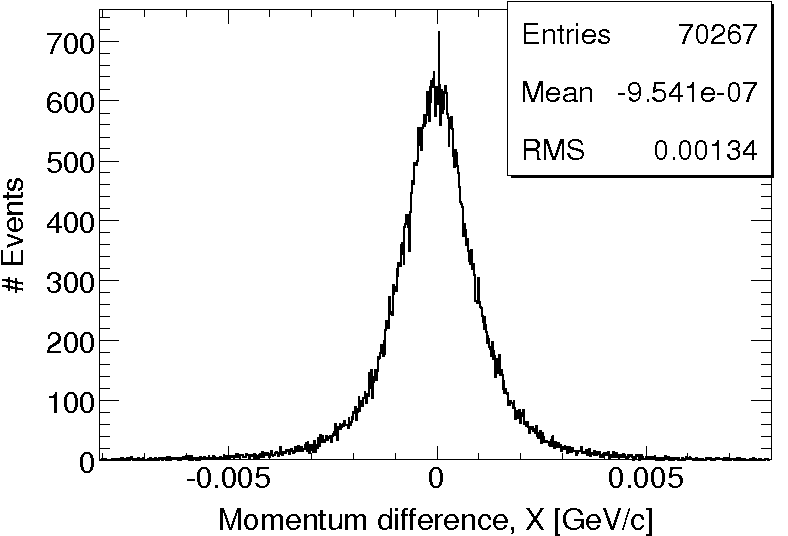}
  \label{fig:diffpix}} 
\subfigure{\includegraphics[bb=0 0 567 386,width=0.6\linewidth,angle=0]{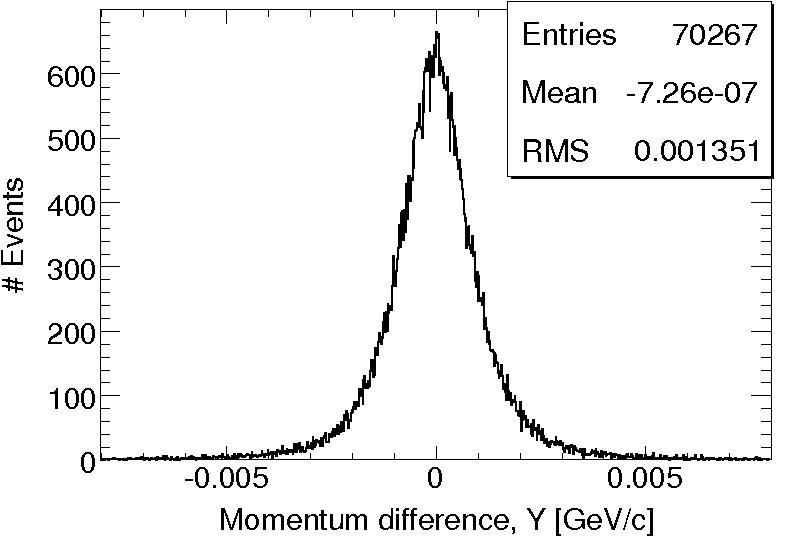}
  \label{fig:diffpiy}} 
\subfigure{\includegraphics[bb=0 0 567 387,width=0.6\linewidth,angle=0]{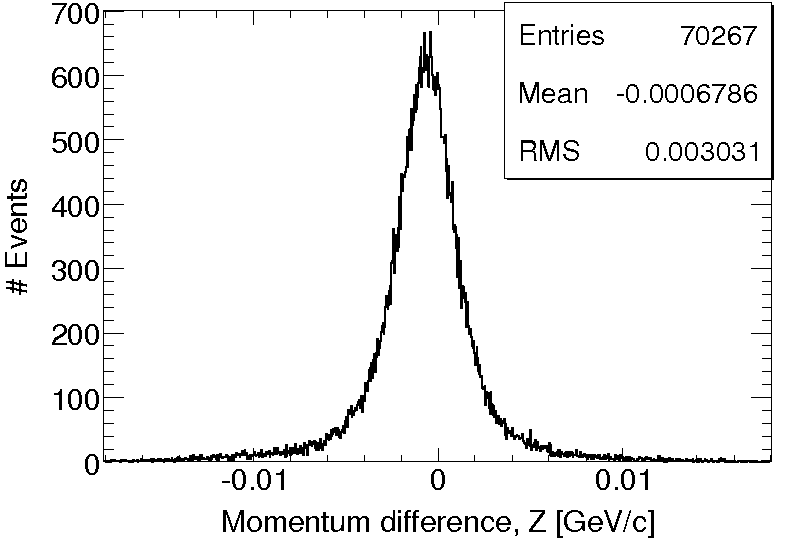}
  \label{fig:diffpiz}} 
\caption{The reconstructed momenta components minus the corresponding
  Monte Carlo truth for the $\pi^+$.}
\label{fig:diffpi}
\end{figure}

\begin{figure}[H]
\centering
\subfigure{\includegraphics[bb=0 0 567 387,width=0.6\linewidth,angle=0]{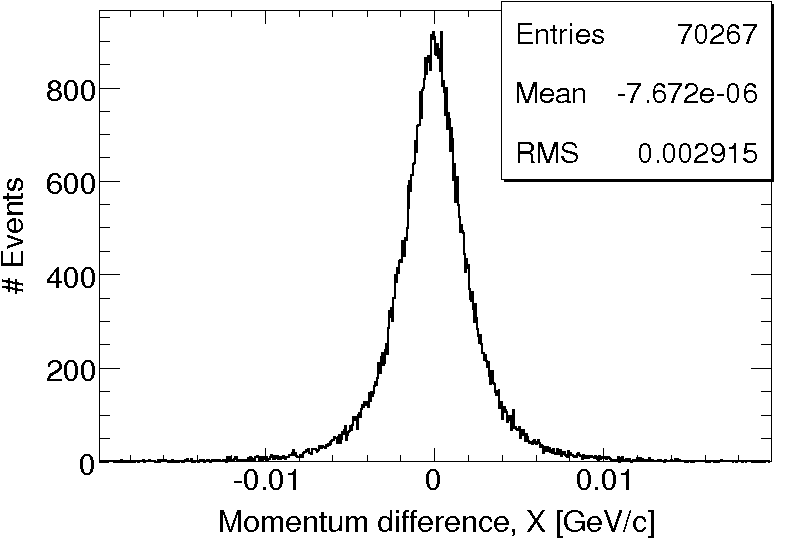}
  \label{fig:diffprx}} 
\subfigure{\includegraphics[bb=0 0 567 387,width=0.6\linewidth,angle=0]{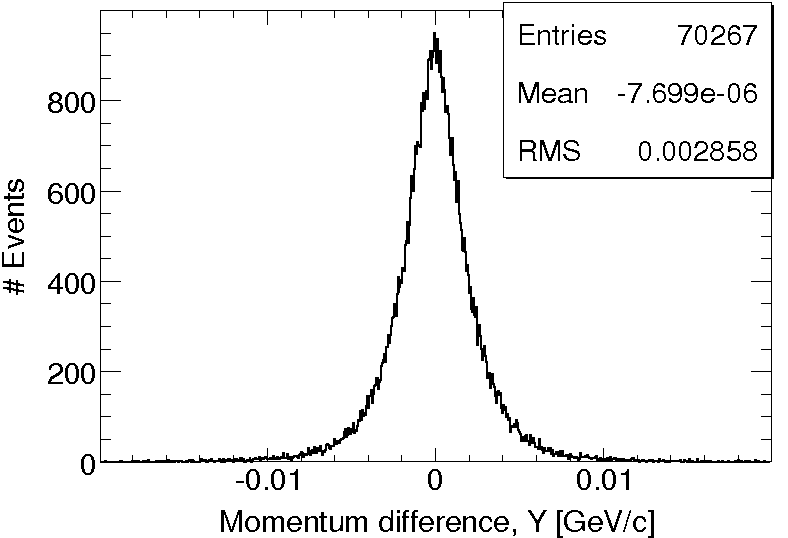}
  \label{fig:diffpry}} 
\subfigure{\includegraphics[bb=0 0 567 385,width=0.6\linewidth,angle=0]{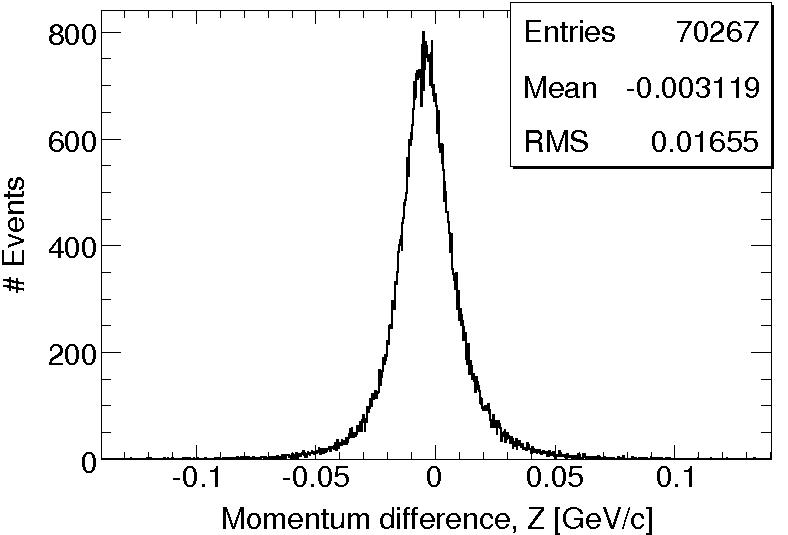}
  \label{fig:diffprz}} 
\caption{The reconstructed momenta components minus the corresponding
  Monte Carlo truth for the $\bar{p}$.}
\label{fig:diffpr}
\end{figure}

The momenta of the decay protons and pions must be well reconstructed
to correctly describe the hyperons. This is also related to how well
the decay point is reconstructed. As the previous study
unveiled, the reconstruction of the slow pions must be improved. An
important follow-up question is how well the reconstructed momenta of
the daughter particles describe their true momenta.\\
\indent
In the case of a reconstructed decay $\mathrm{\bar{p}p}$ system, the
reconstructed momentum for the decay protons and pions in x-, y- and
z-direction were plotted against the Monte Carlo truth for that
specific event. The plots showed a very good correspondence for both
pions and protons, as can be seen in Figures \ref{fig:pirecvsmc} and
\ref{fig:prrecvsmc}.\\
\indent
The differences between reconstructed and generated momenta were also
plotted to display how big the deviations were. The results are presented in Figure~\ref{fig:diffpi} and \ref{fig:diffpr} together with the root mean square (RMS) values.  
\newpage

\begin{figure}[H]
\centering
\subfigure{\includegraphics[bb=0 0 567 387,width=0.6\linewidth,angle=0]{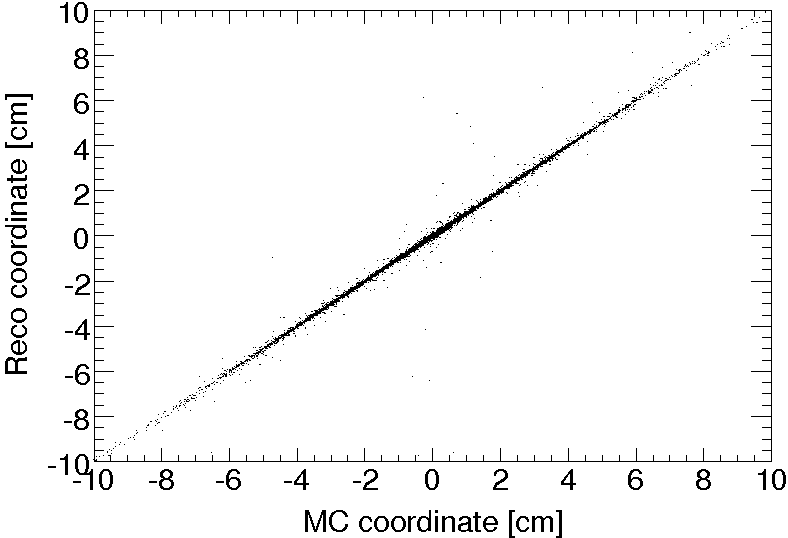}
\label{fig:vtxx}}
\subfigure{\includegraphics[bb=0 0 567 386,width=0.6\linewidth,angle=0]{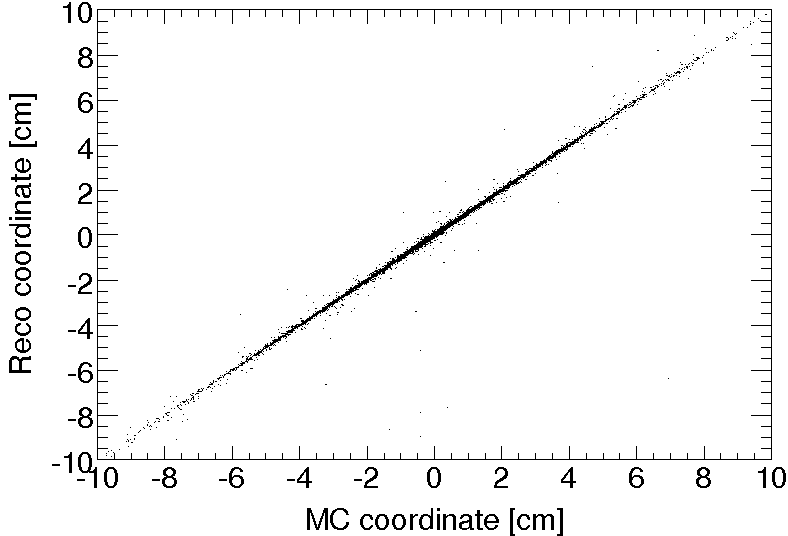}
\label{fig:vtxy}}
\subfigure{\includegraphics[bb=0 0 567 387,width=0.6\linewidth,angle=0]{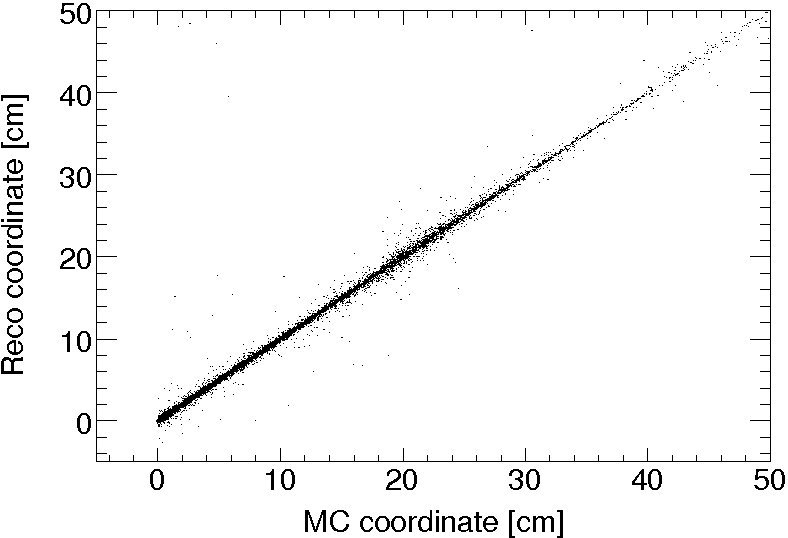}
\label{fig:vtxz}}
\caption{Reconstructed $\mathrm{\bar{\Lambda}}$ decay vertex coordinates
  versus the Monte Carlo coordinates in
  the x-direction \ref{fig:vtxx}, y-direction \ref{fig:vtxy} and
  z-direction \ref{fig:vtxz}.}
\label{fig:vtx}
\end{figure}

Also the reconstructed decay vertex of the $\mathrm{\bar{\Lambda}}$
agrees well with its generated values as can be seen in Figure~\ref{fig:vtx}.

\begin{figure}[H]
\centering
\subfigure{\includegraphics[bb=0 0 567 386,width=0.6\linewidth,angle=0]{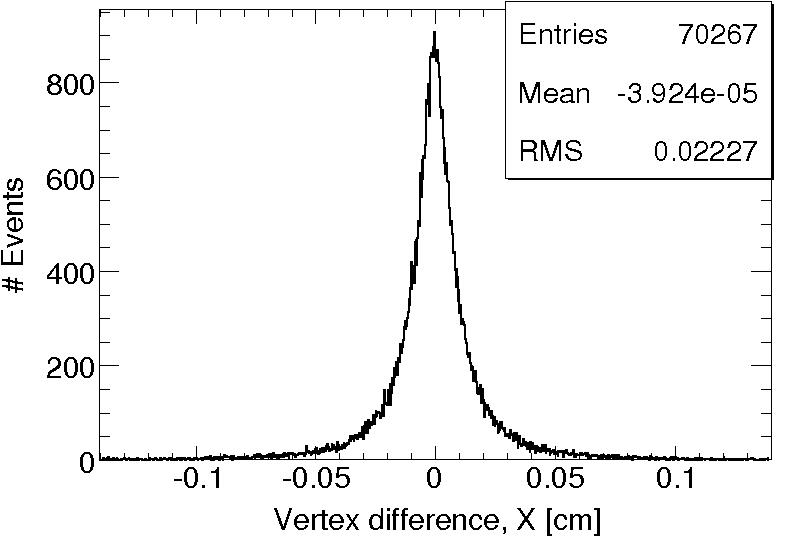}
\label{fig:diffvtxx}}
\subfigure{\includegraphics[bb=0 0 567 386,width=0.6\linewidth,angle=0]{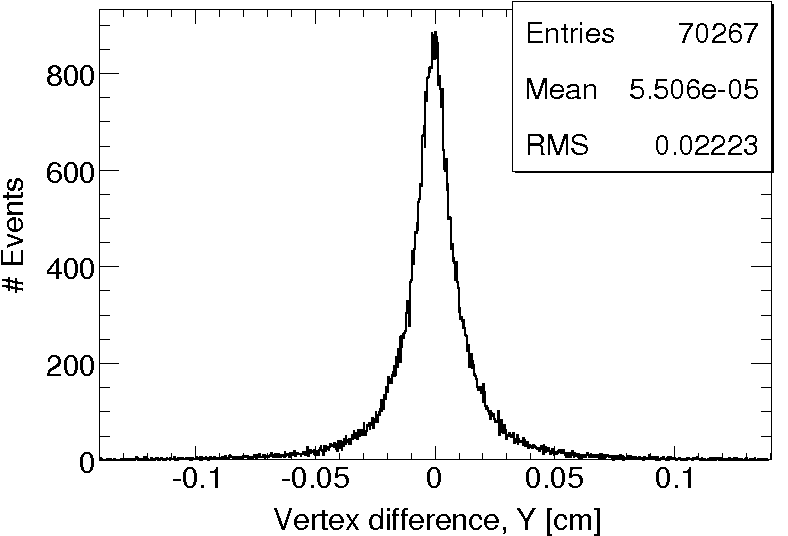}
\label{fig:diffvtxy}}
\subfigure{\includegraphics[bb=0 0 567 387,width=0.6\linewidth,angle=0]{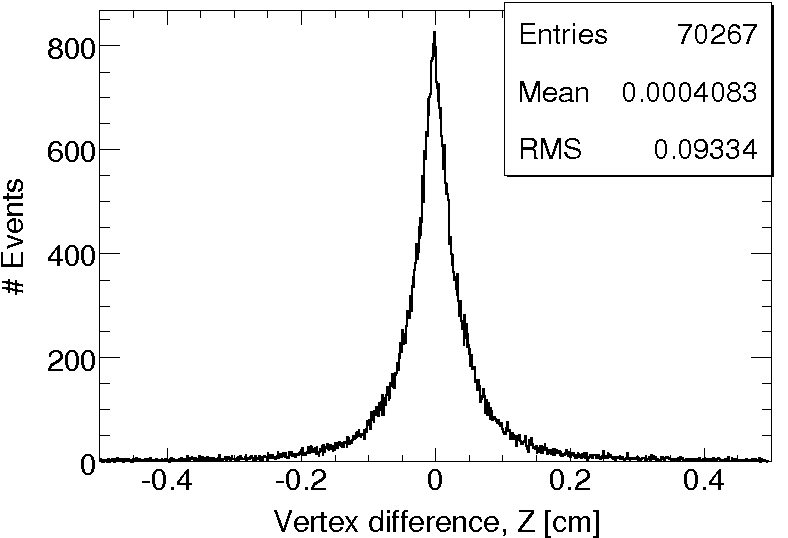}
\label{fig:diffvtxz}}
\caption{The reconstructed vertex components minus the corresponding
  Monte Carlo truth for the $\bar{\Lambda}$.}
\label{fig:diffvtx}
\end{figure}

The difference between the reconstructed vertex coordinates and the
Monte Carlo truth is shown in Figure~\ref{fig:diffvtx}.

\section{Conclusions from the Simulation Results}
The $\bar{p}p \rightarrow \bar{\Lambda}\Lambda$ reaction is
reconstructable in the BaBar-like framework of PANDA. However, there
are obvious problems with reconstructing slow pions coming from the
decay of the hyperons. This will affect the reconstruction of the
hyperon polarisation, as can be seen in the non-isotropic
distributions of $\mathrm{\cos\theta_{\bar{p}}}$ for the anti-protons coming from the unpolarised $\mathrm{\bar{\Lambda}}$ particles. As a consequence, one
must take the detector response into account in order to obtain the correct polarisation. If this is done, a polarisation similar to the Monte Carlo truth is obtained. More detailed studies are necessary to investigate how much of this is due
to the detector response and the software reconstruction respectively.\\
\indent
In those cases when the daughter pion and proton are
reconstructed, the life time and decay vertex of the decaying hyperon
look very similar to the Monte Carlo truth of those events. Also the
reconstructed momentum of the daughter particles correspond well to
the Monte Carlo truth.

\chapter{Conclusion and Outlook}
\label{consim}
\section{Conclusion}
This licenciate thesis has treated the topics of energy resolution and
light yield uniformity measurements of PWO crystals for PANDA, as well
as software reconstruction of the $\bar{p}p \rightarrow
\bar{\Lambda}\Lambda$ reaction including polarisation.\\
\indent
Two energy resolution measurements were done using a 3$\times$3 array of non-tapered crystals, cooled to -15~$^{\circ}$C and -25~$^{\circ}$C, respectively. The tagged photon beam of energies between 19 and 56~MeV at MAX-Lab was used. The energy contributions from all nine crystals were summed and the resulting peaks were fitted with Gaussian distributions to extract the resolution. The relative energy resolution for the September measurements was 0.12 at a photon energy of 18.9~MeV 0.072 at 51.6~MeV. The measured relative energy resolution, expressed as the standard deviation $\sigma$ of the Gaussian distribution divided by the photon energy E was fitted with the parametrisation $\sigma(E)/E=a/\sqrt{E} \oplus b/E \oplus c$. For both the April and the September measurement, the energy resolution expression contained a negative square of the Poisson parameter, $a$. This has not a physical interpretation, but might be understood in terms of the large correlations with the other two parameters. After imposing the condition that the $a$-parameter should correspond to a light yield of 50~phe/MeV, an energy resolution of $\sigma/E= 0.45\%/\sqrt{E_{GeV}} \oplus 0.18\%/E_{GeV}$ $\oplus 8.63\%$ was obtained for the April measurement. For the September data the result became $\sigma/E= 0.45\%/\sqrt{E_{GeV}} \oplus 0.21\%/E_{GeV}$ $\oplus 6.12\%$. Upon combining data points from the September measurement with eight data points for energies ranging between 64 and 715~MeV, it was seen that the energy resolutions agree very well in the overlapping range of 50-70 MeV. The fit over the entire energy interval becomes positive and is given by $\sigma /E= 1.6\% /\sqrt{E_{GeV}}$ $\oplus 0.095\% /E_{GeV} \oplus 2.1\%$. However, there is a very large correlation (anti-correlation) between all three parameters. For the two low energy measurements at MAX-Lab, the magnitude of these correlations are close to 100\%. For the combined fit, they are slightly lower between the parameters $a$ and $c$ as well as for $b$ and $c$. It seems as the standard parameterisation is not ideal in describing the energy resolution at low energies or in explaining it in terms of something that has physical relevance.\\
\indent
The light uniformity measurements were performed with two non-tapered and three tapered PWO crystals and a $^{22}$Na source. Two different crystal wrappings were used: white Teflon covered with aluminium foil and mirror-like VM2000. Pulse height spectra were recorded at different source positions along the crystals and the number of photo electrons per unit energy was determined as a function of interaction point in the crystal. For the tapered crystals the light collection yield increased with the distance between the interaction point and the PM tube. For the non-tapered crystals the light collection yield peaked somewhere close to the PM tube. The shape of the responses was qualitatively the same irrespective of which wrapping was used, but using VM2000 gave a 17\% higher overall light output. Black tape was put on different positions on one of the tapered crystals in an attempt to make the light yield more uniform. Placing black absorbing tape t one or two of the lateral surfaces close to the short end of the crystal decreased the increase of light yield in this region, however, at the expense of an overall decrease in light yield.\\
\indent
A simulation of the $\bar{p}p \rightarrow \bar{\Lambda}\Lambda$
reaction at a momentum of 1.64~GeV/c was done in the BaBar-like software
Framework for PANDA. 198~000 $\mathrm{\bar{\Lambda}\Lambda}$ pairs
were created using a modified generator originally made for the
PS185 experiment at LEAR, CERN. The reconstruction efficiency was found to be
about 35\%. The angular distribution of the $\mathrm{\bar{\Lambda}}$
could be reconstructed correctly. Figures of the reconstructed versus
Monte Carlo momenta of the decay pion and anti-proton agree well, as
does the vertex position of the decaying hyperon. It was discovered that slow
pions are difficult to reconstruct as they tend to spiral in the
solenoid field of the detector without being seen by the
sub-detectors. At 100\% polarisation perpendicular to the scattering
plane and without compensating for the angular detector efficiency,
the reconstructed polarisation of the $\mathrm{\bar{\Lambda}}$ does not reproduce the correct value in the direction of the true polarisation ($\hat{y}$) and is significantly different from 0 in the direction of the outgoing hyperon ($\hat{z}$) and as well along the third axis ($\hat{x}$). By correcting data with the results from unpolarised events, a polarisation which is (2.6$\pm$2.0)\%, (99$\pm$1.8)\% and (4.1$\pm$2.1)\% along $\hat{x}$, $\hat{y}$ and $\hat{z}$, respectively, is obtained.

\section{Outlook}
The next step in energy resolution measurements with PANDA PWO
crystals is to use a 5$\times$5 array of tapered crystals shaped for
the forward end-cap and cool it to -25~$^{\circ}$C to repeat the
measurements described in this thesis. This will ensure that the
electromagnetic showers are contained inside the crystal array, and a
better energy resolution is expected.\\
\indent
Regarding the light yield uniformity measurements, it would be interesting to
develop a masking technique for improved uniformity than just
placing arbitrary amounts of tape on different sides of the
crystals. It is important to develop a technique which is
applicable for larger amounts of crystals and which does not require
individual adaptations for each crystal.\\
\indent
For the simulations of the hyperons the next step is to repeat the
same steps at higher momenta of the incoming anti-protons and to
investigate the influence from background reactions, such as elastic
scattering and $\mathrm{\bar{p}p\rightarrow \bar{p}p\pi^+\pi^-}$. Hyperon channels with photons among the end products, such as $\Sigma^0$ and $\Lambda$(1405), will later also be studied to connect the hyperon channels with the electromagnetic calorimeter.

\chapter*{Acknowledgement}
Firstly I would like to thank my two supervisors professor Tord Johansson at the Department for Physics and Astronomy in Uppsala and professor Per-Erik Tegn\'{e}r at the Department of Physics at Stockholms University for their time, support and useful comments. They have both contributed with valuable ideas on how to make this work meaningful. Tord Johansson has been very helpful in pointing out areas of interest in the hyperon field and in putting the different parts of this thesis together. I am also very happy about the hands-on help I have received from Per-Erik Tegn\'{e}r during all sorts of crystal measurements as well as the patience he has shown me when it has been time to discuss the analysis.\\
\indent
Further, I would like to thank the staff at MAX-Lab in Lund for making things run smoothly during the crystal measurements; Bent Schr\"{o}der, Kurt Hansen, Pawel Golubev, Lennart Isaksson and the accelerator crew. A special thanks goes out to Lennart Isaksson for helping me sort out access to all the data. Also in connection to this, I would like to thank Linda Karlsson for keeping suck good track of the temperature during the April measurements and the two visiting students Jan Schulze and J\"{o}rn Becker for great company in Lund as well as for help with the electronical set-up.\\ 
\indent
I would also like to direct a ``thank you'' to the Bochum group in the PANDA collaboration for being helpful when it comes to generating files for the simulations. Especially, Jan Zhong deserves a big medal for being more than helpful at all times of both the day and the year.\\ 
\indent
Another thanks goes out to my office mate Erik Thom\'e and his long lasting patience and for reading this thesis at an early stage. Thanks for contributing with good company and interesting discussions on anything and everything at all times. Even though you may not always seem to have very strong opinions on different topics, I know you can surprise.\\
\indent
Finally I would like to thank Oscar St\aa l who also found the time to read through this thesis and comment on it, as well as anyone else who has contributed to this work by putting up with my occasional bad moods, cheered me up or helped me formulate tricky paragraphs. Or all three of them.\\
\indent
Thanks to all my colleagues at the department for a pleasant working environment and for help with providing or solving cross words during the coffee breaks! Even if you most likely blame me when things go wrong in the latter case...

\end{document}